\documentclass[a4paper, 12pt]{amsart}

\usepackage{review}
\usepackage[T1]{fontenc}
\usepackage[utf8]{inputenc}
\usepackage{hyperref}
\begin{document}

\title{Maxima of log-correlated fields: some recent developments}

\dedicatory{We dedicate this review to the memory of Fritz Haake, who did so much to understand and explain the connections between Random Matrix Theory and Quantum Chaos, and whose friendship one of us (JPK) was greatly fortunate to enjoy.} 

\abstract
We review recent progress relating to the extreme value statistics of the characteristic polynomials of random matrices associated with the classical compact groups, and of the Riemann zeta-function and other $L$-functions, in the context of the general theory of logarithmically-correlated Gaussian fields.  In particular, we focus on developments related to the conjectures of Fyodorov \& Keating concerning the extreme value statistics, moments of moments, connections to Gaussian Multiplicative Chaos, and explicit formulae derived from the theory of symmetric functions. 
\endabstract

\author{E. C. Bailey}
\email{e.c.bailey@bristol.ac.uk}
\address{School of Mathematics, University of Bristol, Bristol, BS8 1UG, United Kingdom and Heilbronn Institute for Mathematical Research, Bristol, United Kingdom}

\author{J. P. Keating}
\email{keating@maths.ox.ac.uk}
\address{Mathematical Institute, University of Oxford, Oxford, OX2 6GG, United Kingdom}

\maketitle

\tableofcontents


\section{Introduction}\label{sec:introduction}

Our aim in this review is to survey some recent lines of research motivated by conjectures of Fyodorov, Hiary, and Keating~\cite{fyohiakea12} and Fyodorov and Keating~\cite{fyokea14}.  These conjectures relate to the extreme value statistics of the characteristic polynomials of random matrices associated with the classical compact groups, and to the extreme value statistics of the Riemann zeta-function and other $L$-functions.  The past decade has seen considerable progress towards proving these conjectures, although many aspects remain to be understood. Much of this recent progress has been stimulated by establishing connections with ideas and techniques that sit at the interface between statistical mechanics, mathematics physics, probability, combinatorics, and number theory.  For example, links with Gaussian Multiplicative Chaos, the general theory of logarithmically-correlated Gaussian fields, the theory of symmetric functions, and methods from analytic number theory have played an important role.  We have endeavoured to make this overview as accessible as possible and so have attempted to introduce the wide range of ideas and techniques at the level of a colloquium, rather than focusing on the technical details.  

This section introduces the relevant theory so that this review can be as self-contained as possible.  Section~\ref{sec:fk_conj} presents the three main conjectures and discusses progress towards resolving the first two of them.  Different approaches have been taken to understand the third conjecture, and these are the subject of section~\ref{sec:moments}. 

\subsection{Characteristic polynomials of random matrices}

\begin{figure}[!htb]
  \centering
  \begin{tikzpicture}
    \begin{axis}[
        axis lines=center,
        xmin=-1.2,
        xmax=1.2,
        ymin=-1.2,
        ymax=1.2,
        grid=none,
        xtick={-1,1},
        xticklabels={,,},
        ytick={-1,1},
        yticklabels={,,},
        axis equal image
      ]
	  \addplot[
        only marks,
        scatter,
        mark=*,
        mark size=1.5pt,
        scatter src=explicit,
        scatter/use mapped color={fill=black},
      ]
	  table[
        x expr=\thisrowno{0},
        y expr=\thisrowno{1},
        meta expr=\thisrowno{2}
      ]
	  {unitary2_eig.txt};
    \end{axis}
  \end{tikzpicture}
  \caption{Plot of the eigenvalues of a random $50 \times 50$ unitary matrix.}\label{fig:50_unitary}
\end{figure}
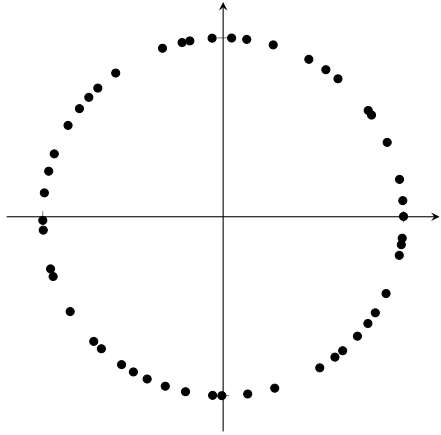

Throughout this review, we will focus on \textit{random matrices} that are drawn from one of the classical compact groups uniformly with respect to the associated Haar measure. The classical compact groups are: the $N\times N$ \textit{unitary} matrices,
\[\U(N)\coloneqq \{A\in \operatorname{GL}(N, \mathbb{C}) : \overline{A}^TA=A\overline{A}^T=I\}, \]
the $N\times N$ \textit{orthogonal} matrices,
\[\operatorname{O}(N)\coloneqq \{A\in \operatorname {GL}(N, \mathbb{R}): AA^T=A^TA=I\},\]
and the $2N \times 2N$ \textit{symplectic} matrices,
\[\Sp(2N)\coloneqq\{A\in \U(2N): A \Omega A^T=\Omega\},\]
where $\Omega$ is the skew-symmetric block matrix
\begin{equation}\label{omega_matrix}
  \Omega \coloneqq
  \begin{pmatrix}
    0 & I_N\\
    -I_N & 0
  \end{pmatrix}.
\end{equation}
The \emph{special orthogonal} subgroup $\SO(N)$ is the set of orthogonal matrices with unit determinant.  The coset\footnote{Clearly, $\operatorname{O}^{-}(N)$ is not a group.} $\operatorname{O}^{-}(N)$ contains the orthogonal matrices with determinant $-1$.  We will focus mainly on random unitary matrices as a canonical example, denoting by $A^*\equiv \overline{A}^T$ the conjugate transpose of $A$. For a more general account of these topics, and many more, see~\cite{mehta04, abd11, meckes19}.  A key property of matrices from the classical compact groups is that their eigenvalues lie on the unit circle in the complex plane, as illustrated in figure~\ref{fig:50_unitary}.

We introduce the following general notation.  Let $A\in G(N)$ be a matrix from one of groups $G(N)\in\{\U(N), \Sp(2N), \SO(2N)\}$. We write for its characteristic polynomial
\begin{equation}\label{def:char_poly}
  P_{G(N)}(A,\theta)\coloneqq \det(I-Ae^{-i\theta}),
\end{equation}
Since the eigenvalues of $A$ lie on the unit circle, we write the polynomial variable as $e^{-i\theta}$ and consider real $\theta$.  Furthermore, if $A\in \Sp(2N)$ or $A\in \SO(2N)$ then, whilst the matrix size is $2N$, the eigenvalues of $A$ come in $N$ complex conjugate pairs, hence the $N$ in the subscript rather than $2N$. Whenever we focus on one particular group, for simplicity we will often omit the label, writing $P_N\equiv P_{G(N)}$.

Since $\U(N)$, $\Sp(2N)$, and $\SO(2N)$ are compact Lie groups, one can endow each with a translation-invariant measure, namely a \textit{Haar} measure (for numerous constructions of the measure see~\cite{meckes19}).  Taking $\U(N)$ together with its Haar measure corresponds to the \textit{Circular Unitary Ensemble}, or $\CUE$.  This ensemble was first introduced by Dyson~\cite{dys62}, together with the Circular Orthogonal and Circular Symplectic Ensembles, $\COE$ and $\CSE$ respectively\footnote{It is important to emphasize that, unlike with the $\CUE$, the $\COE$ and $\CSE$ are not simply $\operatorname{O}(N)$ or $\Sp(2N)$ together with their respective Haar measures. See, for example, Mehta~\cite{mehta04}.}. There is also a more general notion of the Circular $\beta$ ensemble, cf.~\eqref{eq:cbe}.

One way to express the Haar measure $\mu_{\text{Haar}}$ on a particular compact group is via the explicit formulae of Weyl~\cite{weyl46}.  We use the unitary group as an example. Firstly, we define a \textit{class function} on $A\in \U(N)$ by the property that $f(A)\coloneqq f(\theta_1,\dots,\theta_N)$ (where the $\theta_j$ are the eigenphases of $A$) is symmetric in all of its variables.  Then,
\begin{equation}\label{weyl_int_formula}
  \int_{\U(N)}f(A)d\mu_{\text{Haar}}(A) = \frac{1}{(2\pi)^N N!}\int_0^{2\pi}\cdots \int_0^{2\pi}f(\theta_1,\dots,\theta_N)\left|\Delta(e^{i\theta_1},\dots,e^{i\theta_N})\right|^2d\theta_1\cdots d\theta_N,
\end{equation}
where $\Delta(z_1,\dots,z_N)$ is the Vandermonde determinant\footnote{By convention, $\Delta(z_1)=1$.} $\prod_{1\leq i<j\leq N}(z_j-z_i)$. The proof relies on the fact that translation-invariance of the Haar measure also implies invariance under similarity transformations, i.e.
\begin{equation}\label{eq:haar_simple}
  d\mu_{\text{Haar}}(A)=d\mu_{\text{Haar}}(U A U^*)
\end{equation}
for $U\in \U(N)$. As unitary matrices are diagonalizable by such a transformation, integrals with respect to $d\mu_{\text{Haar}}(A)$ can be expressed as integrals over the eigenphases of $A$. Similar formulae hold for symplectic and orthogonal matrices, see Weyl~\cite{weyl46}. Henceforth, for ease of notation, we write
\begin{equation}
  dA\coloneqq d\mu_{\text{Haar}}(A)
\end{equation}
whenever the context allows. 

Moments of characteristic polynomials will form a centrepiece of this review.  For a fixed $\theta\in[0,2\pi)$, the $2\beta$th moments of $P_N\equiv P_{\U(N)}$ is
\begin{equation}\label{eq:unitary_moment}
  M_N(\beta)\coloneqq \int_{\U(N)}|P_N(A,\theta)|^{2\beta}dA,
\end{equation}
where to ensure integrability $\RE(2\beta)>-1$. Due to the rotational invariance of the Haar measure on $\U(N)$, $M_N(\beta)$ is independent\footnote{Such a statement is not true of averages over $\Sp(2N)$ or $\SO(2N)$.} of $\theta$. By the Weyl integration formula \eqref{weyl_int_formula}, the $2\beta$th moment is equivalent to
\begin{equation}\label{eq:moment_weyl_form}
  M_N(\beta)=\frac{1}{(2\pi)^NN!}\int_0^{2\pi}\cdots \int_0^{2\pi}\prod_{j=1}^N\left|1-e^{i(\theta_j-\theta)}\right|^{2\beta}|\Delta(e^{i\theta_1},\dots,e^{i\theta_N})|^2d\theta_1\cdots d\theta_N.
\end{equation}
This integral can be evaluated~\cite{bakfor97, keasna00a}, giving the following. 

\begin{theorem}\label{thm:ks}
  Let $A\in \U(N)$ and $\RE(2\beta)>-1$.  Then
  \begin{equation}\label{eq:thmks}
    M_N(\beta)=\prod_{j=1}^N\frac{\Gamma(j)\Gamma(j+2\beta)}{(\Gamma(j+\beta))^2},
  \end{equation}
  where $\Gamma(z)$ is the usual extension of the factorial function. 
\end{theorem}

Before a brief discussion of proof of theorem~\ref{thm:ks}, we make a few comments. Firstly, note that \eqref{eq:thmks} clearly has an analytic continuation in $\beta$ to the rest of the complex plane. Further, as $N\rightarrow\infty$,
\begin{equation}\label{eq:ks_asympt}
  M_N(\beta)\sim c_U(\beta)N^{\beta^2}
\end{equation}
where 
\begin{equation}\label{unitary_loc}
  c_U(\beta)=\frac{\mathcal{G}^2(1+\beta)}{\mathcal{G}(1+2\beta)}
\end{equation}
and $\mathcal{G}(z)$ is the Barnes $\mathcal{G}$-function (so $\mathcal{G}(z+1)=\Gamma(z)\mathcal{G}(z)$, and $\mathcal{G}(1)=1$)~\cite{keasna00a}.  When $\beta$ is an integer, the statement of theorem~\ref{thm:ks} reads 
\begin{equation}\label{eq:ks-moment-int}
  M_N(\beta)=\prod_{0\leq i<j\leq \beta-1}\left(\frac{N}{i+j+1}+1\right)\sim c_U(\beta)N^{\beta^2},
\end{equation}
and the leading order coefficient simplifies to
\begin{equation}
  c_U(\beta)\coloneqq\prod_{j=0}^{\beta-1}\frac{j!}{(j+\beta)!}.
\end{equation}
By~\eqref{eq:ks-moment-int}, one observes that for $\beta\in\mathbb{N}$, $M_N(\beta)$ is a polynomial in $N$ of degree $\beta^2$. 
    
The key tool in the proof of theorem~\ref{thm:ks} is the celebrated Selberg integral, see for example~\cite{mehta04}.  
\begin{theorem}[Selberg's Integral~\cite{mehta04}]
  Let $n\in\mathbb{N}$. Take $a, b ,\alpha, \beta, \gamma\in\mathbb{C}$ with $\RE(a),\RE(b), \RE(\alpha),\RE(\beta)$ all strictly positive, $\RE(\alpha+\beta)>1$, and
  \begin{equation}
    -\frac{1}{n}<\RE(\gamma)<\min\left(\frac{\RE(\alpha)}{n-1},\frac{\RE(\beta)}{n-1},\frac{\RE(\alpha+\beta+1)}{2(n-1)}\right).
  \end{equation}
  Then
  \begin{align}
    J(a,b,\alpha,\beta,\gamma,n)&\coloneqq \int_{-\infty}^\infty\cdots \int_{\infty}^\infty |\Delta(x_1,\dots,x_n)|^{2\gamma}\prod_{j=1}^n(a+ix_j)^{-\alpha}(b-ix_j)^{-\beta}dx_1\cdots dx_n\\
    &=\frac{(2\pi)^n}{(a+b)^{(\alpha+\beta)n-\gamma n(n-1)-n}}\prod_{j=0}^{n-1}\frac{\Gamma(1+\gamma+j\gamma)\Gamma(\alpha+\beta-(n-1+j)\gamma-1)}{\Gamma(1+\gamma)\Gamma(\alpha-j\gamma)\Gamma(\beta-j\gamma)}.\label{selberg_formula}
  \end{align}
\end{theorem}
    
One motivation for computing $M_N(\beta)$ was to establish the value distribution of the real and imaginary parts of the logarithm of $P_N(A,\theta)$~\cite{bakfor97, keasna00a}. Notice that $M_N(\beta)$ is the generating function\footnote{The moment generating function for the imaginary part $\exp(it\IM(\log P_N(A,\theta)))$ can be similarly constructed, and one can write down the generating function of the joint moments, see~\cite{keasna00a}.} for the moments of $\log|P_N(A,\theta)|$. As $N\rightarrow\infty$, that the real and imaginary parts tend independently to Gaussian random variables:
\begin{theorem}[Central Limit Theorem~\cite{keasna00a}]\label{thm:ks-clt}
  Take any rectangle $B\subset\mathbb{C}$.  For fixed $\theta\in[0,2\pi)$, 
    \begin{equation}\label{eq:ks_clt}
      \lim_{N\rightarrow\infty}\operatorname{meas}\left\{A\in \U(N):\frac{\log P_N(A,\theta)}{\sqrt{\frac{1}{2}\log N}}\in B\right\}=\frac{1}{2\pi}\int\int_Be^{-\frac{1}{2}(x^2+y^2)}dxdy,
    \end{equation}
    where the measure of the set is taken to be the usual Haar measure on $\U(N)$. 
\end{theorem}
\sloppy Hence, on average $\log|P_N(A,\theta)|\sim \sqrt{(1/2)\log N}$.  Both the Gaussian nature of $\log|P_N(A,\theta)|$ and the scaling in theorem~\ref{thm:ks-clt} will be important for the remainder of this review, particularly in section~\ref{sec:rmt_and_nt}.  Large deviations are a focal point of section~\ref{sec:fk_progress}. See also~\eqref{eq:ks_clt_largedeviation} and the surrounding discussion for results concerning large deviations of~\eqref{eq:ks_clt}.

\subsection{The Riemann zeta function}\label{sec:nt_intro}

One of the main applications of random matrix theory over the past twenty years has been to number theory.  We now review the relevant number theoretic concepts, beginning with the definition of one of the central functions of number theory.
\begin{definition}[Riemann zeta function]\label{zeta}
  Let $s\in\mathbb{C}$ with $\RE(s)>1$.  Then the Riemann zeta function is defined by
  \[\zeta(s)\coloneqq\sum_{n=1}^\infty\frac{1}{n^s}.\]
\end{definition}

In general, expressions of the type
\begin{equation}\label{eq:dirichlet_series}
  \sum_{n=1}^\infty \frac{a_n}{n^s}
\end{equation}
for $s, a_n\in\mathbb{C}$ are known as Dirichlet series. Hence, $\zeta(s)$ can be defined by a  Dirichlet series with $a_n=(1,1,1,\dots)$. Equivalently, $\zeta(s)$ can also be expressed as a product over primes, known as an Euler product,
\begin{equation}\label{zeta_euler_prod}
  \zeta(s)=\prod_p\left(1-\frac{1}{p^s}\right)^{-1},
\end{equation}
also for $\RE(s)>1$. Products of the form appearing in \eqref{zeta_euler_prod} are over primes $p$, unless otherwise explicitly stated.  The equality between the Dirichlet series and the Euler product formulation follows from the fundamental theorem of arithmetic. 

One can analytically continue $\zeta(s)$ to all of the complex plane, with the exception of a simple pole at $s=1$ (many proofs of this fact can be found, for example, in~\cite{edwards74}).  A consequence of the meromorphic continuation is the functional equation,
\begin{equation}\label{zeta_functional_equation}
  \zeta(s)=2^s\pi^{s-1}\sin\left(\tfrac{\pi s}{2}\right)\Gamma(1-s)\zeta(1-s).
\end{equation}

Using \eqref{zeta_functional_equation}, it is easy to see that there are zeros at the negative even integers coming from the sine function. These are known as \emph{trivial} zeros.  The functional equation also implies symmetries for the remaining \textit{non-trivial} zeros.  Denoting by $\rho_n$ any zero of $\zeta(s)$ other than those at the negative even integers (i.e.~$\rho_n$ are the non-trivial zeros), then $\rho_n$ must lie in the \emph{critical strip}, $0<\operatorname{Re}(s)<1$. Additionally, if $\rho_n$ is a non-trivial zero of $\zeta(s)$, then so are $1-\rho_n$, $\overline{\rho}_n$, and $1-\overline{\rho}_n$.

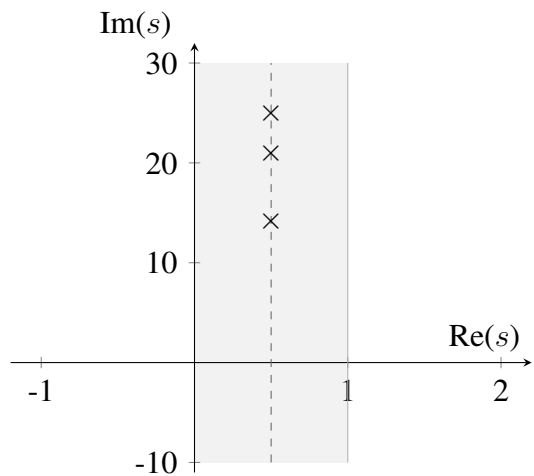
\begin{figure}
  \centering
  \begin{tikzpicture}
    \begin{axis}[
            axis lines=center,
            xmin=-1.2,
            xmax=2.2,
            ymin=-1.1,
            ymax=3.2,
            grid=none,
            y label style={at={(axis description cs:0.24,1.1)},anchor=north},
            xlabel={{Re($s$)}},
            ylabel={{Im($s$)}},
            xtick={-1,1,2},
            xticklabels={-1,1,2},
            ytick={-1,1,2,3},
            yticklabels={-10,10,20,30}
      ]
      \begin{pgfonlayer}{background}
        \fill[color=black!5] (axis cs:0,-1) rectangle (axis cs:1,3);
      \end{pgfonlayer}
      \addplot[color=black!30] coordinates {(1,-1) (1,3)};
      \addplot[color=black!60,dashed] coordinates {(0.5,-1) (0.5,3)};
      \node at (axis cs:0.5,1.4134725) {$\times$};
      \node at (axis cs:0.5,2.102204) {$\times$};
      \node at (axis cs:0.5,2.5010858) {$\times$};
    \end{axis}
  \end{tikzpicture}
  \caption[Plot showing the critical strip for $\zeta(s)$ and the first few non-trivial zeros.]{The critical strip for $\zeta(s)$ and the first three (positive) non-trivial zeros.  All non-trivial zeros of $\zeta(s)$ lie within the shaded region (i.e. $0<\RE(s)<1$), and the Riemann hypothesis states that all should lie on the dashed line (i.e. $\RE(s)=1/2$).}\label{fig:crit_strip}
\end{figure}

The Riemann hypothesis states that the non-trivial zeros lie in the centre of the critical strip, on the \emph{critical line} as illustrated in figure~\ref{fig:crit_strip}.

\begin{conj}[Riemann hypothesis]
  $\operatorname{Re}(\rho_n)=\frac{1}{2} \mbox{   } \forall  \mbox{  } n$. 
\end{conj}

The proportion of non-trivial Riemann zeros that lie on the critical line is currently known to be at least $0.41729$ (`more than $5/12$')~\cite{przz20}.  This improves a result of Conrey~\cite{con89}, which gave `more than $2/5$' of zeros on the critical line, and the subsequent improvement of Bui et al.~\cite{buiconyou11} to `more than 41\%'.

Moving beyond the non-trivial zeros of $\zeta(s)$, a related  question is the size of $\zeta(1/2+it)$ for $t\in\mathbb{R}$, either on average or the exceptionally large values. Selberg proved the following central limit theorem for $\zeta(s)$~\cite{sel46}.

\begin{theorem}[Selberg's Central Limit theorem~\cite{sel46}]\label{thm:selbergclt}
  For any rectangle $B\subset\mathbb{C}$,
  \[\lim_{T\rightarrow\infty}\frac{1}{T}\meas\left\{T\leq t\leq 2T:\frac{\log \zeta(\tfrac{1}{2}+it)}{\sqrt{\frac{1}{2}\log\log \tfrac{t}{2\pi}}}\in B\right\}=\frac{1}{2\pi}\int\int_Be^{-\frac{1}{2}(x^2+y^2)}dxdy.\]
\end{theorem}
Thus, both the real part and the imaginary part of the logarithm of the zeta function independently tend to a Gaussian random variable.   Note that this means that the typical size of $\log|\zeta(1/2+it)|$ is $O(\sqrt{\log\log t})$; the Lindel\"of hypothesis states that $\zeta(1/2+it)=o(t^\vareps)$, for any $\vareps>0$.  The Riemann hypothesis implies the Lindel\"of hypothesis. The similarities between \eqref{eq:ks_clt} and theorem~\ref{thm:selbergclt} are noteworthy.

More generally, one is interested in determining both the size of the moments of $\zeta(s)$ over long stretches of the critical line\footnote{Notice that the Lindel\"of hypothesis is equivalent to the statement that the moments in \eqref{eq:zeta_moments} are $o(T^{\vareps})$ for all $\vareps>0$ and $\beta\in\mathbb{N}$. A more precise form of such a moment conjecture is discussed in section~\ref{sec:rmt_and_nt}, see in particular conjecture~\ref{conj:zeta_moments}.},
\begin{equation}\label{eq:zeta_moments}
  \frac{1}{T}\int_0^T|\zeta(\tfrac{1}{2}+it)|^{2\beta}dt,
\end{equation}
and the extreme values of $\zeta(1/2+it)$. Such questions have motivated much of the research we shall review here.

There are natural generalizations of the Riemann zeta function known as $L$-functions. These will feature in section~\ref{sec:mom_other}. Most broadly, an $L$-function is a Dirichlet series with an Euler product and a functional equation (cf.~\eqref{eq:dirichlet_series}, \eqref{zeta_euler_prod}, and \eqref{zeta_functional_equation}). In order to keep this review self-contained we focus on two canonical extensions.  For a comprehensive account, we refer to~\cite{iwanieckowalski04}.

Clearly, the Riemann zeta function is an $L$-function. The simplest extension to $\zeta(s)$ is the Dirichlet $L$-function for the non-trivial character of conductor $3$, defined as follows.
\begin{definition}\label{def:simplest_dirichlet}
  Define $\chi_{-3}:\mathbb{N}\rightarrow\mathbb{C}$ by
  \begin{equation}\label{eq:simplest_dirichlet}
    \chi_{-3}(n)\coloneqq
    \begin{cases}
      0 & \text{if }n\equiv 0\mod{3}\\
      1 & \text{if }n\equiv 1\mod{3} \\
      -1 & \text{if }n\equiv -1\mod{3}.
    \end{cases}
  \end{equation}
  This is a periodic function with period $3$. The Dirichlet $L$-function corresponding to $\chi_{-3}$ is\footnote{The reason for using $-3$ rather than $3$ in the notation will become apparent shortly.}
  \begin{equation}
    L(s,\chi_{-3})\coloneqq\sum_{n=1}^\infty\frac{\chi_{-3}(n)}{n^s}=1-\frac{1}{2^s}+\frac{1}{4^s}-\frac{1}{5^s}+\cdots
  \end{equation}
   convergent for $\RE(s)>1$. Like $\zeta(s)$ (see \eqref{zeta_euler_prod} and \eqref{zeta_functional_equation}), $L(s,\chi_{-3})$ has an Euler product and a functional equation (see for example~\cite{iwanieckowalski04}).
\end{definition}
One can generalize definition~\ref{def:simplest_dirichlet} to moduli other than $3$. Take an integer $d$ such that
\begin{equation}\label{eq:fund_disc}
  d=
  \begin{cases}
    k & \text{if }k\equiv 1 \mod 4, k\text{ square-free, or},\\
    4m & \text{if }m\equiv 2\text{ or }3 \mod{4},\text{ and $m$ is square-free}.
  \end{cases}
\end{equation}
If $d$ satisfies either of the above conditions then $d$ is called a fundamental discriminant\footnote{The name is due to the fact that such $d$ are the discriminants of quadratic number fields, with $d=1$ being the `degenerate' quadratic field $\mathbb{Q}$.}.  The first few positive fundamental discriminants are $d=1,5,8,12,13,17,21$, and the first negative values are $d=-3,-4,-7,-8$.

Define for such $d$
\begin{equation}\label{eq:character_kronecker}
  \chi_d(n)\coloneqq \left(\frac{d}{n}\right),
\end{equation}
where $\left(\frac{d}{n}\right)$ is the Kronecker symbol, the generalization of the Legendre symbol.  Explicitly, for an integer $n$ with prime decomposition  $n=u\cdot p_1^{e_1}\cdots p_k^{e_k}$, with $u=\pm 1$ and $p_j$ prime, then 
\begin{equation}\label{eq:kronecker_symbol}
  \left(\frac{d}{n}\right)\coloneqq\left(\frac{d}{u}\right)\prod_{j=1}^k\left(\frac{d}{p_j}\right)^{e_j}.
\end{equation}
In the right hand side of \eqref{eq:kronecker_symbol} $\left(\frac{a}{p}\right)$ is the Legendre symbol, which take the values for $p\neq 2$,
\begin{align}
  \left(\frac{a}{p}\right)&\coloneqq
  \begin{cases}
    0 & \text{if }a\equiv 0\mod{p},\\
    1 & \text{if }a\equiv m^2\mod{p}\text{ and } a\not\equiv 0 \mod{p},\\
    -1 & \text{if }a\not\equiv m^2\mod{p}\text{ and }a\not\equiv 0 \mod{p},
  \end{cases}\\
\intertext{and}
  \left(\frac{a}{2}\right)&\coloneqq
  \begin{cases}
    0 & \text{if }a\text{ is even},\\
    1 & \text{if }a\equiv \pm 1 \mod{8},\\
    -1 & \text{if }a\equiv \pm 3 \mod{8}.
  \end{cases}
\end{align}
Finally, $\left(\frac{a}{1}\right)\coloneqq1$, $\left(\frac{a}{-1}\right)\coloneqq-1$ if $a<0$ and $1$ otherwise, and $\left(\frac{a}{0}\right)\coloneqq1$ if $a=\pm 1$ and $0$ otherwise. Then $\chi_d$ is called a \emph{real Dirichlet character}.  When $d=1$, $\chi_d$ is the trivial character (taking the value $1$ for all $n$), and for other fundamental discriminants $d$, $\chi_d$ is a real, primitive, \emph{quadratic} Dirichlet character of modulus $d$. Notice that for $d=-3$ (the first negative fundamental discriminant), $\chi_{-3}(n)=\left(\frac{-3}{n}\right)$ which matches \eqref{eq:simplest_dirichlet}. 

Given $\chi_d$, a real, quadratic Dirichlet character modulo a fundamental discriminant $d$, the associated \emph{Dirichlet $L$-function} is
\begin{equation}\label{dirichlet_l_function}
  L(s,\chi_d)\coloneqq \sum_{n=1}^\infty\frac{\chi_d(n)}{n^s}.
\end{equation}
Such $L$-functions again have an Euler product, a functional equation, and a meromorphic continuation to the full complex plane (see~\cite{iwanieckowalski04}). Further, they have an associated Riemann hypothesis which conjectures that the non-trivial zeros of $L(s,\chi_d)$ are also on the critical line $\RE(s)=1/2$. 

The second example of an $L$-function that we give is associated to elliptic curves. Consider an elliptic curve $E$ defined over $\mathbb{Q}$,
\begin{equation}
  E: y^2=x^3 + ax+b,
\end{equation}
for $a,b\in\mathbb{Z}$ such that the discriminant $\Delta=-16(4 a^3+27 b^2)\neq 0$ (which ensures that $E$ has distinct roots, or equivalently, is non-singular). If the pair $x,y\in\mathbb{C}$ form a solution to the equation defining the curve $E$, then we say they lie on $E$ and sometimes write $(x,y)\in E$.  Given a prime $p\nmid\Delta$, one can consider the number of points on $E$ modulo $p$.  This leads to the following definition
\begin{equation}\label{eq:a_p_def}
  a_p\coloneqq p+1-|\{(x,y)\in E : x,y\in \mathbb{Z}/p\mathbb{Z}\}|.
\end{equation}
These coefficients $a_p$ are used when constructing the $L$-function for the curve $E$, which is defined by its Euler product\footnote{This is formulation is the algebraic convention, though it will mean that the symmetry point is at $s=1$, rather than $s=1/2$. However, one can simply renormalize each local factor to shift the critical line for $L(s,E)$ to the traditional $\RE(s)=1/2$, see~\cite{iwasar00}.}
\begin{equation}\label{elliptic_euler_prod}
  L(s,E)\coloneqq \prod_{p}\left(1-a_pp^{-s}+\mathbbm{1}_{p\nmid \Delta}p^{-2s+1}\right)^{-1},
\end{equation}
where $\mathbbm{1}_{p\nmid\Delta}$ is $1$ for the `good primes' not dividing the discriminant, and $0$ otherwise. Just as with $\zeta(s)$, from the Euler product one can derive the appropriate Dirichlet series, and in turn the meromorphic continuation and the functional equation~\cite{iwasar00}.


\subsection{Random matrix theory and number theory}\label{sec:rmt_and_nt}

The origins of the connection between random matrix theory and number theory can be traced back to a conversation in 1971 between Hugh Montgomery and Freeman Dyson, introduced at tea at the Institute for Advanced Study by Sarvadaman Chowla. The conversation turned to the study of pair correlations of eigenvalues and of Riemann zeta zeros, which turn out to have an identical form.

We shall focus in this section on the unitary case, noting that there are more general connections between more general number theoretic functions -- $L$-functions -- and the other classical compact matrix groups (see~\cite{mezsna05, katsar99a, katsar99b, rudsar96, baikea21a, cfkrs05, keasna00a, keasna00b}), the discussion of which we postpone until later.  For brevity, we temporarily write $P_{N}\equiv P_{\U(N)}$ for the characteristic polynomial of a matrix in the $\CUE$. 
  
The two-point correlation function of the eigenphases of $A\in \U(N)$ has an expression due to Dyson (see~\cite{dys62, mehta04}). Denote by $e^{i\theta_1},\dots,e^{i\theta_N}$ the eigenvalues of $A$, and rescale the eigenphases $\theta_j$ so that on average they have unit spacing,
\begin{equation}\label{eq:rescaled_eigenphases}
  \phi_j\coloneqq\frac{\theta_jN}{2\pi}.
\end{equation}
The two-point correlation function for $A$ is
\begin{equation}\label{eq:rmt_pair_corr}
  R_2(A,x)\coloneqq \frac{1}{N}\sum_{n=1}^N\sum_{m=1}^N\sum_{k=-\infty}^\infty\delta(x+kN-(\phi_n-\phi_m)).
\end{equation}
Dyson established the following result.
\begin{theorem}[Dyson's Pair Correlation~\cite{dys62}]\label{thm:dyson}
  Take $A\in \U(N)$ and let $\phi_1,\dots,\phi_N$ be the normalized eigenphases of $A$, see~\eqref{eq:rescaled_eigenphases}. For test functions $f$ such that $f(x)\rightarrow 0$ as $|x|\rightarrow\infty$,
  \begin{equation}
    \lim_{N\rightarrow\infty}\int_{\U(N)}\int_{-\infty}^\infty f(x)R_2(A,x)\; dx\; dA=\int_{-\infty}^\infty f(x)\left(\delta(x)+1-\left(\frac{\sin(\pi x)}{\pi x}\right)^2\right)dx.
  \end{equation}
\end{theorem}
Hence, taking the test function to be $f(x)=1$ for $x\in[\alpha,\beta]$ and $0$ otherwise one finds
\begin{equation}\label{eigen_pair_corr}
  \lim_{N\rightarrow\infty}\int_{\U(N)} \frac{1}{N}\left|\{\phi_n,\phi_m: \alpha\leq \phi_n-\phi_m\leq \beta\}\right|\ dA=\int_{\alpha}^\beta\left(\delta(x)+1-\left(\frac{\sin(\pi x)}{\pi x}\right)^2\right)dx.
\end{equation}

The equivalent calculation for the non-trivial zeros of the Riemann zeta function was performed by Montgomery~\cite{montgomery73}. 

The non-trivial zeros of $\zeta(s)$ are $\rho_n=1/2+it_n$, with $\RE(t_n)>0$ and where the ordering on the zeros is by height\footnote{If $\rho_n$ is a non-trivial zero of $\zeta(s)$, then so is $\overline{\rho_n}$. Thus, it suffices to only consider zeros with positive imaginary part.}.  Define
\begin{equation}\label{eq:zero_counting_function}
  N(T)\coloneqq|\{n: 0\leq \RE(t_n)\leq T\}|
\end{equation}
to be the number of non-trivial zeros up to height $T$.  Then it can be shown, see for example~\cite{titchmarsh86}, that
\begin{equation}
  N(T)\sim \frac{T}{2\pi}\log\frac{T}{2\pi e},
\end{equation}
as $T\rightarrow\infty$.  This proves that there are infinitely many non-trivial zeros.  Further, the mean density increases logarithmically with height $T$.

We now assume the Riemann Hypothesis to be true, so that $\IM(t_n)=0 \mbox{  }\forall  \mbox{  }n$.  Following the construction for the pair correlation of eigenvalues, we rescale the zeros so that they have unit mean spacing,
\begin{equation}\label{eq:rescaled_zeros}
  w_n\coloneqq \frac{t_n}{2\pi}\log\frac{t_n}{2\pi}.
\end{equation}
Montgomery's pair correlation conjecture is that,
\begin{equation}\label{zeta_pair_corr}
  \lim_{T\rightarrow\infty}\frac{1}{T}\left|\{w_n,w_m\in[0,T]:\alpha\leq w_n-w_m\leq \beta\}\right|=\int_\alpha^\beta\left(\delta(x)+1-\left(\frac{\sin(\pi x)}{\pi x}\right)^2\right)dx.
\end{equation}
This conjecture is motivated by a theorem of Montgomery~\cite{montgomery73}, which can be stated as follows
\begin{equation}\label{thm:mont}
  \lim_{N\rightarrow\infty}\frac{1}{N}\sum_{n,m\leq N}f(w_n-w_m)=\int_{-\infty}^\infty f(x)\left(\delta(x)+1-\left(\frac{\sin(\pi x)}{\pi x}\right)^2\right)dx,
\end{equation}
for a test function $f$ with Fourier transform supported on $(-1,1)$, and such that the left and right sides of \eqref{thm:mont} converge.  The pair correlation conjecture corresponds to choosing $f$ as the indicator function on $[\alpha,\beta)$ (whose Fourier transform does not vanish outside $(-1,1)$).

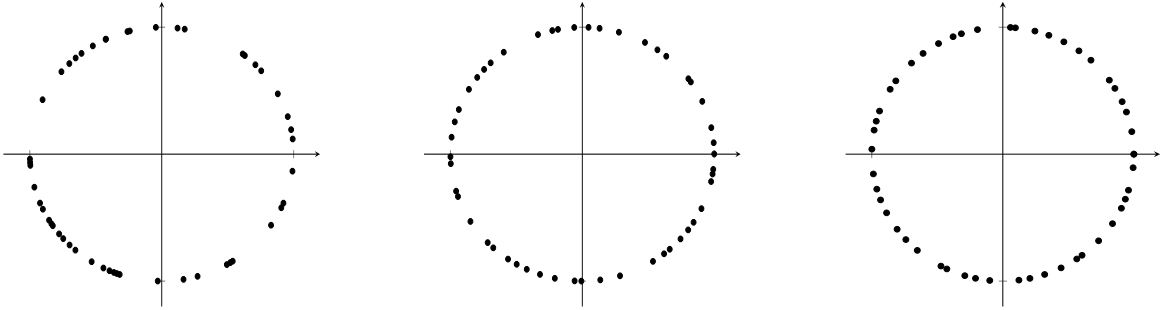
\begin{figure}[!htb]
  \centering
  \begin{subfigure}[t]{.33\textwidth}
    \centering
    \resizebox{4.3cm}{4.05cm}{
    \begin{tikzpicture}[scale=1]
      \begin{axis}[
          axis lines=center,
          xmin=-1.2,
          xmax=1.2,
          ymin=-1.2,
          ymax=1.2,
          grid=none,
          x label style={at={(axis description cs:1,0.5)},anchor=north},
          y label style={at={(axis description cs:0.58,1.03)},anchor=north},
          xtick={-1,1},
          xticklabels={,,},
          ytick={-1,1},
          yticklabels={,,}
        ]
	    \addplot[
          only marks,
          scatter,
          mark=*,
          mark size=1.5pt,
          scatter src=explicit,
          scatter/use mapped color={fill=black}
        ]
	    table[
          x expr=\thisrowno{0},
          y expr=\thisrowno{1},
          meta expr=\thisrowno{2}
        ]
	    {random.txt};
      \end{axis}
    \end{tikzpicture}
    }
    \caption{$50$ points drawn from the uniform distribution on the unit circle.}\label{fig:50unif}
  \end{subfigure}\hfill
  \begin{subfigure}[t]{.33\textwidth}
    \centering
    \resizebox{4.3cm}{4.05cm}{
      \begin{tikzpicture}[scale=1]
        \begin{axis}[
            axis lines=center,
            xmin=-1.2,
            xmax=1.2,
            ymin=-1.2,
            ymax=1.2,
            grid=none,
            x label style={at={(axis description cs:1,0.5)},anchor=north},
            y label style={at={(axis description cs:0.58,1.03)},anchor=north},
            xtick={-1,1},
            xticklabels={,,},
            ytick={-1,1},
            yticklabels={,,}
          ]
	      \addplot[
            only marks,
            scatter,
            mark=*,
            mark size=1.5pt,
            scatter src=explicit,
            scatter/use mapped color={fill=black}
          ]
	      table[
            x expr=\thisrowno{0},
            y expr=\thisrowno{1},
          meta expr=\thisrowno{2}
          ]
	      {unitary2_eig.txt};
        \end{axis}
      \end{tikzpicture}
    }
    \caption{Eigenvalues of a random unitary $50\times 50$ matrix.}\label{fig:50eig}
  \end{subfigure}\hfill
  \begin{subfigure}[t]{.33\textwidth}
    \centering
    \resizebox{4.3cm}{4.05cm}{
      \begin{tikzpicture}[scale=1]
        \begin{axis}[
            axis lines=center,
            xmin=-1.2,
            xmax=1.2,
            ymin=-1.2,
            ymax=1.2,
            grid=none,
            x label style={at={(axis description cs:1,0.5)},anchor=north},
            y label style={at={(axis description cs:0.58,1.03)},anchor=north},
            xtick={-1,1},
            xticklabels={,,},
            ytick={-1,1},
            yticklabels={,,},
            axis equal image
          ]
	      \addplot[
            only marks,
            scatter,
            mark=*,
            mark size=1.5pt,
            scatter src=explicit,
            scatter/use mapped color={fill=black}
          ]
	      table[
            x expr=\thisrowno{0},
            y expr=\thisrowno{1},
            meta expr=\thisrowno{2}
          ]
	      {zeros_scaled.txt};
        \end{axis}
      \end{tikzpicture}
    }
    \caption[$50$ consecutive non-trivial zeros of $\zeta(s)$, scaled.]{$50$ consecutive non-trivial zeros of $\zeta(s)$, from $201^{\text{st}}$ to $250^{\text{th}}$, data from~\cite{lmfdb}, scaled to lie on the unit circle.}\label{fig:50zeros}
  \end{subfigure}
  \caption[Comparison between random points, eigenvalues, and zeta zeros.]{Comparing $50$ points distributed uniformly on the unit circle~\eqref{fig:50unif}, with $50$ eigenvalues of a random unitary matrix~\eqref{fig:50eig}, and with $50$ scaled consecutive non-trivial zeros of $\zeta(s)$~\eqref{fig:50zeros}.}\label{fig:50points}
\end{figure}

  The similarity between \eqref{zeta_pair_corr} and \eqref{eigen_pair_corr} is apparent.  Important numerical evidence for the conjecture has come from the work of Odlyzko~\cite{odlyzko89}, and heuristics and generalizations of Montgomery's theorem have been developed for the general $k$-point correlation function~\cite{rudsar96,hej94, bogkea95, bogkea96}.  A visual comparison can be found in figure~\ref{fig:50points}.  There we have compared plots of $50$ points taken from a uniform distribution on the unit circle, with the eigenvalues of a random matrix drawn from $\U(50)$, and $50$ consecutive non-trivial zeros of $\zeta(s)$, scaled to wrap around the unit circle. The figure shows that the zeta zeros display similar `repulsion' to that seen for the eigenvalues (see~\eqref{weyl_int_formula}), and also that their distribution appears far from uniform on the scale of the mean spacing. 

Further evidence for a connection can be found by comparing theorem~\ref{thm:ks-clt} with Selberg's central limit theorem~\ref{thm:selbergclt} for the Riemann zeta function.  In both cases, the real and imaginary parts of the respective logarithms tend independently to Gaussian random variables. Additionally, if one sets
\begin{equation}\label{n=logt}
  N=\log\tfrac{T}{2\pi}
\end{equation}
then the scalings in both theorems agree. The same identification in the unit mean scaling \eqref{eq:rescaled_zeros} reveals that the average density of zeros matches the average density of eigenvalues in \eqref{eq:rescaled_eigenphases}.

In light of this apparent connection, one may try to model, in a statistical sense, the Riemann zeta function by unitary characteristic polynomials. To this end, we focus on a long-standing number theoretic conjecture. 
\begin{conj}\label{conj:zeta_moments}
When $T\to\infty$
\begin{align}
  \mathscr{M}_{T}(\beta)&\coloneqq\lim_{T\rightarrow\infty}\frac{1}{T}\int_0^T|\zeta(\tfrac{1}{2}+it)|^{2\beta}dt\\
  &\sim c_\zeta(\beta)a_\zeta(\beta)\left(\log\tfrac{T}{2\pi}\right)^{\beta^2},\label{eq:zeta_moment_limit}
\end{align}
where
\begin{equation}\label{zeta_arithmetic}
  a_\zeta(\beta)\coloneqq\prod_p\left[\left(1-\frac{1}{p}\right)^{\beta^2}\left(\sum_{m=0}^\infty\left(\frac{\Gamma(\beta+m)}{m!\Gamma(\beta)}\right)^2p^{-m}\right)\right],
\end{equation}
and where $c_\zeta(\beta)$ depends on the moment parameter $\beta$.
\end{conj}
The only cases where conjecture~\ref{conj:zeta_moments} is known are when $\beta=0,1,2$.  The case of $\beta=0$ is trivial, $\beta=1$ was computed by Hardy and Littlewood~\cite{harlit18}, and Ingham~\cite{ing26} determined $\beta=2$, with, respectively,
\begin{align}
  c_\zeta(0)&=1,\label{eq:zeromoment}\\
  c_\zeta(1)&=1,\\
  c_\zeta(2)&=\frac{2}{4!}.\\
  \intertext{Using number theoretic arguments, Conrey and Ghosh~\cite{congho92} and Conrey and Gonek~\cite{congon01} conjectured the leading order asymptotics of the $6^{th}$ and $8^{th}$ moments, with,}
  c_\zeta(3)&=\frac{42}{9!},\\
  c_\zeta(4)&=\frac{24024}{16!}.\label{eq:eightmoment}
\end{align}
Hence it appears that $c_\zeta(\beta)\cdot (\beta^2)!$ is an integer for $\beta\in\mathbb{N}$.  Additionally, Ramachandra~\cite{ram80} and Heath-Brown~\cite{heabro81} have established a lower bound $\mathscr{M}_{T}(\beta)\gg (\log\frac{T}{2\pi})^{\beta^2}$ for positive, rational $\beta$.  Radziwi{\l}{\l} and Soundararajan~\cite{radsou13} proved lower bounds of the correct order for all real $\beta\geq 1$.  Upper bounds of the correct form $\mathscr{M}_{T}(\beta)\ll (\log\frac{T}{2\pi})^{\beta^2}$ are known, conditionally on the Riemann hypothesis, due to arguments of Soundararajan~\cite{sou09} and Harper~\cite{har13b}.

Let us now compare $\zeta({1/2}+it)$ with $P_N(A,\theta)$ for $A\in \U(N)$, identifying $N=\log\frac{T}{2\pi}$. Hence, $\mathscr{M}_{T}(\beta)$ should be modelled by 
\begin{equation}\label{eq:rmt_moment_restated}
  M_N(\beta)=\int_{\U(N)}|P_N(A,\theta)|^{2\beta}dA.
\end{equation}
Using theorem~\ref{thm:ks}, Keating and Snaith \cite{keasna00a} conjectured that
\begin{equation}\label{conj:ks-coefficient}
  c_\zeta(\beta)=\frac{\mathcal{G}^2(1+\beta)}{\mathcal{G}(1+2\beta)},
\end{equation}
which matches with the all known and conjectural cases~\eqref{eq:zeromoment}--\eqref{eq:eightmoment}.   

In \eqref{eq:zeta_moment_limit}, the leading order asymptotics of the moments involves the product of $c_\zeta(\beta)$, which is conjecturally given by a random matrix calculation, and the arithmetic factor $a_\zeta(\beta)$.  That it should split in such a way is motivated by a formulae due to Gonek, Hughes and Keating~\cite{gonhugkea07} according to which $\zeta(s)$ can be written a truncated product over the primes multiplied by a product over zeros close to $s$.  

As commented above, Keating and Snaith's work implies, for integer $\beta$, that the $2\beta$th moment of $\zeta(1/2+it)$ is of the order of $(\log\frac{T}{2\pi})^{\beta^2}$. Conrey et al.~\cite{cfkrs05} extended this conjecture to
\begin{equation}\label{eq:cfkrs_mci_first}
  \frac{1}{T}\int_{0}^T|\zeta(\tfrac{1}{2}+it)|^{2\beta}dt= Q_\beta(\log\tfrac{T}{2\pi})+o(1)
\end{equation}
for integer $\beta$, where $Q_\beta(x)$ is a polynomial in $x$ of degree $\beta^2$.  This polynomial has a multiple contour integral representation of a kind that we shall discuss in more detail in section~\ref{sec:moments}. We note in passing that the conjectures of Keating and Snaith  and of Conrey et al.\ extend to other $L$-functions~\cite{keasna00b, cfkrs05}.

The classical number theoretic approach to understanding moments of the Riemann zeta function uses Dirichlet polynomial approximations of $\zeta(s)$ and higher powers. Analysis in this direction, as discussed above, gives the second and fourth moments and strong conjectural forms for the sixth and the eight moments.  However, this method fails for $\beta\geq 5$ since it predicts negative values.

Recently, Conrey and Keating~\cite{conkea15a, conkea15b, conkea15c, conkea16, conkea19} have demonstrated heuristically why the Dirichlet polynomial method in its traditional form fails: for higher $\beta$ it is essential that one uses much longer Dirichlet polynomials than is conventional to avoid missing important terms. When one does so, the resulting formulae match exactly the polynomials $Q_\beta(x)$ conjectured in~\cite{cfkrs05} for all $\beta\in\mathbb{N}$. 

The recipe that they derive is based on divisor sums and can be viewed as a type of `multi-dimensional Hardy-Littlewood circle method', or `Manin-type stratification'. Consider
\begin{equation}\label{eq:conkea1}
  \int_0^\infty \left(\prod_{\alpha\in A}\zeta(\tfrac{1}{2}+it + \alpha)\right)\left(\prod_{\beta\in B}\zeta(\tfrac{1}{2}-it+\beta)\right)\psi\left(\frac{t}{T}\right)dt,
\end{equation}
where $A, B$ are sets of size $\beta$ of `shifts' and $\psi$ is a smooth function of compact support.  Note that by letting all $\alpha,\beta\rightarrow0$ we recover the desired $2\beta$th moment. Conrey and Keating demonstrate that \eqref{eq:conkea1} can be evaluated heuristically either by multiple contour integrals of the type appearing in \eqref{eq:cfkrs_mci_first}, or by examining the Dirichlet series 
\begin{equation}\label{eq:multi_point_zeta}
  \prod_{\alpha\in A}\zeta(\tfrac{1}{2}+it+\alpha)=\sum_{n=1}^\infty \frac{\tau_A(n)}{n^s},
\end{equation}
where the arithmetic divisor function $\tau_A(n)$ is defined by the Euler product expansion of the left hand side of \eqref{eq:multi_point_zeta}.  

\subsection{Extreme value theory}\label{sec:evt}

We will be interested here in the extreme values of various functions (for example, characteristic polynomials, or $\zeta(s)$). The introduction provided within this section is tailored to support the ideas of section~\ref{sec:fk_conj}; for a general introduction to the topic see Leadbetter, Lindgren, and Rootz~\cite{llr12}, and de Haan and Ferreira~\cite{dehfer07}.  A central result within the field is the Fisher-Tippett-Gnedenko Theorem.

\begin{theorem}[Fisher-Tippett-Gnedenko Theorem~\cite{gne43}]\label{thm:ftg}
  Let $X_1,X_2,\dots,X_n$ be independent and identically distributed (\iid) random variables.  Define 
  \[M_n\coloneqq\max\{X_1,\dots,X_n\}.\]
  If there exists $a_n>0$ for all $n$, and $b_n$ such that 
  \[\lim_{n\rightarrow\infty}\mathbb{P}\left(\frac{M_n-b_n}{a_n}\leq x\right)=F(x),\] where $F$ is non-degenerate, then $F$ is an extreme value cumulative distribution function and belongs to one of three classes:
  \begin{align*}
    (\operatorname{I})&\quad \Lambda(x)=e^{-e^{-x}}\quad\text{for }x\in\mathbb{R},\\
    (\operatorname{II})&\quad \Phi_\alpha(x)=
    \begin{dcases}
      0&\text{if }x\leq 0\\
      e^{-x^{-\alpha}}&\text{if }x> 0,
    \end{dcases}\text{ for some }\alpha>0,\\
    (\operatorname{III})&\quad \Psi_\alpha(x)=
    \begin{dcases}
      e^{-(-x)^\alpha}&\text{if }x<0\\
      1&\text{if }x\geq 0,
    \end{dcases}\text{ for some }\alpha>0.
  \end{align*}
\end{theorem}

Type (I), Type (II), and Type (III) are known as Gumbel, Fr\'echet, and Weibull distributions respectively. Of particular interest within the context of this exposition is the following result (see for example Leadbetter et al.~\cite{llr12}, Theorem 1.5.3). 

\begin{theorem}\label{maxofnormals}
  Let $\{Z_1,Z_2,\dots\}$ be independent and identically distributed standard Gaussian random variables.  As in theorem~\ref{thm:ftg}, let $M_n\coloneqq\max\{Z_1,\dots,Z_n\}$.  Then define
  \begin{equation}\label{extreme_gaussian_var}
    a_n\coloneqq \frac{1}{\sqrt{2\log n}},
  \end{equation}
  and
  \begin{equation}\label{extreme_gaussian_mean}
    b_n\coloneqq \sqrt{2\log n}-\frac{\log\log n+\log 4\pi}{2\sqrt{2\log n}}.
  \end{equation}
  Then
  \begin{equation}
    \lim_{n\rightarrow\infty}\mathbb{P}\left(\frac{M_n-b_n}{a_n}\leq x\right)=e^{-e^{-x}}.
  \end{equation}
  Thus, after rescaling, the maximum of a collection of standard Gaussian random variables has a Gumbel (Type (I)) distribution. 
\end{theorem}

Theorem~\ref{maxofnormals} shows that for standard Gaussian random variables the approximate size of the maximum is
\begin{equation}
  M_n \approx b_n+a_n\mathcal{M}
\end{equation}
where $\mathcal{M}$ is a Gumbel random variable. It will soon be useful to consider a dyadic number of variables with non-unit variance.  Take $Y_1,\dots,Y_{2^n}$ independent Gaussian random variables, centred and with variance $\sigma^2n$ and call their maximum $M_{2^n}$.  By rescaling the resulting $a_{2^n}$ from theorem~\ref{maxofnormals}, one has
\begin{equation}\label{general_max_gaussian}
  M_{2^n}\approx cn-\frac{\sigma^2}{c}\log n+\mathcal{M},
\end{equation}
where $c=\sqrt{2\sigma^2\log 2}$ and $\mathcal{M}$ has a Gumbel distribution. For comparative purposes, we remark that the leading order is linear in $n$, and the subleading term is logarithmic in $n$ with a coefficient of $1/2$.

Notice that both theorems~\ref{thm:ftg} and~\ref{maxofnormals}, and the subsequent discussion regarding the form of the maxima, importantly assume \textit{independence} of the random variables. The central question of this review, however, concerns variables that instead are \emph{logarithmically correlated}.

\subsection{Log-correlated fields}\label{sec:logcorrintro}

Here we introduce key results and ideas that will underpin many of the areas covered in this review. For a more thorough overview of the wider research area see Duplantier et al.~\cite{drsv17}, and Arguin~\cite{arg16} for an excellent survey with similar aims to this exposition.

One can define a stochastic process $X_n=\{X_n(l): l\in L_n\}$ on a metric space $L_n$ with a distance $|\cdot|$, so that the dimension of the space $\dim L_n$ depends on $n$. Then, the defining feature of log-correlated fields is the form of the covariance\footnote{Here we write `$\approx$' to encompass any covariance structure which has a logarithmic singularity at $l=l^\prime$.},
\begin{equation}
  \mathbb{E}[X_n(l)X_n(l^\prime)]\approx -\log |l-l^\prime|,\label{eq:log_correlations}
\end{equation}
for $l,l^\prime\in L_n$.

As in section~\ref{sec:evt}, choosing $L_n$ to be a discrete field with $2^n$ points will prove to be a particularly instructive example. Take a log-correlated field $X_n=\{X_n(l), l\in L_n\}$ with $|L_n|=2^n$, so that $X_n$ has covariance as defined by the right hand side of \eqref{eq:log_correlations}, and such that the $X_n(l)$ are centred with variance $\mathbb{E}[X_n(l)]^2= \sigma^2n$. 

In general in this case, the maximum $M_{2^n}=\max_{l\in L_n}X_n(l)$ has the following form
\begin{equation}\label{eq:log_corr_max}
  M_{2^n}\approx cn-\frac{3}{2}\frac{\sigma^2}{c}\log n+\mathcal{M},
\end{equation}
where $c=\sqrt{2\sigma^2\log2}$. The distribution of $\mathcal{M}$ is expected \textit{no longer to conform to the Gumbel distribution\footnote{For one of the key examples of a log-correlated field that we will consider, it is in fact believed that the density of $\mathcal{M}$ is
\begin{equation*}\label{eq:sum_of_gumbels}
  2e^{-x}K_0(2e^{-\frac{x}{2}}),
\end{equation*}
where $K_0(z)$ is the modified Bessel function of the second kind \cite{fyohiakea12, fyokea14}.  As observed by Kundu et al.~\cite{kunmajsch13}, the density in \eqref{eq:sum_of_gumbels} matches that for the sum of two independent Gumbel random variables.}}.

We here highlight the similarities and differences between the maximum of log-correlated processes~\eqref{eq:log_corr_max} and the maximum of independent Gaussian random variables, see~\eqref{general_max_gaussian}.  For both, the leading order of the maximum is linear in $n$, and the subleading term is of the order $\log n$, but the subleading coefficient differs between the cases: $-1/2$ versus $-3/2$.  If the random variables are all independent, then the maximum isn't `pulled down' as much as when the variables are log-correlated. Such behaviour is expected to be universal within each class.

Important examples of log-correlated processes are branching random walks, the logarithm of the characteristic polynomial of a random unitary matrix, and the Gaussian free field in two dimensions. 

We now discuss in more detail two models exhibiting such logarithmic correlations: the generalized Random Energy Model (GREM) and branching random walks. These will lay the groundwork for the branching model that will play a central role in section~\ref{sec:fk_conj}.

\subsubsection*{The Random Energy Model}

The `Random Energy Model' (REM) was introduced by Derrida~\cite{der81} in the study of spin glasses. The REM is a stochastic process on the hypercube $\{-1,1\}^n$. For each point in the state space, one associates the independent random variable $X_n(l)\in\cN(0,n)$.

Translating in to the language of log-correlated fields, we have the process $X_n=\{X_n(l): l\in\{-1,1\}^n\}$.  The partition function for the REM is defined as
\begin{equation}\label{partition_function}
  Z_n(\beta)\coloneqq\sum_{l\in\{-1,1\}^{n}}\exp(-\beta X_n(l)),
\end{equation}
where $\beta>0$ represents the inverse temperature of the system and $X_n(l)$ plays the role of the energy. Note that the maximum of the REM will follow~\eqref{general_max_gaussian} with $\sigma^2=1$ due to the assumption of independence of the $X_n(l)$.

Additionally, the \textit{free energy} associated to $Z_n(\beta)$ is
\begin{equation}\label{free_energy}
  f_n(\beta)=-\frac{1}{\beta}\log Z_n(\beta).
\end{equation}
The free energy is related to the extreme values of the energy, $M_{2^n}\coloneqq\max_{l\in\{-1,1\}^n}X_n(l)$, via
\begin{equation}
  \lim_{\beta\rightarrow\infty}f_n(\beta)=-\lim_{\beta\rightarrow\infty}\frac{1}{\beta}\log Z_n(\beta)=-M_{2^n}.
\end{equation} 

One can generalize the REM so that the random variables $X_n(l)$, $l\in\{-1,1\}^{n}$ are no longer independent, but instead depend on the distance $|l-l^\prime|$ for $l,l^\prime\in\{-1,1\}^{n}$. This is a \emph{generalized} random energy model (GREM). Adaptations in particular which introduce \textit{logarithmic} correlations of the form \eqref{eq:log_correlations} have inspired considerable research interest; see for example~\cite{der85,carled01,fyobou08, fyokea14}. Once again, a similar process to that described above yields a connection between the free energy for the GREM and its maximum, which should instead now follow~\eqref{eq:log_corr_max}. 

\subsubsection*{Branching Random Walks}

\begin{figure}[htb]
  \centering
  \begin{tikzpicture}[
      level distance=1.5cm,
      level 1/.style={sibling distance=4cm},
      level 2/.style={sibling distance=2cm},
      level 3/.style={sibling distance=1cm},
      level 4/.style={sibling distance=.5cm},
      every node/.style = {shape=circle, inner sep=1.5pt, draw, align=center, fill=black}]
    \draw[->,ultra thick] (-4.3,-6)--(5,-6);
    \draw[<-,ultra thick] (-4.5,-6)--(-4.5,0.8);
    \draw[-,ultra thick] (4.4,-5.9)--(4.4,-6.1);
    \node[draw=none,fill=none] at (4.5,-6.4) {$1$};
    \draw[-,ultra thick] (-4.3,-5.9)--(-4.3,-6.1);
    \node[draw=none,fill=none] at (-4.3,-6.4) {$0$};
    \node[draw=none,fill=none] at (-4.8,.7) {$m$};
    \node[draw=none,fill=none] at (2,-.3) {\footnotesize{$Y_1(l)\sim\cN(0,\sigma^2)$}};
    \node[draw=none,fill=none] at (4,-2.2) {\footnotesize{$Y_2(l)\sim\cN(0,\sigma^2)$}};
    \node[draw=none,fill=none] at (4.8,-3.8) {\footnotesize{$Y_3(l)\sim\cN(0,\sigma^2)$}};
    \node[draw=none,fill=none] at (5.1,-5.2) {\footnotesize{$Y_4(l)\sim\cN(0,\sigma^2)$}};
    \node[draw=none,fill=none] at (3.8,-6.4) {$l$};
    \node[draw=none,fill=none] at (1.3,-6.4) {$l^\prime$};
    \node {}
    child {node {}
      child {node {}
        child {node {}
          child {node {}}
          child {node {}}
        }
        child {node {}
          child {node {}}
          child {node {}}
        }
      }
      child {node {}
        child {node {}
          child {node {}}
          child {node {}}
        }
        child {node {}
          child {node {}}
          child {node {}}
        }
      }
    }
    child[blue!90!white,dashed,thick] {node[solid,draw=red,circle,fill=white,thick] {}
      child[green!70!black,dashdotted,thick] {node[solid,draw=black,circle] {}
        child[black,solid,thin] {node[solid,draw=black,circle] {}
          child {node[solid,draw=black,circle] {}}
          child {node[solid,draw=black,circle] {}}
        }
        child {node[solid,draw=black,circle] {}
          child {node[solid,draw=black,circle] {}}
          child[black,solid,thin] {node[solid,draw=black,circle] {}}
        }
      }
      child[orange!70!black,dotted,thick] {node {}
        child[black,solid,thin] {node {}
          child {node {}}
          child {node {}}
        }
        child {node {}
          child[black,solid,thin] {node {}}
          child {node {}}
        }
      }
    };
  \end{tikzpicture}
  \caption[Example of a random walk on a binary tree.]{An example of random walks on a binary tree of depth $n=4$, from root to leaves $l$ and $l^\prime$.  Some weightings $Y_j(l)$ are highlighted, where $Y_j(l)\sim \cN(0,\sigma^2)$. The last common ancestor of leaves $l, l^\prime$ is illustrated by the `hollow' (red) node and occurs at level $1$.}\label{fig:binary_tree}
\end{figure}
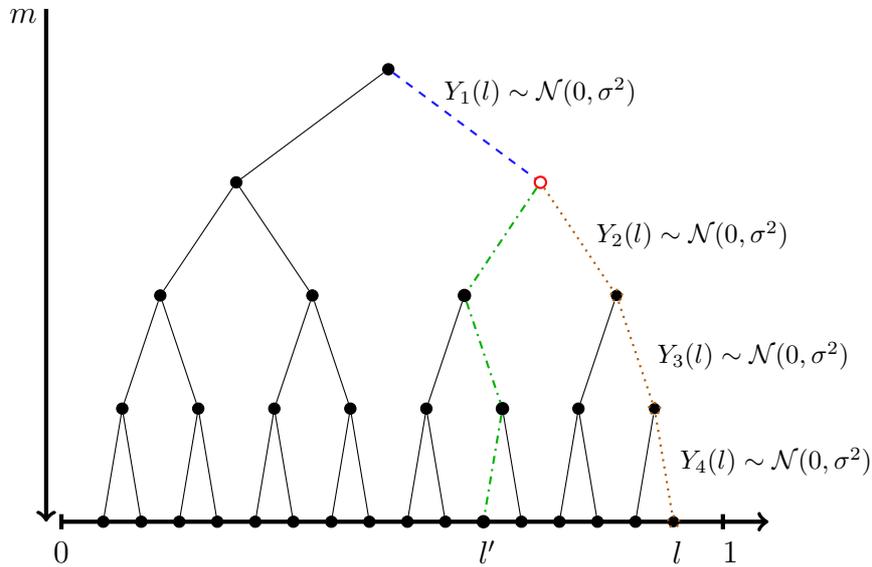

Perhaps the simplest example where the `$3/2$' coefficient can be proven to appear is the case of a Gaussian random walk on a binary tree. This is also sometimes referred to as the hierarchical Gaussian field.  Take a rooted binary tree of depth $n$ and, in the language introduced at the start this section, let $L_n$ be the leaves of such a tree.  For a fixed leaf $l\in\{1,\dots, 2^n\}$, define the random variable $X_n(l)$ by
\begin{equation}
  X_n(l)=\sum_{m=1}^nY_m(l),
\end{equation}
where the $Y_m(l)\sim\cN(0,\sigma^2)$ are independent and identically distributed. Thus, $X_n(l)$ is a random walk from root to leaf $l$. Clearly, the distribution of $Y_m(l)$ does not depend on the level $m$, nor the leaf $l$, but we retain the notation to make the connection with the binary tree clear. Note also for a comparison with the (G)REM that $X_n(l)\sim\cN(0,n\sigma^2)$ and, whilst the $Y_m(l)$ are independent, two walks $X_n(l)$ and $X_n(l^\prime)$ generally will \emph{not} be independent for $l\neq l^\prime$. Figure~\ref{fig:binary_tree} illustrates the process. 

We shall see that the binary tree model is central to the progress detailed within sections~\ref{sec:fk_conj} and~\ref{sec:moments}

To demonstrate that $X_n$ is log-correlated, one determines the covariance structure.  Firstly, one has
\begin{align}
  \mathbb{E}[X_n(l)]&=0\\
  \mathbb{E}[X_n(l)^2]&=\sigma^2n.
\end{align}
To determine the covariance, one requires the notion of the \textit{last common ancestor} of two leaves $l, l'$.  We also henceforth consider the tree embedded in the interval $[0,1]$ with the $2^n$ leaves equally spaced, cf. figure~\ref{fig:binary_tree}.

\begin{definition}[Last common ancestor]\label{def:lca}
  The \textit{last common ancestor} of two leaves $l$ and $l'$ of a binary tree is the first point at which the paths from root to $l$ and from root to $l'$ diverge. This is illustrated in figure~\ref{fig:binary_tree}.  The level of the least common ancestor of $l$ and $l'$ is denoted by $\lca(l,l')$.
\end{definition}

Then, one calculates that 
\begin{align}
  \mathbb{E}[Y_j(l)Y_j(l')]&=
  \begin{dcases}
    \sigma^2,&\text{if }|l-l'|\leq 2^{-j}\\
    0,&\text{if }|l-l'|>2^{-j}.
  \end{dcases}\\
\intertext{Thus, the covariance of $X_n(l)$ and $X_n(l')$ depends on the last common ancestor of $l,l'$:}
  \mathbb{E}[X_n(l)X_n(l')]&=\sigma^2\cdot \lca(l,l').
\end{align}
One concludes that $X_n(l)$ displays the properties of a log-correlated process. Further, take some $0\leq r\leq 1$ such that $r n\in\mathbb{N}$. Then consider, for a fixed leaf $l\in\{1,\dots,2^n\}$, the proportion of neighbours $l^\prime$ whose covariance with $l$ is at least $\sigma^2nr$.  It is precisely those $l^\prime$ whose last common ancestor is at level $r n$:
\begin{equation}
  \frac{1}{2^n}\left|\left\{l^\prime\in\{1,\dots,2^n\}:\frac{\mathbb{E}[X_n(l)X_n(l^\prime)]}{\mathbb{E}[X_n(l)^2]}\geq r\right\}\right|=\frac{1}{2^{rn}}.
\end{equation}
 Such a property is indicative of a log-correlated system, see for example~\cite{arg16}.

Given that the branching random walk has a log-correlated structure, a natural question is to investigate its maximum.  Bramson~\cite{bra78} was the first to identify to subleading order the maximum of the above field $X_n$ with, notably, the subleading constant $3/2$. 

\begin{theorem}[Bramson~\cite{bra78}]\label{thm:bramson}
  Let $X_n(l)$ be a branching random walk from root to leaf $l$ on a binary tree of depth $n$, with independent, centred, Gaussian increments of variance $\sigma^2$. Set $M_{2^n}\coloneqq \max_{l\in\{1,\dots,2^n\}}X_{n}(l)$. Then
  \begin{equation}\label{eq:bramson_theorem}
    M_{2^n}=cn-\frac{3}{2}\frac{\sigma^2}{c}\log n+x
  \end{equation}
  where $c=\sqrt{2\sigma^2\log 2}$ and $x$ is a bounded fluctuating term.
\end{theorem}

One sees the `log-correlated' constant $3/2$ appearing; the importance of the canonical example of branching random walks on binary trees is evident. Such processes will make frequent reappearances hereafter. In particular, we will often set $\sigma^2=(1/2)\log 2$.  Given this choice, \eqref{eq:bramson_theorem} becomes
\begin{align}\label{eq:logcorrmax_quarter}
  M_{2^n}=n\log 2-\frac{3}{4}\log n+x.
\end{align}
Given the regularity with which we will set $\sigma^2=(1/2)\log 2$, we may equally refer to the `log-correlated' constant as being $3/4$, (making the  `independent' constant $1/4$, cf.~\eqref{general_max_gaussian}).

\subsection{Gaussian multiplicative chaos}\label{sec:gmc}

Within this section, we give an overview of the ideas and results from the theory of Gaussian multiplicative chaos ($\GMC$) that will be relevant in the present context.  For an excellent general review of the topic, we direct the reader to the paper of Rhodes and Vargas~\cite{rhovar14}.

The origins of $\GMC$ trace back to the work of Kahane~\cite{kah85}, who introduced the theory for understanding the exponential of a Gaussian field whose covariance has a logarithmic singularity.  To this end, take $D\subset \mathbb{R}^d$ a subdomain, and $X=\{X(v):v\in D\}$ a Gaussian field so that
\begin{equation}
  \mathbb{E}[X(v)]=0,
\end{equation}
and
\begin{align}
  \mathbb{E}[X(v)X(w)]&=\max\{-\log|v-w|,0\}+g(v,w)\\
  &\sim-\log|v-w|,\label{log_covariance}
\end{align}
as $v\rightarrow w$, and for $g$ some bounded function over $D\times D$. Clearly the covariance \eqref{log_covariance} implies a connection to the log-correlated fields discussed previously.

The log-singularity present in \eqref{log_covariance} is precisely the cause of the difficulty when constructing the measure associated with the exponential of the field,
\begin{equation}\label{eq:gmc_measure}
  e^{\gamma X(v)-\frac{\gamma^2}{2}\mathbb{E}[X(v)^2]}dx,
\end{equation}
for some $\gamma\in\mathbb{R}$. A natural solution is to `regularize' the field $X$: introduce a smooth cut-off $X_n(v)$ so that in the large $n$ limit, $X_n(v)\rightarrow X(v)$. For such a cut-off, if one is able evaluate
\begin{equation}
  e^{\gamma X_n(v)-\frac{\gamma^2}{2}\mathbb{E}[X_n(v)^2]}dx\overset{n\rightarrow\infty}{\longrightarrow}\mu_\gamma(dx),
\end{equation}
for some limiting measure $\mu_\gamma(dx)$, then \eqref{eq:gmc_measure} is defined to be said measure. Kahane constructed such an $X_n$ and showed that the limiting measure $\mu_\gamma$ is non-trivial for a certain range of $\gamma$.

\begin{theorem}[Kahane~\cite{kah85}]
  Let $v, w\in D$ and assume that there exists a continuous and bounded function $g:D\times D\rightarrow \mathbb{R}$ such that
  \begin{equation}
    \mathbb{E}[X(v)X(w)]=\max\{-\log|v-w|,0\}+g(v,w),
  \end{equation}
  and further that the covariance has a decomposition
  \begin{equation}
    \mathbb{E}[X(v)X(w)]=\sum_{i=1}^\infty K_i(v,w),
  \end{equation}
  for $K_i$ continuous and positive definite covariance kernels. Let $Y_i$ be the Gaussian field with mean $0$ and covariance given by $K_i$, such that $Y_i$ is independent from $Y_{i'}$ for $i\neq i'$. Set $X_n=Y_1+\cdots+Y_n$.  Then for $\gamma\in\mathbb{R}$, the measures
  \begin{equation}
    \mu_{\gamma,n}(dx)\coloneqq e^{\gamma X_n(x)-\frac{\gamma^2}{2}\sum_{l=i}^nK_i(x,x)}dx,
  \end{equation}
  converge almost surely in the space of Radon measure (for topology given by weak convergence) to some random measure $\mu_\gamma(dx)$.  This convergence is independent of the regularization of $X$. The measure $\mu_\gamma$ is the zero measure for $\gamma^2\geq 2d$ (where recall our field is defined with respect to $D\subset \mathbb{R}^d$), and non-trivial for $\gamma^2<2d$.
\end{theorem}

Frequently, the range $[0,2d)$ (where the measure is non-trivial) is broken up in to two sections, known as the $L^1-$ and $L^2-$phase:
\begin{alignat}{2}
  \text{$L^2-$phase}:&\quad0&&\leq \gamma^2<d\label{eq:gmc_phase_l2},\\
  \text{$L^1-$phase}:&\quad d&&\leq\gamma^2<2d.\label{eq:gmc_phase_l1}
\end{alignat}
For the cases considered in this review, $d=1$. When demonstrating convergence to the $\GMC$ measure for a particular field $X$, it is often the case that the two ranges require different techniques (see section~\ref{sec:fk_progress}). Additionally we remark that, as formulated above, the measure $\mu_\gamma(dx)$ is trivial for $\gamma^2\geq 2d$.  It is possible to construct a $\GMC$ measure for $\gamma^2\geq 2d$. The case of $\gamma^2=2d$ yields a phase transition and is known as \textit{critical chaos}. Conjecturally, see for example~\cite{rhovar14}, all constructions for the critical chaos measure are the same. In the \textit{super-critical} regime of $\gamma^2>2d$, one can define a new class of chaos known as \textit{atomic multiplicative chaos}.  We refer the reader to~\cite{rhovar14}, and the references therein, for further details.

\subsection{Symmetric function theory}

A function $f$ in $n$ variables is \emph{symmetric} if it is invariant under permutations of its arguments.  That is, if $\sigma\in S_n$ (the group of permutations on $n$ symbols) and if
\begin{equation}
  f(x_1,x_2,\dots,x_n)=f(\sigma(x_1),\sigma(x_2),\dots,\sigma(x_n)),
\end{equation}
then $f$ is a symmetric function. 

\begin{definition}[Partition]
  Let $l\in\mathbb{N}$. A \emph{partition} $\lambda$ of length $l=l(\lambda)$ is a sequence of non-increasing non-negative integers with $l$ non-zero elements. Thus, if $\lambda=(\lambda_i)_{i=1}^l$ is an $l$-long partition, then
  \begin{equation}\label{eq:partition}
    \lambda_1\geq \lambda_2\geq \cdots \geq \lambda_l
  \end{equation}
  with $\lambda_i\in\mathbb{N}$ for $i\in\{1,\dots,l\}$. It is sometimes useful to extend partitions with finitely many zeros.  In this case, we identify all partitions which share the same non-zero portion,
  \begin{equation}
    \lambda=(\lambda_1,\dots,\lambda_l)=(\lambda_1,\dots,\lambda_l,0,\dots,0).
  \end{equation}
  Finally,  if $\lambda=(\lambda_i)_{i=1}^l$ has $m_i=m_i(\lambda)$ elements equal to $i$, then we can write $\lambda$ in \emph{multiplicative} notation,
  \begin{equation}
    \lambda=\langle 1^{m_1}2^{m_2}\cdots\lambda_1^{m_{\lambda_1}}\rangle=(\underbrace{\lambda_1,\dots,\lambda_1}_{m_{\lambda_1}},\dots,\underbrace{2,\dots,2}_{m_2},\underbrace{1,\dots,1}_{m_1}).
  \end{equation}
\end{definition}

Given a partition $\lambda$, one can pictorially represent $\lambda$ using a Young diagram (sometimes called a Ferrers diagram).

\begin{figure}
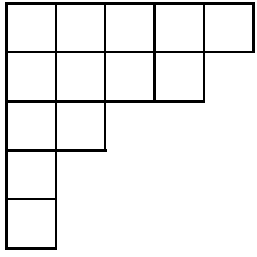
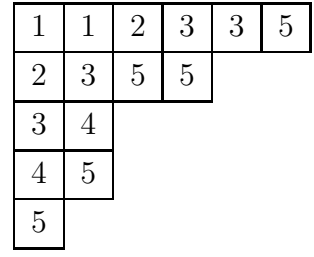

  \begin{subfigure}[t]{0.45\textwidth}
    \centering
    \begin{ytableau}
      \none & & & & & & \none \\
      \none & & & & & \none \\
      \none & & & \none \\
      \none & & \none \\
      \none & & \none
    \end{ytableau}
    \caption[Example of Young diagram.]{The Young diagram for $\lambda=(5,4,2,1,1)$.}\label{fig:young_diagram}
  \end{subfigure}\hfill
  \begin{subfigure}[t]{0.45\textwidth}
    \centering
    \begin{ytableau}
      \none & 1 & 1 & 2 & 3 & 3 & 5 & \none \\
      \none & 2 & 3 & 5 & 5 & \none \\
      \none & 3 & 4 & \none \\
      \none & 4 & 5 & \none \\
      \none & 5 & \none 
    \end{ytableau}
    \caption[Example of a semistandard Young tableau.]{A semistandard Young tableau for $\lambda=(6,4,2,2)$ with entries in $\{1,\dots,5\}$.}\label{fig:ssyt}
  \end{subfigure}
  \caption[Examples of Young diagrams and tableaux]{Examples of a Young diagram and a semistandard Young tableau.}\label{fig:young}
\end{figure}

\begin{definition}[Young diagram]
  Let $\lambda=(\lambda_i)_{i=1}^{l(\lambda)}$ be a partition. The Young diagram of $\lambda$ is a collection of $|\lambda|$ boxes arranged in $l=l(\lambda)$ left-justified rows. The first row has $\lambda_1$ boxes, the second has $\lambda_2$, and so on until the $l$th row which has $\lambda_{l}$ boxes.  Figure~\ref{fig:young_diagram} shows an example of a Young diagram.
\end{definition}

\begin{definition}[Semistandard Young tableau]
  Given $\lambda$, a semistandard Young tableau (SSYT) is a Young tableau of shape $\lambda$  with the additional requirement that entries must strictly increase down columns and weakly increase across rows.  This means that the alphabet must have at least $\max\{\lambda_1,l(\lambda)\}$ symbols. An example is given by figure~\ref{fig:ssyt}.  We denote by $\SSYT_n(\lambda)$ the set of semistandard Young tableaux of shape $\lambda$ with entries in $\{1,\dots,n\}$. 
\end{definition}

\begin{definition}[Schur function]\label{schur_function}
  Let $\lambda$ be a partition and take $n\in\mathbb{N}$, with $n\geq l(\lambda)$. Then the \emph{Schur function} (or Schur polynomial) is
  \begin{align}
    s_\lambda(x_1,\dots,x_n)
    &\coloneqq \sum_{T\in \SSYT_n(\lambda)}x^T\\
    &=\sum_{T\in \SSYT_n(\lambda)}x_1^{t_1}\cdots x_n^{t_n},\label{eq:schur_def}
  \end{align}
  where $t_j=t_j(T)$ is the number of times $j$ appears in $T$.  Note that if $n>l(\lambda)$ then we extend $\lambda$ with zeros until it has length $n$. 
\end{definition}


\section{The  conjectures of Fyodorov \& Keating}\label{sec:fk_conj}

We will now turn our attention to a series of conjectures due to Fyodorov, Hiary, and Keating~\cite{fyohiakea12}, and, in more detail, Fyodorov and Keating~\cite{fyokea14}.  These tie together the statistical mechanics of log-correlated fields (introduced in section~\ref{sec:introduction}), the characteristic polynomials of random unitary matrices, and properties of the Riemann zeta function.  They form the platform for much of the material we shall review here.   

We first state the conjectures in section~\ref{sec:fk_conj_stated}, and then we explain their motivation in section~\ref{sec:fk_conj_recap}. The remaining part of the section is dedicated to reviewing recent progress towards proving them.  

This section only concerns unitary characteristic polynomials.  Thus, until we explicitly state otherwise, $P_N(A,\theta)\equiv P_{\U(N)}(A,\theta)$ will represent the characteristic polynomial of $A\in \U(N)$.

\subsection{Statement of the conjectures}\label{sec:fk_conj_stated}

All of the following are, directly or indirectly, related to maxima of log-correlated fields.  For context, we recall \eqref{eq:log_corr_max} here specialized with $\sigma^2=\frac{1}{2}\log 2$. Specifically, we take $V_n$ to be a metric space and $X_n=\{X_n(v), v\in V_n\}$ to be a log-correlated field with mean zero and $\mathbb{E}[X_n(v)^2]=\sigma^2 n=\frac{n}{2}\log 2$.  We further write $N=2^n$. The maxima of the field is
\begin{equation}\label{eq:log_corr_max_restated}
  \max_{v\in V_n}X_n(v)\approx \log N-\frac{3}{4}\log\log N+\mathcal{M},
\end{equation}
where $\mathcal{M}$ is an $O(1)$ random variable. With these choices of $\sigma^2$ and $N$, the `log-correlated constant' is renormalized to $3/4$, versus the `independent constant' of $1/4$ (compared to the $3/2$ vs. $1/2$ between \eqref{eq:log_corr_max} and \eqref{general_max_gaussian}). 

The first, most general, conjecture is the following~\cite{fyokea14}. 
\begin{conj}\label{conjecturemaxpoly1}
  For $\theta\in[0,L)$, $L\in(0,2\pi]$, and a matrix $A$ sampled uniformly from $\U(N)$,
  \begin{equation}\label{eq:fk_rmt_general}
    \max_{\theta\in[0,L)}\log|P_N(A,\theta)|\sim a_{N_L}+b_{N_L}x_{A,N_L},
  \end{equation}
  as $N_L\rightarrow\infty$, where,
  \begin{equation}\label{eq:fk_rmt_general_coeffs}
    a_{N_L}=\log N_L-\frac{3}{4}\log\log N_L +o(1)\quad\text{and}\quad b_{N_L}=1+O\left(\frac{1}{\log N_L}\right),
  \end{equation}
  and where $N_L\coloneqq NL/2\pi$ is the average number of eigenvalues of the associated $N\times N$ unitary matrix $A$ in the interval $[0,L)$.   The random variable $x_{A,N_L}$ has probability density $p(x_{A,N_L})$. 
\end{conj}

The similarity between conjecture~\ref{conjecturemaxpoly1} and \eqref{eq:log_corr_max_restated} is related to the fact that $\log|P_N(A,\theta)|$ exhibits structure similar to the log-correlated fields described above with respect to $\theta$.  Similar links have been discussed for the imaginary part of $\log P_N(A,\theta)$, see for example Fyodorov \& Le Doussal~\cite{fyoled20}, but we shall concentrate here on the real part. 

The `full circle' case (i.e. $L=2\pi$) is arguably the most interesting, and is where the most progress has been made.  Conjecture~\ref{conjecturemaxpoly1} then takes the following form.  
\begin{conj}\label{conj:fk_rmt}
  For $A\in \U(N)$ sampled uniformly, 
  \begin{equation}\label{eq:fk_rmt1}
    -2\max_{\theta \in[0,2\pi)}\log|P_N(A,\theta)|=-2\log N+\frac{3}{2}\log\log N+y_{A,N},
  \end{equation}
  where $(y_{A,N}, N\in\mathbb{N})$ is a sequence of random variables which converge in distribution\footnote{The reason for introducing a factor of $-2$ in \eqref{eq:fk_rmt1} is explained in section~\ref{sec:fk_conj_recap}.}.  
\end{conj}
Fyodorov and Keating further conjecture that the random variable $y_{A,N}$ should converge in distribution to $y=\mathscr{G}_1+\mathscr{G}_2$, a sum of two independent Gumbel random variables. Thus $p(y)$, the probability density for $y$, decays like $y\exp(-y)$ as $y\rightarrow\infty$:
\begin{equation}\label{eq:gumbel_prob}
  p(y)=2e^{-y}K_0\left(2e^{-\frac{y}{2}}\right)\overset{y\rightarrow-\infty}{\sim} -ye^{y},
\end{equation}
where $K_0$ is a modified Bessel function of the second kind.

In light of section~\ref{sec:rmt_and_nt}, a link between conjecture~\ref{conj:fk_rmt} and the maximum of the Riemann zeta function might be expected.  
\begin{conj}\label{conj:fk_nt}
  Let $t\sim U[T,2T]$ (i.e. $t$ is taken uniformly from the interval $[T,2T]$).  Then  
  \begin{equation}\label{eq:fk_nt1}
    \max_{h\in[0,2\pi)}\log|\zeta(\tfrac{1}{2}+i(t+h))|=\log\log T -\frac{3}{4}\log\log\log T + x_t.
  \end{equation}  
\end{conj}
\noindent Here the random variable $x_t$ is expected to have a limiting value distribution as $t\to\infty$ similar\footnote{Since unlike characteristic polynomials, the $\zeta$-function is not $2\pi$-periodic.} to~\eqref{eq:gumbel_prob}. 

In section~\ref{sec:fk_conj_recap}, we review the heuristic calculation that is used to justify conjecture~\ref{conj:fk_rmt} (and hence also conjecture~\ref{conj:fk_nt}). The final conjecture of Fyodorov and Keating we shall focus on in this review is also based on the same heuristic calculation (and, if established in full generality, would provide a method for proving conjectures~\ref{conj:fk_rmt} and~\ref{conj:fk_nt}). We define first the moment,
\begin{equation}\label{eq:moments_fixed_matrix}
  g_N(\beta;A)\coloneqq\frac{1}{2\pi}\int_0^{2\pi}|P_N(A,\theta)|^{2\beta}d\theta.
\end{equation}
Note, this differs from the moments \eqref{eq:unitary_moment} in that the average is over the spectrum for a fixed matrix $A$.  Taking $A$ now to be a random matrix, $g_N(\beta;A)$ becomes a random variable. Thus, we may additionally define the moments of $g_N$ with respect to the Haar measure on $\U(N)$.  Performing this double average is called taking the \emph{moments of moments}:
\begin{equation}\label{def:mom_lit_review}
  \mom_{\U(N)}(k,\beta)\coloneqq \int_{\U(N)}g_N(\beta;A)^k dA.
\end{equation}
The final conjecture of Fyodorov and Keating~\cite{fyokea14} that we consider gives the asymptotic behaviour of $\mom_{\U(N)}(k,\beta)$.
\begin{conj}\label{conj:fk_mom}
  For $k\in\mathbb{N}$, as $N\to\infty$
  \begin{equation}\label{eq:fk_mom}
    \mom_{\U(N)}(k,\beta)\sim
    \begin{dcases}
      \left(\frac{\mathcal{G}^2(1+\beta)}{\mathcal{G}(1+2\beta)}\right)^k\frac{\Gamma(1-k\beta^2)}{\Gamma^k(1-\beta^2)} N^{k\beta^2}, &\text{if }k<\tfrac{1}{\beta^2},\\
      \gamma_{k,\beta}N^{k^2\beta^2-k+1},&\text{if }k>\tfrac{1}{\beta^2}.
    \end{dcases}
  \end{equation}
\end{conj}
\noindent and at the transition point $k=\beta^{-2}$, the moments of moments grow like $N\log N$, see section~\ref{sec:moments}.

Conjectures~\ref{conj:fk_rmt} and~\ref{conj:fk_nt} will be the focus of the rest of this section.  Justification for them is given in section~\ref{sec:fk_conj_recap}. Significant progress has been made on both fronts and is covered in section~\ref{sec:fk_progress}. Conjecture~\ref{conj:fk_mom} is the main motivation behind the work of section~\ref{sec:moments}. 

\subsection{Justification for conjectures~\ref{conj:fk_rmt},~\ref{conj:fk_nt}, and~\ref{conj:fk_mom}}\label{sec:fk_conj_recap}

Since conjecture~\ref{conj:fk_rmt} concerns the maximum of the real part of the logarithm of characteristic polynomials, it will be convenient to define
\begin{equation}\label{eq:fk_rePN}
  V_N(A,\theta)\coloneqq -2\log|P_N(A,\theta)|.
\end{equation}
The reasons for the $-2$ coefficient will become clear in what follows. Theorem~\ref{thm:ks-clt} implies that $V_N(A,\theta)$ satisfies a central limit theorem,
\begin{equation}\label{eq:vn_clt}
  V_N(A,\theta)\sim \cN(0,2\log N).
\end{equation}
Furthermore, Hughes et al.\cite{hugkeaoco01} show that
\begin{equation}\label{eq:expansion_log}
 \log|P_N(A,\theta)|=-\RE\sum_{n=1}^\infty \frac{1}{\sqrt{n}}\frac{\Tr A^n}{\sqrt{n}}e^{-i n\theta}.
\end{equation}
The following result\footnote{Their result can also be deduced from the Strong Szeg\" o theorem~\cite{sze15, sze52}.} of Diaconis and Shahshahani~\cite{diasha94}  concerning powers of traces of $A$ (see as well~\cite{hkssz96}) is also useful. 
\begin{theorem}[Diaconis and Shahshahani~\cite{diasha94}]\label{thm:diasha}
  Take a sequence $(X_j)_{j=1}^\infty$ of independent and identically distributed complex random variables whose real and imaginary parts are centred Gaussians with variance $1/2$.  Then, for any fixed $k$ and $A\in \U(N)$, as $N\rightarrow\infty$,  
  \begin{equation}
    \left(\Tr A, \tfrac{1}{\sqrt{2}}\Tr A^2, \dots, \tfrac{1}{\sqrt{k}}\Tr A^k\right)\overset{d}{\longrightarrow}\left(X_1,\dots,X_k\right).
  \end{equation}
\end{theorem}

Hence, the coefficients $\Tr A^n/\sqrt{n}$ in \eqref{eq:expansion_log} tend to independent and identically distributed complex Gaussian variables.  By \eqref{eq:expansion_log} together with theorem~\ref{thm:diasha}, see for example~\cite{fyokea14}, one can show that as $N\rightarrow\infty$,
\begin{equation}\label{eq:vn_corr}
  \mathbb{E}[V_N(A,\theta_1)V_N(A,\theta_2)]\approx
  \begin{dcases}
    -2\log|\theta_1-\theta_2|, & \text{for }\frac{1}{N}\ll |\theta_1-\theta_2|\ll 1\\
    2\log N, & \text{for }|\theta_1-\theta_2|\ll \frac{1}{N}.
  \end{dcases}
\end{equation}
Hence $V_N(A,\theta)$ (and so $P_N(A,\theta)$) behaves like a log-correlated Gaussian random function, and so one might expect its maximum to fall into the corresponding universality class. 

Similarly, for the Riemann zeta function, one defines
\begin{equation}\label{eq:fk_rez}
  V_{\zeta}(t,h)\coloneqq-2\log\left|\zeta\left(\tfrac{1}{2}+i(t+h)\right)\right|,
\end{equation}
for fixed $t\in\mathbb{R}$. By Selberg's central limit theorem (cf. theorem~\ref{thm:selbergclt}) we have that $V_{\zeta}(t,h)$ converges to a Gaussian distribution with a dependence on the shift $h$, as $t\rightarrow\infty$
\begin{equation}\label{eq:selberg_clt_restated}
  V_\zeta(t,h)\sim \cN\left(0,2\log\log(t+h)\right).
\end{equation}
Moreover, the correlation for two points $h_1,h_2$ can be calculated in the following way (see~\cite{bou10} or the appendix of~\cite{fyokea14} for details).  Firstly, using the Euler product for zeta to expand each $V_\zeta(t,h_j)$,
\begin{align}
  \mathbb{E}[V_\zeta(t,h_1)&V_\zeta(t,h_2)]\nonumber\\
  &\quad=4\sum_{p_1,p_2}\sum_{n_1,n_2=1}^\infty \frac{1}{n_1}\frac{1}{n_2}\frac{1}{p_1^{n_1/2}}\frac{1}{p_2^{n_2/2}}\mathbb{E}[\cos(n_1(t+h_1)\log p_1)\cos(n_2(t+h_2)\log p_2)]\label{eq:expanded_zeta_corr}
\end{align}
where the expectation is the average over $[t-h/2,t+h/2]$ for some $h$ satisfying $1/\log t\ll h\ll t$. Note that in the large $t$ limit this interval will contain an increasing number of zeros. Then by expanding the product of cosines in \eqref{eq:expanded_zeta_corr} and using that the main term comes from the diagonal contribution ($p_1=p_2$, $n_1=n_2$) with standard prime estimates, one finds
\begin{equation}\label{eq:vz_corr}
  \mathbb{E}[V_\zeta(t,h_1)V_\zeta(t,h_2)]\approx
  \begin{dcases}
    -2\log|h_1-h_2|,&\text{for }\frac{1}{\log t}\ll|h_1-h_2|\ll1\\
    2\log\log t, &\text{for }|h_1-h_2|\ll\frac{1}{\log t}.
 \end{dcases}
\end{equation}
The similarity to the covariance of $V_N(A,\theta)$ is evident.  Once again there is a dependence on the distance between the points $h_1, h_2$.  If $h_1$ is very close to $h_2$ on the scale of mean zero spacing, then essentially $V_\zeta(A,h_1)$ and $V_\zeta(t,h_2)$ are perfectly correlated.  However, if they are separated on the same scale as $\theta_1$ and $\theta_2$ must be in \eqref{eq:vn_corr} (i.e. making the usual identification $N\sim \log t$), then instead one sees the logarithmic correlation.

The fact that $V_\zeta$ and $V_N$ are both log-correlated has important ramifications. Recall that Selberg's central limit theorem reveals that on average $\log|\zeta(1/2+it)|$ is on the order of $\sqrt{(1/2)\log\log t}$. The Lindel\"of Hypothesis states that
\begin{equation}\label{eq:lindelof}
  |\zeta(\tfrac{1}{2}+it)|=o(t^\vareps)
\end{equation}
for any $\vareps>0$.  Under the Riemann hypothesis, one has that
\begin{equation}\label{osizeofzeta}
  |\zeta(\tfrac{1}{2}+it)|=O\left(\exp\left(\frac{c_1\log t}{\log\log t}\right)\right),
\end{equation}
for some constant $c_1$ (see for example~\cite{titchmarsh86}). However, it is also known (unconditionally) that\footnote{In \eqref{omegasizeofzeta}, $f(x)=\Omega(g(x))$ means $f(x)$ takes the value $g(x)$ infinitely often.}
\begin{equation}\label{omegasizeofzeta}
  |\zeta(\tfrac{1}{2}+it)|=\Omega\left(\exp\left(\sqrt{\frac{\log t}{\log\log t}}\right)\right).
\end{equation}
Hence, the extreme values of the zeta function must lie between \eqref{osizeofzeta} and \eqref{omegasizeofzeta}.

If $\log|\zeta(1/2+i(t+h_1))|$ and $\log|\zeta(1/2+i(t+h_2))|$ were independent of each other, then the Fisher-Tippett-Gnedenko theorem (see section~\ref{sec:introduction}, section~\ref{sec:evt}, and theorem~\ref{maxofnormals}) could be applied.  Such an assumption was examined by Montgomery in the context of estimating the typical size of $\log|\zeta(1/2+it)|$.  If one sets $X_1,\dots,X_n$ to be $n$ local maxima of $\log|\zeta(1/2+it)|$, assumed to be independent, then theorem~\ref{maxofnormals} would imply that 
\begin{equation}\label{eq:max_indep_zeta}
  \max_{t\in[1,T]}|\zeta(\tfrac{1}{2}+it)|=O\left(\exp\left(c_2\sqrt{\log T\log\log T}\right)\right)
\end{equation}
for some constant $c_2$. The log of the right-hand side here is considerably larger than $\sqrt{(1/2)\log\log t}$, the typical value of $\log|\zeta(1/2+it)|$, and is closer to \eqref{omegasizeofzeta} than \eqref{osizeofzeta}.  However, as demonstrated, the assumption of independence is incorrect. 

Bondarenko and Seip~\cite{bonsei17, bonsei18} have shown that
\begin{equation}
  \max_{t\in[1,T]}|\zeta(\tfrac{1}{2}+it)|\geq \exp\left((1+o(1))\sqrt{\frac{\log T\log\log\log T}{\log\log T}}\right).
\end{equation}
This has recently been improved to
\begin{equation}\label{eq:large_zeta_nt}
  \max_{t\in[1,T]}|\zeta(\tfrac{1}{2}+it)|\geq \exp\left((\sqrt{2}+o(1))\sqrt{\frac{\log T\log\log\log T}{\log\log T}}\right)
\end{equation}
by de la Bret\`eche and Tenenbaum~\cite{breten19}. Both results are unconditional, and more generally cover intervals of length $[T^\alpha,T]$ for $\alpha\in[0,1)$.

Using techniques from random matrix theory, Farmer, Gonek and Hughes~\cite{fargonhug07} have conjectured that
\begin{equation}\label{eq:large_zeta_rmt}
  \max_{t\in[1,T]}|\zeta(\tfrac{1}{2}+it)|=\exp\left(\Bigl(\tfrac{1}{\sqrt{2}}+o(1)\Bigr)\sqrt{\log T\log\log T}\right).
\end{equation}

Comparatively, the maximum in short intervals (ranges of length $O(1)$) is more tractable, both theoretically and numerically. In particular, since the ranges of \eqref{eq:fk_rmt1} and \eqref{eq:fk_nt1} are considerably shorter than the $O(T)$ lengths considered in \eqref{eq:max_indep_zeta}--\eqref{eq:large_zeta_rmt}, experimental calculations are feasible~\cite{fyognukea18, aabhr21}. 

We now focus our attention on the techniques used by Fyodorov and Keating to construct the precise form of conjectures~\ref{conj:fk_rmt},~\ref{conj:fk_nt}, and~\ref{conj:fk_mom}. This technique is inspired by a class of problems from statistical physics.

In a statistical mechanics context, Fyodorov and Bouchaud~\cite{fyobou08} studied a related \emph{circular-logarithmic} model.  Rather than defining $V_N(A,\theta)$ or $V_\zeta(t,h)$ (cf.~\eqref{eq:fk_rePN} and \eqref{eq:fk_rez}), one takes a sequence $\{V_1,\dots, V_M\}$ of random, centred Gaussian variances sampled from the two-dimensional Gaussian Free Field equidistantly along the unit circle.  Once again, this is a log-correlated process.  In~\cite{fyobou08}, Fyodorov and Bouchard showed that in the large $M$ limit, the (positive, integer) moments\footnote{The expectation in \eqref{eq:fb_partition} is with respect to the Gaussian random variables.} of the partition function
\begin{equation}\label{eq:fb_partition}
  \mathbb{E}\left[\left(\frac{1}{M}\sum_{i=1}^M e^{-\beta V_i}\right)^k\right]
\end{equation}
can be calculated using the Dyson-Morris version of the Selberg integral.  Explicitly, they show that
\begin{equation}\label{eq:mom_fyobou}
  \mathbb{E}\left[\left(\frac{1}{M}\sum_{i=1}^M e^{-\beta V_i}\right)^k\right]\sim
  \begin{dcases}
    \frac{\Gamma(1-k\beta^2)}{\Gamma^k(1-\beta^2)}M^{k\beta^2},&\text{for }1<k<1/\beta^2\\
    \alpha_{k,\beta}M^{k^2\beta^2-k+1},&\text{for }k>1/\beta^2
  \end{dcases}
\end{equation}
as $M\rightarrow\infty$, for some $O(1)$ leading order coefficient $\alpha_{k,\beta}$.  Fyodorov and Bouchard were also able to use the moments~\eqref{eq:mom_fyobou} to construct the probability density of the partition function in the `high-temperature' range $|\beta|<1$. It is instructive to compare~\eqref{eq:fk_mom} to~\eqref{eq:mom_fyobou}.

A trivial rewriting of $g_N(\beta;A)$ from \eqref{eq:moments_fixed_matrix} reveals
\begin{equation}\label{eq:fixed_a_moment}
  g_N(\beta;A)\coloneqq\frac{1}{2\pi}\int_0^{2\pi}|P_N(A,\theta)|^{2\beta} d\theta\equiv\frac{1}{2\pi}\int_0^{2\pi}e^{-\beta V_N(A,\theta)}d\theta,
\end{equation}
where $\beta>0$. In light of the above discussion, $g_N$ can be viewed as a partition function. Then, $V_N(A,\theta)$ is the `energy' for the system and $\beta$ is the inverse temperature. The reason for including $-2$ in the definition of $V_N(A,\theta)$ and $V_\zeta(t,h)$ is now also apparent.

A related, important function (cf. section~\ref{sec:logcorrintro}) is the corresponding \emph{free energy} for the system:
\begin{equation}\label{eq:free_energy}
  \mathcal{F}(\beta)\coloneqq -\frac{1}{\beta}\log g_N(\beta;A).
\end{equation}
The maximum of $\log|P_N(A,\theta)|$ can be recovered as the large-$\beta$ limit
\begin{equation}\label{free_energy_maximum}
  \lim_{\beta\rightarrow\infty}\mathcal{F}(\beta)=-2\max_{\theta\in[0,2\pi)}\log|P_N(A,\theta)|.
\end{equation}
A similar construction can be made for $\log|\zeta(1/2+it)|$.

Fyodorov and Keating demonstrate that the free energy~\eqref{eq:free_energy} has a `freezing' property (related to the temperature parameter $\beta$). Define the \emph{normalized} free energy to be
\begin{equation}
  \mathscr{F}(\beta)\coloneqq -\frac{1}{\beta\log N}\log\left(Ng_N(\beta;A)\right).
\end{equation}
By considering the average of $\mathscr{F}(\beta)$ with respect to the Haar measure, they argue (see~\cite{fyokea14}) that for $\beta$ small (i.e. high temperature) the average of $\mathscr{F}(\beta)$ is governed by the typical values of $P_N(A,\theta)$.  By theorem~\ref{thm:ks}, the moments behave like $N^{\beta^2}$. Hence, for small $\beta$ one expects
\begin{equation}
   -\mathbb{E}\left[\mathscr{F}(\beta)\right]\overset{N\rightarrow\infty}{\sim} \beta+\frac{1}{\beta}.
\end{equation}

However, as $\beta$ grows large (i.e. as temperature decreases), the free energy will instead be governed by the extreme values (cf.\eqref{free_energy_maximum}), which by conjecture~\ref{conj:fk_rmt} scale as $\log N$ to leading order.  Hence in the large $N$ limit for large $\beta$, instead
\begin{equation}
  -\mathbb{E}\left[\mathscr{F}(\beta\rightarrow\infty)\right]\rightarrow 2.
\end{equation}

This is the meaning of \emph{freezing} for this system: as the temperature moves from small to large $\beta$ (i.e. temperature decreases), the free energy reaches a critical temperature and thereafter remains constant.  Here, the critical temperature is $\beta=1$:
\begin{equation}\label{eq:log-corr_freezing}
  -\mathbb{E}\left[\mathscr{F}(\beta)\right]\sim
  \begin{dcases}
    \beta+\frac{1}{\beta}&\text{if }\beta\leq 1\\
    2&\text{if }\beta>1.
  \end{dcases}
\end{equation}

The moments of the random variable $g_N(\beta;A)$ with respect to the Haar measure on $\U(N)$ are, by \eqref{free_energy_maximum}, related to the maximum of $\log|P_N(A,\theta)|$. The exact quantity of interest is
\begin{equation}\label{eq:first_mom}
  \mathbb{E}[g_N(\beta;A)^k]=\mathbb{E}\left[\left(\frac{1}{2\pi}\int_0^{2\pi}|P_N(A,\theta_j)|^{2\beta}d\theta\right)^k\right],
\end{equation}
where $\mathbb{E}[\cdot]$ is the Haar measure on $\U(N)$.

Provided that $k\in\mathbb{N}$, \eqref{eq:first_mom} can be written the $k$-fold integral of a Toeplitz determinant $D_N(f)$ with symbol $f(z)=\prod_{j=1}^k|z-e^{i\theta_j}|^{2\beta}$. The values $z=e^{i\theta_j}$ correspond to so-called `Fisher-Hartwig' singularities. Using Widom's result~\cite{wid73} on the Fisher-Hartwig asymptotic formula, the integrand\footnote{For $k\in\mathbb{N}$, and with the order of integration switched.} of \eqref{eq:first_mom} can be seen to be 
\begin{equation}
  \mathbb{E}\left[\prod_{j=1}^k|P_N(A,\theta_j)|^{2\beta}\right]\propto |\Delta(e^{i\theta_1},\dots,e^{i\theta_k})|^{-\beta^2}
\end{equation}
as $N\rightarrow\infty$.  Thus, provided that the Fisher-Hartwig singularities at $e^{i\theta_1},\dots,e^{i\theta_k}$ can be treated as distinct, one can apply Selberg's integral and find that
\begin{equation}
  \mathbb{E}[g_N(\beta;A)^k]\sim \left(\frac{\mathcal{G}^2(1+\beta)}{\mathcal{G}(1+2\beta)}\right)^k\frac{\Gamma(1-k\beta^2)}{\Gamma^k(1-\beta^2)} N^{k\beta^2}.
\end{equation}

In order to ensure that any coalescences between the Fisher-Hartwig singularities do not influence the leading order asymptotics, one requires the restriction $k<1/\beta^2$, see for example~\cite{fyokea14}. This is the justification (and indeed, an outline of the proof) for this regime of conjecture~\ref{conj:fk_mom}.  Outside of this range, i.e. $k>1/\beta^2$, the Fisher-Hartwig singularities coalesce, and so computing $\mathbb{E}[g_N(\beta;A)^k]$ proves to be much more difficult.  In fact, one either requires a uniform Fisher-Hartwig asymptotic formula valid for when the singularities coalesce, or an alternative approach. It is precisely the contributions from the coalescing singularities that leads to the different leading order behaviour in conjecture~\ref{conj:fk_mom}. See section~\ref{sec:moments} for a more detailed discussion of the moments of $g_N(\beta;A)$. 

This concludes the review of the heuristics behind conjectures~\ref{conj:fk_rmt}, ~\ref{conj:fk_nt}, and ~\ref{conj:fk_mom}. We now summarize recent advances towards proving the two `maxima' Fyodorov-Keating conjectures.  The final conjecture (conjecture~\ref{conj:fk_mom}) concerning the leading order of the \emph{moments of moments} is the subject of section~\ref{sec:moments}.  

\subsection{Progress towards conjectures~\ref{conj:fk_rmt} and~\ref{conj:fk_nt}}\label{sec:fk_progress}

There has been a concerted effort to prove the conjectures outlined in the previous section, resulting in considerable progress in both the random matrix and number theoretic cases.

Typically problems in random matrix theory are more tractable than their equivalent formulations in number theory. Thus, one often uses random matrix results to inform the number theoretic calculation.  However, in this case, a model of the Riemann zeta function, introduced by Arguin, Belius, and Harper~\cite{argbelhar17}, was the catalyst for much of the subsequent progress.  We begin by presenting this model and the main results relating to it. We sketch how these results are proved, focusing in particular on the identification of the approximate branching structure. These techniques are fundamental not only to the result of Arguin, Belius, and Harper~\cite{argbelhar17}, but also to the proofs of the ensuing random matrix and number theoretic results, a discussion of which forms the remaining part of this section.

\subsubsection*{Identification of a branching structure in a model of $\zeta(1/2+it)$}

The primes behave in many respects like a random set of integers.  To try to understand the Riemann zeta function, and in particular the behaviour of its maximum in short intervals, one might try to exploit this pseudo-random structure.  Using work of Soundararajan~\cite{sou09}, Harper~\cite{har13a} constructed a randomized model of the zeta function which captures many of its main statistical features, especially those relating to extreme values.  Subsequently, Arguin et al.~\cite{argbelhar17} developed this model further and established to subleading order an adaptation of the conjecture of Fyodorov and Keating.  A crucial part of this work is the identification of an approximate tree structure.  As will be discussed, it is also possible to demonstrate such structure within the Riemann zeta-function itself, and characteristic polynomials.  

\begin{figure}[!htb]
  \centering
  \begin{subfigure}[t]{.47\textwidth}
    \centering
    \resizebox{4.3cm}{4.05cm}{
      \begin{tikzpicture}[scale=1]
        \begin{axis}[
            axis lines=center,
            xmin=-1.2,
            xmax=1.2,
            ymin=-1.2,
            ymax=1.2,
            grid=none,
            x label style={at={(axis description cs:1,0.5)},anchor=north},
            y label style={at={(axis description cs:0.58,1.03)},anchor=north},
            xtick={-1,1},
            xticklabels={,,},
            ytick={-1,1},
            yticklabels={,,},
            axis equal image
          ]
	      \addplot[
            only marks,
            scatter,
            mark=*,
            mark size=1.5pt,
            scatter src=explicit,
            scatter/use mapped color={fill=black}
          ]
	      table[
            x expr=\thisrowno{0},
            y expr=\thisrowno{1},
          meta expr=\thisrowno{2}
          ]
	      {steinhaus_unitcircle.txt};
        \end{axis}
      \end{tikzpicture}
    }
    \caption{Plotting $50$ evaluations of a Steinhaus random variable.}\label{fig:steinhaus}
  \end{subfigure}\hfill
  \begin{subfigure}[t]{.47\textwidth}
    \centering
    \resizebox{4.3cm}{4.05cm}{
      \begin{tikzpicture}[scale=1]
        \begin{axis}[
            axis lines=center,
            xmin=-1.2,
            xmax=1.2,
            ymin=-1.2,
            ymax=1.2,
            grid=none,
            x label style={at={(axis description cs:1,0.5)},anchor=north},
            y label style={at={(axis description cs:0.58,1.03)},anchor=north},
            xtick={-1,1},
            xticklabels={,,},
            ytick={-1,1},
            yticklabels={,,},
            axis equal image
          ]
	      \addplot[
            only marks,
            scatter,
            mark=*,
            mark size=1.5pt,
            scatter src=explicit,
            scatter/use mapped color={fill=black}
          ]
	      table[
            x expr=\thisrowno{0},
            y expr=\thisrowno{1},
            meta expr=\thisrowno{2}
          ]
	      {primes_unitcircle.txt};
        \end{axis}
      \end{tikzpicture}
    }
    \caption{Plotting $p^{-100i}$ for the first fifty primes $p$.}\label{fig:primes_unitcircle}
  \end{subfigure}
  \caption[Steinhaus random variables compared to $p^{-i \tau}$.]{Comparing $50$ evaluations of a Steinhaus random variable with $p^{-i\tau}$, for the first $50$ primes and $\tau=100$.}\label{fig:primes_v_steinhaus}
\end{figure}
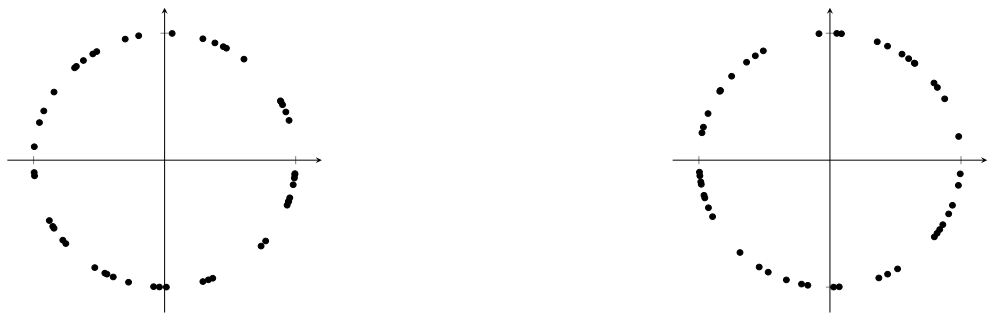

In~\cite{har13a}, under the Riemann hypothesis, Harper proves that there exists a set $H$ of measure at least $0.99$ with $H\subseteq[T,T+1]$ such that
\begin{equation}\label{eq:abh11}
  \log|\zeta(\tfrac{1}{2}+i\eta)|=\operatorname{Re}\Biggl\{\sum_{p\leq T}p^{-(\frac{1}{2}+i\eta)}\frac{\log\tfrac{T}{p}}{\log T}\Biggr\} + O(1),
\end{equation}
for all $\eta\in H$. The set of small measure $[T,T+1]\backslash H$ where the result fails essentially covers those points close to the zeros of zeta (where the logarithm is not well-defined). In order to capture the `quasi-randomness' of the primes, Harper introduces the following random variables. Let $(U_p, p \text{ prime})$ be a sequence of independent random variables distributed uniformly on the unit circle\footnote{In the literature, these are sometimes referred to as \emph{Steinhaus} random variables.}. Heuristically speaking\footnote{One can make the argument more rigorous, see~\cite{har13a}}, $U_p$ models $p^{-i\tau}$, see also figure~\ref{fig:primes_v_steinhaus}.

It is then easy to justify, see for example~\cite{argbelhar17}, that the random variable 
\begin{equation}\label{eq:random_model}
  X_T(h)\coloneqq\sum_{p\leq T}\frac{\operatorname{Re}(p^{-ih}U_p)}{\sqrt{p}}
\end{equation}
for $h\in[0,1]$ is a good model\footnote{In fact in \eqref{eq:random_model}, one could for example replace $U_p$ with $G_p$, a Gaussian random variable; see the discussion near~\eqref{eq:gaussian_model}.}  for $\log|\zeta(1/2+i(\tau+h))|$.

Arguin et al.~\cite{argbelhar17} prove the following. 
\begin{theorem}[Arguin, Belius, Harper~\cite{argbelhar17}]\label{thm:abh}
Let $(\Omega,\mathcal{F},\mathbb{P})$ be a probability space and $(U_p, p\text{ prime})$ be independent random variables distributed uniformly on the unit circle. Then 
\[\max_{h\in[0,1]}\sum_{p\leq T}\frac{\operatorname{Re}(p^{-ih}U_p)}{\sqrt{p}}=\log\log T-\frac{3}{4}\log\log\log T+o_\mathbb{P}(\log\log\log T).\]
The error term $o_\mathbb{P}(\log\log\log T)$ converges to zero in probability when divided by $\log\log\log T$.
\end{theorem}

In other words, the maximum over $h\in[0,1]$ of the model for $\log|\zeta(1/2+i(\tau+h))|$ matches conjecture~\ref{conj:fk_nt} to both leading and subleading order.

We now outline the key technique that enabled Arguin et al.\ to prove the above result, concentrating in particular on the identification of the approximate branching structure. We then discuss the general method of proof that would be employed if one instead had an \emph{exact} branching structure. It is this recipe that proves key to the success of the proof of theorem~\ref{thm:abh}, and to the progress towards conjectures~\ref{conj:fk_rmt} and ~\ref{conj:fk_nt} detailed later within this section.

In order to make the comparison with the branching random walk clear, set without loss of generality $T=e^{2^n}$ for some large $n\in\mathbb{N}$, where $T$ is the height at which the interval is situated up the critical line. Now set 
\begin{align}
  M_{2^n}
  &=n\log 2 -\frac{3}{4}\log n + O(1),\\
  M_{2^n}(\vareps)&=n\log2-\frac{3}{4}\log n+\varepsilon\log n,
\end{align}
so $M_{2^n}$ is the expected maximum of a log-correlated process with $ 2^n\equiv \log T$ points. 

Define the random process
\begin{equation}\label{zetamodel}
  \left(X_n(h)\coloneqq \sum_{p\leq e^{2^n}}\frac{\operatorname{Re}(p^{-ih}U_p)}{\sqrt{p}}, h\in[0,1]\right).
\end{equation}
Thus, theorem~\ref{thm:abh} follows if one can show
\begin{equation}\label{altthm:abh}
  \lim_{n\rightarrow\infty}\mathbb{P}\left(M_{2^n}(-\varepsilon)\leq\max_{h\in[0,1]}X_n(h)\leq M_{2^n}(\varepsilon)\right)=1,\quad\forall\varepsilon>0.
\end{equation}

To unveil the branching structure within $X_n(h)$, denote the summands in \eqref{zetamodel} by
\begin{equation}
  W_p(h)\coloneqq\frac{\RE(U_p p^{-ih})}{\sqrt{p}}.
\end{equation}
Straightforwardly,
\begin{align}
  \mathbb{E}[X_n(h_1)]&=0\label{eq:model_mean}\\
  \mathbb{E}\left[W_p(h_1)W_p(h_2)\right]&=\frac{1}{2p}\cos(|h_1-h_2|\log p)\label{eq:w_p_covariance}\\
  \mathbb{E}\left[X_n(h_1)X_n(h_2)\right]&=\frac{1}{2}\sum_{p\leq e^{2^n}}\frac{\cos(|h_1-h_2|\log p)}{p}\label{model_covariance}
\end{align} 
for $h_1,h_2\in[0,1]$. The first calculation \eqref{eq:model_mean} follows by symmetry.  For \eqref{eq:w_p_covariance}, one first rewrites $\RE(U_{p_j} p_j^{-ih_j})$ for $j=1,2$ using $U_p$ and $\overline{U_p}$ and uses the fact that all off-diagonal terms are zero in expectation. The final equality follows immediately from \eqref{eq:w_p_covariance} and \eqref{zetamodel}.

Thus the covariance of $X_n(h)$ depends on the distance between the points $h_1$ and $h_2$ in a \emph{logarithmic} way:
\begin{equation}
  \mathbb{E}\left[X_n(h_1)X_n(h_2)\right]\approx
  \begin{dcases}
    \frac{1}{2}\log|h_1-h_2|^{-1}, &\text{if }|h_1-h_2|\geq 2^{-n},\\
    \frac{1}{2}\log 2^n, &\text{if }|h_1-h_2|<2^{-n}.
  \end{dcases}
\end{equation}
That is, if $|h_1-h_2|\geq 2^{-n}$  (the scale of separation on average) then one can estimate $\frac{1}{2}\log|h_1-h_2|^{-1}$ from \eqref{model_covariance}. However, if the two points lie closer together than $2^{-n}$, i.e. they lie `unusually' close, then effectively they are almost perfectly correlated
\begin{equation}
  \mathbb{E}\left[X_n(h_1)X_n(h_2)\right]=\frac{1}{2}\sum_{p\leq e^{2^n}}\frac{1}{p}\approx\frac{1}{2}\log2^n.
\end{equation}
The above calculation will be compared to the situation with exact branching shortly.

To exhibit the source of the branching structure, break the sum in \eqref{zetamodel} in to dyadic-like partitions, defining
\begin{equation}\label{eq:multiscale}
  Y_m(h)\coloneqq\sum_{2^{m-1}<\log p\leq2^m}W_p(h)=\sum_{2^{m-1}<\log p\leq2^m}\frac{\operatorname{Re}(U_pp^{-ih})}{\sqrt{p}}.
\end{equation}
Hence, by \eqref{eq:w_p_covariance},
\begin{align}
  \mathbb{E}[Y_m(h)^2]&=\sum_{2^{m-1}<\log p\leq 2^m}\frac{1}{2p},\label{eq:abh1}\\
  \mathbb{E}[Y_m(h_1)Y_m(h_2)]&=\frac{1}{2}\sum_{2^{m-1}<\log p\leq 2^m}\frac{\cos(|h_1-h_2|\log p)}{p}\label{eq:abh2}.
\end{align}
Thus the model of the Riemann zeta function can be written in the form of a branching random walk with increments $Y_m$,
\begin{equation}\label{eq:zeta_branching}
  X_n(h)=\sum_{m=0}^n Y_m(h).
\end{equation}

The following lemma of Arguin et al.\ gives the distance at which two walks $X_n(h_1)$ and $X_n(h_2)$ become essentially uncorrelated. For ease of notation write 
\begin{equation}\label{wedgedef}
  h_1\wedge h_2\coloneqq\lfloor\log_2 |h_1-h_2|^{-1}\rfloor,
\end{equation}
cf. also definition~\ref{def:lca}.  

\begin{lemma}[Arguin et al.~\cite{argbelhar17}]\label{lemma:abh1}
  For $h_1,h_2\in\mathbb{R}$, $m\geq1$, and some constant $c$,
  \begin{align}
    \mathbb{E}[Y_m(h_1)^2]&=\frac{1}{2}\log2+O\left(e^{-c\sqrt{2^m}}\right),\label{eq:abh3}\\
    \mathbb{E}[Y_m(h_1)Y_m(h_2)]&=
    \begin{dcases}\label{eq:abh4}
      \frac{1}{2}\log 2+O\left(2^{-2(h_1\wedge h_2-m)}\right) +O\left(e^{-c\sqrt{2^m}}\right),& \text{if }m\leq h_1\wedge h_2,\\
      O\left(2^{-(m-h_1\wedge h_2)}\right), & \text{if }m>h_1\wedge h_2.
    \end{dcases}
  \end{align}
\end{lemma}
The proof follows from a strong form of the prime number theorem and integration by parts. Often, one has to handle the case of $m=0$ (i.e. the contribution from small primes) separately, which is the cause of the requirement $m\geq 1$ in the statement of lemma~\ref{lemma:abh1}. 

Lemma~\ref{lemma:abh1} shows that the $Y_m(h)$, which act as the increments of the branching random walk, are essentially perfectly correlated when $h_1$ and $h_2$ lie close, relative to $m$ (which is the analogue of depth in the binary tree).  Otherwise, effectively, they are perfectly uncorrelated.

The proof of theorem~\ref{thm:abh} is inspired by the techniques that one would use if $X_n(h)$ were an exact branching random walk. Thus, we now outline the method in this precise setting, and comment how Arguin et al.\ are able to adapt this proof to the approximate situation for the model of the Riemann zeta function.  As mentioned above, this method is also used in a number of other key results relating to conjectures~\ref{conj:fk_rmt} and~\ref{conj:fk_nt}.

\subsubsection*{General method of proof}

Take a binary tree of depth $n$ so that its $2^n$ leaves are equally spaced within the interval $[0,1]$. To each branch attach an independent, centred Gaussian random variable with variance $\sigma^2=\frac{1}{2}\log 2$. The $m$ levels, $m=1,\dots,n$ correspond to the `dyadic' decomposition of the primes for the Riemann zeta model~\eqref{eq:multiscale} (again, the contribution from small primes, $m=0$, is handled separately).

Usually we write $\lca(l_1,l_2)$ for the level of the lowest common ancestor of two leaves $l_1,l_2$ (cf. definition~\ref{def:lca}), but to emphasize the analogy with the model of $\zeta(s)$, we here write $l_1\wedge l_2$ (cf. \eqref{wedgedef}).

The general method used to prove theorem~\ref{thm:abh} is based on the following recipe, applied to the exact branching random walk (see also~\cite{arg16} and the introduction of~\cite{argbelhar17}). 

One first identifies a branching structure (either exact, or approximate).  For the branching random walk, this is trivially
\begin{equation}
  X_n(l)=\sum_{m=1}^n Y_m(l),
\end{equation}
where $Y_m(l)\sim\cN(0,\frac{1}{2}\log 2)$.

Consider now the number of exceedances,
\begin{equation}
  Z(t)\coloneqq \left|\{1\leq l\leq 2^n: X_n(l)\geq t\}\right|.
\end{equation}
The relationship between maxima and $Z(t)$ is clearly the following
\begin{equation}
  \max_{l\in\{1,\dots,2^n\}}X_n(l)\geq t \quad\iff\quad Z(t)\geq 1.
\end{equation}
For questions relating to maxima over a continuous range, it is often necessary at this stage to additionally show that one can discretize the interval. The discretization will depend on the (approximate) branching structure. In order to identify the correct size of the maximum of $X_n(h)$, one finds $t$ so that $\mathbb{P}(Z(t)\geq 1)=o(1)$ (cf. theorem~\ref{thm:bramson}). 

Thus, one proceeds to bound $\mathbb{P}(Z(t)\geq 1)$. An upper bound is attained through a union bound,
\begin{equation}\label{eq:union_upper_bound}
  \mathbb{P}(Z(t)\geq1)\leq 2^n\mathbb{P}(X_n\geq t),
\end{equation}
where $X_n\sim\cN(0,\frac{1}{2}\log 2^n)$. Standard Gaussian tail estimates give
\begin{align}
  \mathbb{P}(X_n\geq t)&\approx \frac{\sqrt{n}}{t} \exp\left(-\frac{t^2}{\log 2^n}\right)
\end{align}
implying that $\mathbb{P}(Z(t)\geq 1)\leq o(1)$ when $t=t(n)=\log 2^n-(\frac{1}{4}-\vareps)\log n$.  Recall from section~\ref{sec:evt}, that this is approximately the correct order of the maximum for \emph{independent} Gaussian random variables.  Instead here the random variables are log-correlated and hence the maximum should be lower. Shortly, we discuss an adaptation to this upper bound that delivers the correct size of the maximum. 
The Paley-Zygmund inequality delivers a lower bound,
\begin{equation}\label{eq:paley_zygmund}
  \mathbb{P}(Z(t)\geq 1)\geq\frac{\mathbb{E}[Z(t)]^2}{\mathbb{E}[Z(t)^2]}.
\end{equation}
One can show for the branching random walk that the second moment is exponentially larger than the first moment squared (see~\cite{argbelhar17, argbelbou17}); it is inflated by those exceeding walks `pulling up' neighbours.

Altogether, this implies that $Z(t)$ is not the right quantity to consider. Instead, one alters the definition of $Z(t)$ to take into account the additional structure underpinning the model.  In~\cite{argbelhar17}, it is shown that with high probability a walk $X_m(l)$ up to level $1\leq m\leq n$ lies below a \emph{linear barrier} $\log 2^m+B$ for some $B$ growing slowly with $n$. Hence, if instead one works with
\begin{equation}\label{eq:new_event}
  \tilde{Z}(t)\coloneqq |\{1\leq l\leq 2^n: X_n\left(l\right)\geq t, \text{ and }X_m\left(l\right)\leq \log 2^m+B, \forall m\leq n\}|
\end{equation}
then (with some modifications) calculating the upper and lower bounds \eqref{eq:union_upper_bound} and \eqref{eq:paley_zygmund} with $\tilde{Z}(t)$ replacing $Z(t)$ is precisely the right approach.

Making such a method rigorous for functions with \emph{approximate} branching structure, such as the model of the Riemann zeta function described above, is where the technicalities lie.  

\subsubsection*{Progress Towards Conjecture~\ref{conj:fk_rmt}}
 
Conjecture~\ref{conj:fk_rmt} states that, for $A\in \U(N)$ sampled uniformly with respect to the Haar measure, we expect
\begin{equation}
  \max_{\theta \in[0,2\pi)}\log|P_N(A,\theta)|=\log N-\frac{3}{4}\log\log N+x_{A,N},
\end{equation}
where $(x_{A,N},N\in\mathbb{N})$ is a sequence of random variables which converge in distribution.

The first step towards a proof of this conjecture was made by Arguin, Belius, and Bourgade~\cite{argbelbou17}.  The following is their main theorem (theorem 1.2 in~\cite{argbelbou17}), establishing the conjecture to leading order. 

\begin{theorem}[Arguin, Belius, Bourgade~\cite{argbelbou17}]\label{thm:abb}
For $N\in \mathbb{N}$, let $A\in \U(N)$ be sampled uniformly with respect to the Haar measure. Then 
\[\lim_{N\rightarrow\infty}\frac{\max_{\theta\in[0,2\pi)}\log |P_N(A,\theta)|}{\log N}=1\quad\text{in probability.}\]
\end{theorem}

Soon after, the work of Paquette and Zeitouni~\cite{paqzei17} verified that the subleading term in the conjecture is correct. Precisely, their main result (theorem 1.2 in~\cite{paqzei17}) is as follows.
 
\begin{theorem}[Paquette and Zeitouni~\cite{paqzei17}]\label{thm:pz}
For $N\in \mathbb{N}$, let $A\in \U(N)$ be sampled uniformly. Then
\[\lim_{N\rightarrow\infty}\frac{\max_{\theta\in[0,2\pi)}\log |P_N(A,\theta)|-\log N}{\log\log N}=-\frac{3}{4}\quad\text{in probability.}\]
\end{theorem}

Importantly, their work establishes the constant $-3/4$, the characteristic coefficient in the subleading term for processes with logarithmic correlations.  Finally, the best result currently is due to Chhaibi, Madaule, and Najnudel~\cite{chhmadnaj18} who prove the conjecture up to tightness (theorem 1.2 in~\cite{chhmadnaj18}).
\begin{theorem}[Chhaibi, Madaule, Najnudel~\cite{chhmadnaj18}]\label{thm:cmn}
If $A\in \U(N)$ is chosen uniformly with respect to the Haar measure, and $\{\theta_1,\dots,\theta_N\}$ is the set of eigenphases of $A$, then the family of random variables
\begin{equation}\label{eq:cmn}
\left(\max_{\theta\in[0,2\pi)\backslash\{\theta_1,\dots,\theta_N\}}\log |P_N(A,\theta)|-\left(\log N-\frac{3}{4}\log\log N\right)\right)_{N\geq 2}
\end{equation}
is tight.
\end{theorem}
Chhaibi et al.\ in fact prove a stronger statement concerning the $\operatorname{C\beta E}$, the \emph{Circular $\beta$ Ensemble}, the probability distribution on $N$ points on the unit circle with density,
\begin{equation}\label{eq:cbe}
  \frac{\Gamma\left(1+\tfrac{\beta}{2}\right)^N}{(2\pi)^N\Gamma(1+N\tfrac{\beta}{2})}\prod_{1\leq j<k\leq N}|e^{i\theta_j}-e^{i\theta_k}|^\beta d\theta_1\cdots d\theta_N.
\end{equation}
Theorem~\ref{thm:cmn} is a specialization to $\beta=2$ (the $\CUE$) of a more general result that holds for all $\beta$\footnote{Associated matrix models for general $\beta$ have been constructed~\cite{kilnen04}.}.  In~\cite{chhmadnaj18}, the imaginary counterpart to \eqref{eq:cmn} is also handled. 

Hence only the identification of the distribution of the fluctuating term of conjecture~\ref{conj:fk_rmt} remains. Recall that this is conjectured to be a sum of two independent Gumbel random variables.  Such a statement has been proved~\cite{rem19} for yet another related model, which we discuss shortly, but not for characteristic polynomials.  At time of writing, the state of the art for the number-theoretic maxima conjecture is stronger than theorem~\ref{thm:cmn} with the correct form of the right-tail being identified (cf. theorem~\ref{thm:abr} and the subsequent discussion). 

\subsubsection*{Branching structure for $\log|P_N(A,\theta)|$}

Following the progress made by identifying the approximate branching structure in a model of the Riemann zeta function by Arguin et al.~\cite{argbelhar17}, a crucial part of proving the various results towards conjecture~\ref{conj:fk_rmt} is the identification of a quasi-tree structure within the logarithm of the characteristic polynomial.

The proof of theorem~\ref{thm:abb}  follows a similar procedure to that outlined in~\eqref{eq:multiscale}. The authors show that the tree-like structure emerges from a multiscale decomposition.  Define $f:\mathbb{R}\rightarrow\mathbb{R}$ by  $f(\theta)=\log|1-e^{i\theta}|$, and observe that the Fourier series of $f$ is
\begin{equation}
  -\sum_{j=1}^\infty\frac{\RE(e^{-ij\theta})}{j}.
\end{equation}
 This means that 
\begin{equation}\label{eq:abbdecomp}
  \log |P_N(A,\theta)|=-\sum_{j=1}^\infty\frac{\RE(\Tr(A^j)e^{-ij\theta})}{j},
\end{equation}
where\footnote{When $\theta$ is an eigenphase of $A$, define both sides of \eqref{eq:abbdecomp} to be $-\infty$.} $\theta\in[0,2\pi)$.  Compare this to~\eqref{eq:expansion_log}, where such a decomposition was used to justify the log-correlated nature of $\log|P_N(A,\theta)|$.

We make two remarks on traces of powers of unitary matrices.  Firstly, due to orthogonality of characters of the unitary group (equivalently, the rotational invariance of the Haar measure), traces are uncorrelated (see for example~\cite{diasha94}):
\begin{equation}
  \mathbb{E}_{A\in \U(N)}\left[\Tr(A^j)\overline{\Tr(A^k)}\right]=\delta_{j,k}k,
\end{equation}
provided that $k\leq N$.  Secondly, theorem~\ref{thm:diasha} (see as well~\cite{hkssz96}) described the convergence of powers of traces,
\begin{equation}
  \left(\frac{\Tr(A^j)}{\sqrt{j}}\right)_{j\geq 1}\overset{N\rightarrow\infty}{\longrightarrow}\left(\cN_j^{\mathbb{C}}\right)_{j\geq 1}.
\end{equation}
Further, the speed of the convergence is superexponential~\cite{joh97}. 

Hence, one may truncate the sum in \eqref{eq:abbdecomp} at $N$, just gaining an error at the level of $O(1)$.  Decomposing the now finite sum, one defines
\begin{equation}\label{eq:abbdecomp2}
  X_n(\theta)\coloneqq\sum_{m=1}^{n}W_m(\theta)=-\sum_{m=1}^n\sum_{e^{m-1}\leq j<e^m}\frac{\RE(\Tr(A^j)e^{-ij\theta})}{j},
\end{equation}
for $n\in\{0,\dots, \log N\}$.  $X_n(\theta)$ should be compared to~\eqref{eq:zeta_branching} under the usual dictionary $N\equiv \log \frac{T}{2\pi}$.  The increments $W_m(\theta)$ are, by the above discussion, uncorrelated (for different $m$) and have variance approximately $\frac{1}{2}$. If one takes two points $\theta_1, \theta_2$, then the covariance of the increments for a fixed `level' is calculated (see~\cite{argbelbou17}) to be
\begin{equation}\label{eq:abb_increments}
  \mathbb{E}\left[W_m(\theta_1)W_m(\theta_2)\right]=\frac{1}{2}\sum_{e^{m-1}\leq j<e^m}\frac{\cos(|\theta_1-\theta_2|j)}{j}=
  \begin{dcases}
    \frac{1}{2}+O\left(e^{m-\theta_1\wedge\theta_2}\right),&\text{if }m\leq \theta_1\wedge\theta_2,\\
    O\left(e^{-2(m-\theta_1\wedge\theta_2)}\right),&\text{if }m>\theta_1\wedge\theta_2,
  \end{dcases}
\end{equation}
where
\begin{align}
  \theta_1\wedge\theta_2
  &\coloneqq-\log(\min\{|\theta_1-\theta_2|,2\pi-|\theta_1-\theta_2|\})
\end{align}
(cf. \eqref{wedgedef}).  The decoupling occurs at level $\theta_1\wedge\theta_2$, the logarithm of the inverse distance between the points on the circle. Therefore, $X_n(\theta)$ is approximately a branching random walk with $N$ leaves and one can restrict one's attention to $\theta$ in the discrete set
\begin{equation}\label{eq:discrete_set}
  T_N\coloneqq\left\{0,\frac{2\pi}{N},\dots,(N-1)\frac{2\pi}{N}\right\}.
\end{equation}
By~\eqref{eq:abb_increments}, if $\theta_1\wedge\theta_2\leq e^{-m}$, then the walks $X_m(\theta_1)$ and $X_m(\theta_2)$ differ only by an order-one term.  Thus, the appropriate analogous branching structure is a tree with $e^m$ particles at level $m$; i.e. not a binary tree but a tree with branching rate $e$. 

It turns out that it is first easier to work with a shorter walk
\begin{equation}
  X_{(1-\frac{1}{K})\log N}(\theta)=\sum_{m=1}^{(1-\frac{1}{K})\log N}W_m(\theta)
\end{equation}
for some large integer $K$. By creating room at the top of the sum, Arguin et al.\ prove sharp large deviation estimates.

The upper bound on $X_{(1-\frac{1}{K})\log N}$ follows from a union bound and the expected number of exceedances,
\begin{equation}
  Z(\vareps)=|\{\theta\in T_N: X_{(1-\tfrac{1}{K})\log N}(\theta)>(1-\vareps)\log N\}|.
\end{equation}
Since the result establishes leading order, a modification of $Z$ as in~\eqref{eq:new_event} is not necessary.  This is not the case for the lower bound. In particular, via a Chebyshev inequality, they show that a Gaussian-type bound holds for the truncated walk
\begin{equation}
  \mathbb{P}\left(X_{(1-\frac{1}{K})\log N}(\theta)\ge x\right)\leq C\exp\left(-\frac{x^2}{2\sigma^2}\right),
\end{equation}
where $C$ is some constant, and $\sigma^2=\frac{1}{2}\sum_{j=1}^{N^{1-\frac{1}{K}}}\frac{1}{j}$.  Such a bound is proved using a Riemann-Hilbert analysis.  The top of the sum is handled separately (again using Riemann-Hilbert techniques), and the upper bound follows by taking the large $N$ limit.

The lower bound requires more work and in particular a truncated second moment argument is required (a modification of the Paley-Zygmund inequality described previously). This controls the second moment, which as discussed following \eqref{eq:paley_zygmund}, would otherwise dominate the square of the first moment. Once again, Riemann-Hilbert techniques are used to determine the second moment of the appropriate counting function. Combining the upper and lower bound delivers the claimed leading order.  

Swiftly following the result of Arguin, Belius, and Bourgade, conjecture~\ref{conj:fk_rmt} was verified to subleading order by Paquette and Zeitouni. The key new techniques are a careful comparison between the field $\log|\det(I-zA)|$ and a centred Gaussian field inside the unit circle on optimal scales, and an adaptation of the barrier akin to~\eqref{eq:new_event}.  Here, the lower bound is determined by studying the field $\log|\det(I-zA)|$ inside the unit circle, i.e. $|z|<1$.  It turns out that this is particularly convenient since there $\log|\det(I-zA)|$ is harmonic, so almost surely
\begin{equation}
  \sup_{|z|<1} \log|\det(I-zA)|=\max_{z\in\{z\in\mathbb{C}:|z|=1\}}\log|\det(I-zA)|.
\end{equation}
Any value in the interior will provide a lower bound for the maximum on the boundary. 

Conversely, for any $M>0$, Paquette and Zeitouni show that there exists an $\tilde{N}(M)$ sufficiently large such that for all integers $N>\tilde{N}(M)$ and any $\{\theta_1,\dots,\theta_N\}\in[0,2\pi)$ one has 
\begin{equation}
  \max_{|z|<1}\log|\det(I-zA)|\leq \max_{|z|=1-\tfrac{M}{N}} \sum_{j=1}^N\log\big|1-ze^{-i\theta_j}\big|+M.
\end{equation}
One is then motivated to understand the behaviour of the field $\log|\det(I-zA)|$ at $N$ equally spaced points (similar to the discrete set $T_N$,~\eqref{eq:discrete_set}), shifted to lie just inside the unit circle:
\begin{equation}\label{eq:pq_discretisation}
  \left\{(1-\tfrac{M}{N})e^{2\pi i \tfrac{j}{N}}\right\}_{j=1}^N.
\end{equation}

The estimates involved in the proof of the upper and lower bounds are delicate.  The route that Paquette and Zeitouni take is to prove that $\log|\det(I-zA)|$ is very close to a real-valued Gaussian field.  Hence, one can work directly with the Gaussian process and reap the associated benefits. 

Determining the upper bound follows from a first moment argument.  In particular they show that the probability that the field lies below an adapted barrier for most of the `walk' and yet achieves a large value at the end of the walk is small. In fact for this argument, a full barrier is not needed.  In comparison, more control is required for the lower bound (including the full barrier).  Concealing many of the technical details, the argument broadly follows the usual lines: applying a truncated second moment argument.  

Finally, the most recent improvement towards fully establishing conjecture~\ref{conj:fk_rmt} is due to Chhaibi, Madaule, and Najnudel~\cite{chhmadnaj18} (though see theorem~\ref{thm:abr} for a comparatively stronger result in the number theoretic case). The result shows that the conjectured family of random variables is tight. The key additions are orthogonal polynomials on the unit circle (OPUC) and Verblunsky coefficients (cf. Szeg\" o~\cite{szego39}).

Let $\mathbb{D}$ be the unit circle in $\mathbb{C}$, and $\mu$ a probability measure on $\mathbb{D}$. By applying the Gram-Schmidt procedure to $\{z^n:n=0,1,\dots\}$, one may create a sequence of monic polynomials $\{\Phi_k(z),k=0,1,\dots\}$ orthogonal with respect to $\mu$. These polynomials can be generated using the Szeg\" o recurrence relation, 
\begin{equation}\label{eq:orthogonal_polynomials}
  \Phi_{k+1}(z)=z\Phi_k(z)-\overline{\alpha_k}\Phi^*_k(z),
\end{equation}
where 
\begin{equation}
  \Phi^*_k(z)=z^k\overline{\Phi_k\left(\overline{z}^{-1}\right)}
\end{equation}
and $\Phi_0(z)=1$. The $*$ operator reverses the order of the polynomial coefficients. The numbers $\alpha_k$ are known as \emph{Verblunsky coefficients}.  OPUC can be used to understand the behaviour of unitary characteristic polynomials.
\begin{lemma}[Chhaibi, Madaule, Najnudel~\cite{chhmadnaj18}]\label{lemma:opuc}
  The following family of random variables is tight,
  \[\left(\sup_{\theta\in[0,2\pi)\backslash\{\theta_1,\dots,\theta_N\}}\log|P_N(A,\theta)|-\sup_{\theta\in[0,2\pi)}\log|\Phi_{N-1}^*(e^{i\theta})|\right)_{N\geq 1}.\]
\end{lemma}

Hence, theorem~\ref{thm:cmn} is equivalent to the tightness of the family
\begin{equation}\label{thm:equivalentthm}
  \left(\max_{\theta\in[0,2\pi)}\log |\Phi_N^*(e^{i\theta})|-\left(\log N-\frac{3}{4}\log\log N\right)\right)_{N\geq 2}.
\end{equation}

One can express $\log \Phi_N^*(e^{i\theta})$ as a sum of logarithms of a function of Verblunsky coefficients and continuous real functions called Pr\"ufer phases.  This introduces a martingale structure which is particularly useful when determining the extreme values of polynomials $(\Phi_k^*)_{k\geq 0}$. With some careful construction, one can then define a new field
\begin{equation}\label{eq:cmn_altfield}
  Z_k(\theta)\coloneqq \sum_{j=0}^{k-1}\frac{X_j^{\mathbb{C}}e^{i\psi_j(\theta)}}{\sqrt{j+1}}
\end{equation}
for $\theta\in\mathbb{R}$, where $X_j^{\mathbb{C}}$ is a complex Gaussian of variance $1$, and $\psi_j(\theta)$ are so-called Pr\"ufer phases.  The problem is thus reduced to showing that the family
\begin{equation}\label{eq:newtightness}
\left(\sup_{\theta\in[0,2\pi)}\left|\log \Phi_k^*(e^{i\theta})-Z_k(\theta)\right|\right)_{k\geq 0}
\end{equation}
is tight.  Finally, the authors show that
\begin{equation}\label{eq:reducedtheorem}
\sup_{\theta\in[0,2\pi)}\RE(Z_N(\theta))=\log N-\frac{3}{4}\log\log N+O(1),
\end{equation}
where $O(1)$ is a tight family of random variables, and hence conclude.

As with the previous two results, in essence the proof follows from applying a discretization result, an asymptotic upper bound (using a first moment argument) and an asymptotic lower bound (by Paley-Zygmund). For the upper bound, after showing that the usual discretization holds, the authors use a (full, increasing, non-constant) barrier.  The lower bound, as usual, requires more careful handling and the authors move from \eqref{eq:cmn_altfield} to a new process in order to achieve more independence to make the first and second moment estimates more tractable.

At time of writing, it remains to identify the distribution of the fluctuating term (conjectured to be the sum of two independent Gumbel random variables).  Remy~\cite{rem19} considered a related model, discussed below, and there proved the conjectured fluctuations.  Alternatively, recently Arguin, Bourgade and Radziwi{\l}{\l}~\cite{argbourad20} have established an upper bound on the right tail for the Riemann zeta function of the predicted size, see theorem~\ref{thm:abr} and the discussion thereafter.  It is possible that their techniques could be adapted to the random matrix setting. 

\subsubsection*{Gaussian Multiplicative Chaos and Gumbel random variables}

In~\cite{rem19}, Remy establishes the sum of two independent Gumbel random variables in an analogous problem.  Before discussing the result, we briefly survey some relevant background.  Gaussian multiplicative chaos measures were introduced in section~\ref{sec:logcorrintro}.

In two successive works~\cite{web15, niksakweb20}, it was determined that for $\alpha\in(-1/2,2)$ and as $N\rightarrow\infty$,
\begin{equation}\label{eq:char_pol_gmc}
  \frac{|P_N(A,\theta)|^\alpha}{\mathbb{E}[|P_N(A,\theta)|^\alpha]}\frac{d\theta}{2\pi}\overset{\text{law}}{\longrightarrow}e^{\alpha X(\theta)-\tfrac{\alpha^2}{2}\mathbb{E}[X(\theta)^2]}\frac{d\theta}{2\pi}.
\end{equation}
Thus, if one considers the left hand side as a sequence of measures on the unit circle, by \eqref{eq:char_pol_gmc} the sequence converges in law to the $\GMC$ measure found on the right hand side of the statement. Recall that we write $X(\theta)$ for a centred and logarithmically correlated Gaussian field, where explicitly\footnote{Whilst the normalization constant of $1/2$ appearing here is not standard within the log-correlated literature, its role is to mimic the random matrix setting.}
\begin{equation}
  \mathbb{E}[X(\theta_1)X(\theta_2)]=-\frac{1}{2}\log|e^{i\theta_1}-e^{i\theta_2}|.
\end{equation}

It was first shown by Webb~\cite{web15} that the convergence~\eqref{eq:char_pol_gmc} is true in the $L^2-$phase $\alpha\in\left(-\tfrac{1}{2},\sqrt{2}\right)$ (cf. \eqref{eq:gmc_phase_l2}).  In fact, Webb proves a more general version allowing for certain twists of the characteristic polynomial.  In a subsequent work due to Nikula, Saksman, and Webb~\cite{niksakweb20}, the convergence~\eqref{eq:char_pol_gmc} was extended to include the $L^1-$phase $\alpha\in[\sqrt{2},2)$ (cf. \eqref{eq:gmc_phase_l1}).  Motivated by the theory of multiplicative chaos, it is conjectured that the limiting object for $\alpha>2$ will be zero, thus the interesting behaviour has now been categorized. Similar results have been established for the Riemann zeta function~\cite{sakweb16}. 

In~\cite{rem19}, Remy also works with the field $X(\theta)$, though with a different normalization in the covariance.  For consistency, we have translated his results to the notation of this paper.  Define
\begin{equation}
  Y_\alpha=\frac{1}{2\pi}\int_0^{2\pi}e^{\alpha X(\theta)}d\theta,
\end{equation}
for $\alpha\in(0,2)$.  As usual, $Y_\alpha$ is rigorously defined as a limit using an appropriate cut-off $X_\vareps(\theta)$ of $X(\theta)$,
\begin{equation}\label{eq:GMC_again}
  e^{\alpha X(\theta)}\coloneqq\lim_{\vareps\rightarrow0}e^{\alpha X_\vareps(\theta)-\frac{\alpha}{2}\mathbb{E}[X_{\vareps}(\theta)^2]}
\end{equation}
(see~\cite{rem19} or section~\ref{sec:logcorrintro}). It is instructive to compare this to \eqref{eq:fixed_a_moment}.  The main theorem of Remy proves a conjecture of Fyodorov and Bouchaud~\cite{fyobou08}.
\begin{theorem}\label{thm:remy}
  Take $\alpha\in(0,2)$.  For all $\rho\in\mathbb{R}$ such that $\rho\alpha^2<4$, one has
  \begin{equation}\label{eq:remy}
    \mathbb{E}[Y_\alpha^\rho]=\frac{\Gamma\left(1-\tfrac{\rho\alpha^2}{4}\right)}{\Gamma\left(1-\tfrac{\alpha^2}{4}\right)^\rho}.
  \end{equation}
\end{theorem}

Setting $\alpha=2\beta$ and $\rho=k$, one may combine theorem~\ref{thm:remy} with the convergence to the $\GMC$ measure.  Hence, one expects that for $k\beta^2<1$, and $2\beta\in(0,2)$, and as $N\rightarrow\infty$,
\begin{align}
  \mathbb{E}\left[\left(\frac{1}{2\pi}\int_0^{2\pi}\frac{|P_N(A,\theta)|^{2\beta}}{\mathbb{E}[|P_N(A,\theta)|^{2\beta}]}d\theta\right)^k\right]&\rightarrow \mathbb{E}\left[\left(\frac{1}{2\pi}\int_0^{2\pi} e^{2\beta X(\theta)-2\beta^2\mathbb{E}[X(\theta)^2]} d\theta\right)^k\right]\label{remy_conv_to_gmcB}\\
  &=\frac{\Gamma(1-k\beta^2)}{\Gamma(1-\beta^2)^k}.\label{remy_conv_to_gmc} 
\end{align}
Thus, by theorem~\ref{thm:ks}, (assuming that the convergence \eqref{remy_conv_to_gmc} holds),
\begin{equation}\label{eq:gmc_mom}
  \mathbb{E}
  \left[\left(\frac{1}{2\pi}\int_0^{2\pi}|P_N(A,\theta)|^{2\beta}d\theta\right)^k\right]\sim \left(\frac{\mathcal{G}^2(1+\beta)}{\mathcal{G}(1+2\beta)}\right)^k\frac{\Gamma(1-k\beta^2)}{\Gamma(1-\beta^2)^k}N^{k\beta^2},
\end{equation}
which is precisely the first regime of conjecture~\ref{conj:fk_mom}.

Additionally, Remy considers the `critical' $\GMC$ at $\alpha=2$ (i.e. $\beta=1$). In this case, the measure is denoted $-X(\theta)e^{X(\theta)}d\theta$ and is found via
\begin{equation}
  -X(\theta)e^{2 X(\theta)}d\theta\coloneqq -\lim_{\vareps\rightarrow 0}\left(X_\vareps(\theta)-2\mathbb{E}[X_\vareps(\theta)^2]\right)e^{2X_\vareps(\theta)-2\mathbb{E}[X_\vareps(\theta)^2]}d\theta,
\end{equation}
again for a suitable cut-off $X_\vareps$. Such a construction gives a non-trivial random positive measure, see~\cite{drsv14a, drsv14b, pow18}.  Now define
\begin{equation}
  Y'\coloneqq -\int_0^{2\pi}X(\theta)e^{2X(\theta)}d\theta.
\end{equation}
It was shown by Aru, Powell, and Sep\'ulveda~\cite{arupowsep19} that $Y'$ is related to the limit as $\alpha\rightarrow 2$ of $Y_\alpha$,
\begin{equation}
  Y'=\lim_{\alpha\rightarrow 2}\frac{Y_\alpha}{2-\alpha}
\end{equation}
in probability. One can hence deduce (see~\cite{rem19}) the density $f_{Y'}$ of $Y'$,
\begin{equation}\label{density_of_y}
  f_{Y'}(y)=
  \begin{dcases}
    y^{-2}e^{-y^{-1}} & \text{if }y\geq 0\\
    0 &\text{if }y<0,
  \end{dcases}
\end{equation}
so $\log Y'$ has a standard Gumbel law. Finally, recent results (see~\cite{bislou16, dinroyzei17}), have shown that for suitable cut-offs $X_\vareps$, there exists a constant $C$ such that
\begin{equation}\label{max_of_gmc}
  \max_{\theta\in[0,2\pi)}X_\vareps(\theta)-\log\frac{1}{\vareps}+\frac{3}{4}\log\log\frac{1}{\vareps}\rightarrow \mathscr{G}+\log Y'+C,
\end{equation}
where $\mathscr{G}$ is a Gumbel random variable independent from $Y'$.

Hence, combining \eqref{max_of_gmc}, \eqref{density_of_y}, and matching $N$ with $1/\vareps$ and $\log|P_N(A,\theta)|$ with $X_\vareps(\theta)$, this analogy would precisely imply conjecture~\ref{conj:fk_rmt}.  This approach also suggests that it may be easier to instead establish 
\begin{equation}
  \max_{\theta\in[0,2\pi)}\log|P_N(A,\theta)|-\log N +\frac{3}{4}\log\log N\rightarrow \mathscr{G}_1+\log Y'+C.
\end{equation}

\subsubsection*{Progress towards conjecture~\ref{conj:fk_nt}}

The identification of the branching structure within Harper's model~\cite{har13a} of $\log|\zeta(1/2+it)|$ and (approximately) within $\log|P_N(A,\theta)|$ was been crucial to progress towards conjecture~\ref{conj:fk_nt} and~\ref{conj:fk_rmt}.  This theme continues when working directly with the Riemann zeta function. Within this section, we sketch some of the key aspects of the relevant proofs; for further details we direct the reader to the respective papers and the excellent survey~\cite{har19a}. 

Conjecture~\ref{conj:fk_nt} states that for $\tau\sim U[T,2T]$, as $T\rightarrow\infty$,
\begin{equation}\label{eq:fk_nt2}
  \max_{h\in[0,2\pi)}\log|\zeta(\tfrac{1}{2}+i(\tau+h))|=\log\log T -\frac{3}{4}\log\log\log T + x_T,
\end{equation}
where $x_T\rightarrow x$ in distribution\footnote{For more detail on the distribution $x$, see theorem~\ref{thm:abr} and the subsequent discussion.}.  For $\zeta$, the range for the maximum need not be constrained to $[0,2\pi)$ (since unlike with unitary matrices, one has no periodicity).  The conjecture~\eqref{eq:fk_nt2} should hold for any interval that is $O(1)$ as $T$ grows.  For intervals that grow or shrink with $T$, see theorems~\ref{thm:aor} and~\ref{thm:adh}.

Najnudel~\cite{naj18} established the conjecture to leading order under the Riemann hypothesis (and also proves an analogous result for the imaginary part of $\log\zeta(1/2+it)$). 
\begin{theorem}[Najnudel~\cite{naj18}]\label{thm:naj}
  Take $\vareps>0$. Assuming the Riemann hypothesis and as $T\rightarrow\infty$
  \begin{equation}
   \frac{1}{T}\text{meas.}\left\{T\leq t\leq 2T: (1-\vareps)\log\log T< \max_{|t-h|\leq 1}\log|\zeta(\tfrac{1}{2}+ih)|< (1+\vareps)\log\log T\right\}\rightarrow 1.
  \end{equation}
\end{theorem}
The random walk structure originates in the Euler product of zeta. In particular, for $\RE(s)>1$, 
\begin{equation}\label{eq:naj_decomp}
  \log\zeta(s+it)=\sum_p \sum_{k=1}^\infty\frac{p^{-k(s+it)}}{k}=\sum_{n\geq 1}\frac{l(n)}{n^{s+it}},
\end{equation}
for $t\in \mathbb{R}$, and where $l(n)=1/k$ if $n=p^k$ and $0$ otherwise.

Establishing an upper bound, i.e. $\max_{|t-h|\leq 1}\log|\zeta(1/2+ih)|< (1+\vareps)\log\log T$ is straightforward and doesn't require the Riemann hypothesis.  Najnudel uses a method inspired by~\cite{chhmadnaj18} followed by an application of Markov's inequality and classical estimates on the second moment of $\zeta$.  An alternative (simpler) proof can be found in section 2 of~\cite{abbrs19}. For both and subsequent approaches, one first discretizes the interval.  An important observation is that the zeta function fluctuates with approximate frequency $1/\log T$ (seen for example via its Fourier series).  This permits one to discretize the order-one interval into approximately $\log T$ points\footnote{Compare this to the RMT discretization of $[0,2\pi)$ in to $N$ points, e.g. \eqref{eq:discrete_set} and \eqref{eq:pq_discretisation}.}.  

For the lower bound, Najnudel works with a Dirichlet series related to~\eqref{eq:naj_decomp}:
\begin{equation}\label{eq:naj_lambda}
  \Lambda_\psi(\tau,H)\coloneqq \sum_{n\geq 1}\frac{l(n)}{n^{\frac{1}{2}+i\tau}}\psi\left(\tfrac{\log n}{H}\right),
\end{equation}
where $\psi$ is chosen to be the Fourier transform of a function $\phi:\mathbb{R}\rightarrow\mathbb{R}$ satisfying, for $\RE(s)>1$,
\begin{equation}\label{eq:naj_anacon}
  \int_{-\infty}^\infty\log\zeta(s+it)\phi(t)dt=\sum_{n\geq 1}\frac{l(n)}{n^s}\hat{\phi}(\log n).
\end{equation}
In order to conclude information from \eqref{eq:naj_anacon} about the values of $\log\zeta$ inside the critical strip, Najnudel assumes the Riemann hypothesis in order to define an analytic continuation.  Hence, under the Riemann hypothesis and control on the regularity of $\phi$, Najnudel shows for $\tau\in\mathbb{R}$, $H>0$,
\begin{equation}\label{eq:naj_prop3.2}
  \int_{-\infty}^\infty \log\zeta\left(\tfrac{1}{2}+i\left(\tau+\tfrac{t}{H}\right)\right)\phi(t)dt=\Lambda_\psi(\tau,H)+O_\phi\left(1+\tfrac{e^{O_\phi(H)}}{1+|\tau|}\right).
\end{equation}

Najnudel then proves (propositions 5.1, 5.2 in~\cite{naj18}) that, without too big an error term, and by making careful choices for $\tau, H$ in \eqref{eq:naj_lambda}, the maximum of $\RE(\Lambda_{\psi}(\tau, H))$ provides a lower bound on the maximum of $\log|\zeta(1/2+i\tau)|$ in the relevant short intervals.

Working with $\Lambda_\psi$ allows one to apply the branching techniques discussed previously.  Indeed, after applying a multiscale-type decomposition one sees that $\Lambda_\psi(\tau,H)$ has an approximate branching structure. By comparing $\Lambda_\psi$ to a randomized version (\`a la Harper's model~\eqref{eq:random_model}) one can, via many technical lemmas, apply the usual Paley-Zygmund approach. 

Soon after Najnudel's result, Arguin, Belius, Bourgade, Radziwi{\l}{\l} and Soundararajan were able to remove the assumption of the Riemann hypothesis. 
\begin{theorem}[Arguin et al.~\cite{abbrs19}]\label{thm:abbrs}
  For any $\varepsilon>0$, as $T\rightarrow\infty$, we have 
  \begin{equation}
    \frac{1}{T}\meas\left\{T\leq t\leq 2T:(1-\vareps)\log\log T<\max_{|t-h|\leq 1}\log|\zeta(\tfrac{1}{2}+ih)|<(1+\vareps)\log\log T\right\}\rightarrow 1.
  \end{equation}
\end{theorem}
By examining $\zeta$ off-axis, Arguin et al.\ avoid assuming the Riemann hypothesis.

As mentioned above, the upper bound is much simpler to prove.  For the lower bound, the authors establish that large values just off the critical line imply large values lying on the critical line, thus permitting them to work slightly to the right of the line $1/2+it$. They then construct a `mollifier' (i.e. a function $M(s)$ so that, just to the right of the critical line, $M(s)\zeta(s)\approx 1$). This allows them to show that for almost all $t\in[T, 2T]$, one instead may work with a significantly shorter Dirichlet polynomial
\begin{equation}\label{eq:abbrs_short_dirichlet}
  \RE\left(\sum_{p\leq X}\tfrac{1}{p^{\frac{1}{2}+\vareps+ih}}\right)
\end{equation}
for an $X$ \emph{much smaller} than $T$.  In particular they show that a large value of the maximum of \eqref{eq:abbrs_short_dirichlet} (over $|t-h|\leq 1$) implies a large value of $\max_{|t-h|\leq 1}\log|\zeta(1/2+ih)|$. The sum \eqref{eq:abbrs_short_dirichlet} provides the approximate branching structure.

By breaking up the length of the `walk' $X$ via the usual style of multiscale decomposition\footnote{The contribution from the very small and very large primes has to be handled separately.}, one applies the Paley-Zygmund inequality (which requires precise large deviation estimates on the Dirichlet polynomials in question) to prove the lower bound. 

Improving further on the leading order, Harper provided a nearly sharp upper bound. 
\begin{theorem}[Harper~\cite{har19b}]\label{thm:har}
  For any real function $g(T)$ tending to infinity with $T$,
  \begin{equation}
    \max_{|h|\leq 1/2}\log|\zeta(\tfrac{1}{2}+i(t+h))|\leq \log\log T-\frac{3}{4}\log\log\log T+\frac{3}{2}\log\log\log\log T+g(T),
  \end{equation}
  for a set of $t\in[T,2T]$ with measure $(1+o(1))T$. 
\end{theorem}

The analytic proof still follows the same road-map from the probabilistic approach. Rather than an explicit discretization, Harper shows that the maximum question can be replaced using Cauchy's Integral Formula by an average over a small rectangle of height $1/\log T$) which will eventually be summed over all shifts (the analogue of a union bound).  This is essentially a `moments of moments' style result (see section~\ref{sec:moments}).  Harper's proof also differs in how $\zeta(s)$ is approximated by Dirichlet polynomials. Now working with an integral, the zeta function is approximated in mean square by the product of two Dirichlet polynomials: one over smooth numbers and one over rough numbers.  Typically, one discards the latter in favour of short Dirichlet series, see for example~\eqref{eq:abbrs_short_dirichlet}.  However, in order to achieve the desired subleading term, one should work with the full range. Additionally, Harper requires estimates on the fourth moment of Dirichlet polynomials to achieve the subleading bound.  Compare this to the leading order upper bound which only required the second moment.  This approach delivers the claimed bound. 

Finally, the strongest result to date is that of Arguin, Bourgade and Radziwi{\l}{\l}~\cite{argbourad20}.  There they determine the upper bound of conjecture~\ref{conj:fk_nt}, including the predicted tail of the random variable~\eqref{eq:gumbel_prob}.  Recall from~\eqref{eq:gumbel_prob} that the right tail of the fluctuating term should be like $xe^{-2x}$ (using the notation of~\eqref{eq:fk_nt2}).   This is stronger than what is currently known for the random matrix case.

\begin{theorem}[Arguin, Bourgade, Radziwi{\l}{\l}~\cite{argbourad20}]\label{thm:abr}
  There exists $C>0$ such that for any $T\geq 3$ and $x\geq 1$ one has
  \begin{equation}\label{eq:thm_abr}
    \frac{1}{T}\meas\left\{ t\in[T,2T]: \max_{h\in [0,1]}\log|\zeta(\tfrac{1}{2}+i(t+h))|>\log\log T-\frac{3}{4}\log\log\log T+x\right\}\leq Cxe^{-2x}.
  \end{equation}
\end{theorem}

Clearly, the expected sharpness of theorem~\ref{thm:abr} depends on the size of $x$. For $x=O(\sqrt{\log\log T})$, the bound in theorem~\ref{thm:abr} is expected to be sharp.  For $x=O(\log\log T)$ however (i.e. $x$ can grow at the same rate as the leading order term), it is expected that instead the sharp decay rate should also feature the Gaussian behaviour:
\begin{align}\label{eq:precise_righttail}
  \frac{1}{T}\meas\left\{ t\in[T,2T]: \max_{h\in [0,1]}\log|\zeta(\tfrac{1}{2}+i(t+h))|\right.&>\left.\log\log T+x-\frac{3}{4}\log\log\log T\right\}\\
  &\ll xe^{-2x}\exp\left(-\tfrac{x^2}{\log\log T}\right).\nonumber
\end{align}
Notice that \eqref{eq:thm_abr} can be rephrased in terms of a particular Lebesgue measure of points.

As is now routine, the first step is to discretize the interval, focusing on $\log T$ discrete points. The authors then make a comparison between the logarithm of the Euler product and a random walk
\begin{equation}\label{eq:rw_abr}
  S_k(h)\coloneqq \sum_{e^{1000}\leq \log p\leq e^k}\RE\left(\frac{1}{p^{\frac{1}{2}+i(\tau+h)}}+\frac{1}{2p^{2(\frac{1}{2}+i(\tau+h))}}\right),
\end{equation}
for $k\leq \log\log T$. In~\eqref{eq:rw_abr}, one has essentially the first two terms in the Taylor series of the $\log$ Euler product. Taking the length of the walk close to $k=\log\log T$ is a source of most of the difficulty of the proof.  The reason is that one requires precise large deviation estimates.  To achieve these, one needs high moments of long Dirichlet sums.  Current number theoretic techniques produce the estimate
\begin{equation}
  \mathbb{E}[(S_k)^{2\alpha}]\approx \frac{(2\alpha)!}{2^\alpha(\alpha !)}\left(\frac{k}{2}\right)^{2\alpha}+O\left(\frac{\exp(2\alpha e^k)}{T}\right).
\end{equation}
For this situation, one needs $\alpha\approx\log\log T$. Notice that if $k>\log\log T-c\log\log\log T$ (the level of the maximum), then the error becomes unmanageable.  Therefore Arguin et al.\ instead reduce the number of $h$s to be considered (i.e. rule out many paths). They achieve this by introducing both an upper barrier (\`a la Bramson, see also the discussion prior to~\eqref{eq:new_event}) and a new \emph{lower} barrier.  Call the space in between the barriers the corridor which, by construction, narrows as one approaches $k=\log\log T$. Arguin et al.\ iteratively (but with a finite number of steps) build this corridor, progressively discarding escaping paths.  By reducing the number of walks, they are able to circumvent the aforementioned large deviation problem. In their estimates, like Harper, they require estimates on (twisted) fourth moments of Dirichlet polynomials. 

A natural extension to conjecture~\ref{conj:fk_nt} would be to consider maxima over intervals which grow or shrink with $T$.  It turns out that the branching analogy extends to the case with intervals of length $O(\log^\theta(T))$ for some real $\theta>-1$.  
\begin{theorem}[Arguin, Ouimet, Radziwi{\l}{\l}~\cite{argouirad19}]\label{thm:aor}
  Take $\theta>-1$. Let $\tau$ be uniformly distributed on $[T,2T]$. Then 
  \begin{equation}
    \lim_{T\rightarrow\infty}\frac{1}{\log\log T}\max_{h\in[-\log^\theta T,\log^\theta T]}\log|\zeta(\tfrac{1}{2}+i(\tau+h))|=
    \begin{cases}
      1+\theta,&\theta\leq 0\\
      \sqrt{1+\theta},&\theta>0.
    \end{cases}
  \end{equation}
  One requires the Riemann hypothesis for $\theta>3$. 
\end{theorem}
Notice that for $\theta\neq 0$, even at leading order, the asymptotic behaviour of the maximum has changed compared to conjecture~\ref{conj:fk_nt}. 

The reason that the theorem requires the Riemann hypothesis for $\theta>3$ is that $2\sqrt{1+\theta}$ corresponds (via an application of Markov's inequality) to the $\beta$th moment of the Riemann zeta function.  Since precise unconditional moment asymptotics are only available for $\beta\leq 4$ (i.e. $\theta\leq 3$), one must assume the Riemann hypothesis for higher moments, see for example~\cite{sou09}. 

In the same work, Arguin, Ouimet and Radziwiłł also consider moments of $\zeta$ over intervals of length $O(\log^\theta T)$, (akin to~\eqref{eq:fixed_a_moment} for $\theta=0$). They demonstrate that above a certain temperature (i.e. some critical moment), the moments exhibit freezing (cf.~\eqref{eq:log-corr_freezing}).

The interpretation of theorem~\ref{thm:aor} is as follows.  For $\theta\leq0$, one is in the situation discussed at length within this section already:  the branching tree picture and relevant theory still holds for a tree with $(1+\theta)\log\log T$ increments and $\log^{1+\theta}T$ leaves on an interval of width $2\log^\theta T$, with branching rate $e$. Using the usual heuristic, this would imply that the maximum at leading order grows like the logarithm of the number leaves, i.e. $(1+\theta)\log\log T$.  This matches theorem~\ref{thm:aor} for $\theta\leq 0$ (and clearly theorems~\ref{thm:naj},~\ref{thm:abbrs},~\ref{thm:har},~\ref{thm:abr} corresponding to $\theta=0$).

However, for $\theta>0$  (i.e. intervals larger than those considered in conjecture~\ref{conj:fk_nt}), one has to alter the branching picture.  Instead of considering just one tree, the correct analogy is to consider $\log^\theta T$ independent copies of the branching trees on intervals of order one.  This leads to a leading order like $\sqrt{1+\theta}\log\log T$.  Similar models, known as \emph{Continuous} Random Energy Models, studied by Bovier and Kurkova~\cite{bovkur04} led to the following generalization.

\begin{conj}[Arguin, Ouimet, Radziwi{\l}{\l}~\cite{argouirad19}; Arguin, Dubach, Hartung~\cite{argdubhar21}; Arguin et al.~\cite{aabhr21}]\label{conj:zeta_meso}
  Take $\theta\geq0$. Let $\tau$ be uniformly distributed on $[T,2T]$.  Then
  \begin{equation}\label{eq:zeta_meso_conj}
    \max_{h\in[-\pi\log^\theta T,\pi\log^\theta T]}\log |\zeta(\tfrac{1}{2}+i(\tau+h))|=\sqrt{1+\theta}\log\log T - r(\theta)\log\log\log T+y_{\theta,T}
  \end{equation}
  where
  \begin{equation}\label{eq:subleading_function}
    r(\theta)\coloneqq
  \begin{cases}
    \frac{3}{4},&\text{if }\theta= 0,\\
    \frac{1}{4\sqrt{1+\theta}},&\text{if }\theta>0,\\
    \frac{1+2\gamma}{4},&\text{if }\theta\sim(\log\log T)^{-\gamma}.
  \end{cases}
  \end{equation}
  For $\theta>0$, the sequence of random variables $(y_{\theta,T},T\geq 1)$ converges in distribution to a Gumbel random variable $G_\theta$ with $\mathbb{P}(G_\theta \leq x)=\exp(-e^{-\frac{1}{\beta}(x-m)})$ and parameters
  \begin{align}
    \beta&=\beta(\theta)=\frac{1}{2\sqrt{1+\theta}}\\
    m&=m(\theta)=\mathcal{C}+\beta^2\log f_{\sqrt{1+\theta}}-\frac{\beta^2}{2}\left(\log(1+\theta)-\log(4\pi)\right).\label{eq:m_para}
  \end{align}
\end{conj}
In~\cite{aabhr21}, the authors provide numerical evidence in support of~\eqref{eq:zeta_meso_conj}.

The parameter $m$ in \eqref{eq:m_para} contains two constants of interest.  Firstly, $4\mathcal{C}$ is the Meissel-Mertens constant.  Secondly, $f_{k}$ is the leading order coefficient of the $2k$th moment of $\zeta$~\eqref{eq:zeta_moment_limit}. The appearance of the moment coefficient is related to large deviations of Selberg's central limit theorem~\cite{rad11, fermelnik18, aabhr21}. Indeed, one can show\footnote{See~\cite{fermelnik18} for the original statement, and~\cite{aabhr21} for an alternative proof.} that the equivalent moment parameter appears in the large deviations of the Keating and Snaith central limit theorem~\ref{thm:ks-clt}.  For $\beta\geq 0$, any $\theta\in[0,2\pi)$, and $V\sim k\sqrt{2\log N}$,
\begin{equation}\label{eq:ks_clt_largedeviation}
  \mathbb{P}(\log|P_N(A,\theta)|>\sqrt{Q_2(N)}\cdot V)\sim \frac{\mathcal{G}^2(1+\beta)}{\mathcal{G}(1+2\beta)}\int_V^\infty e^{-\frac{x^2}{2}}\frac{dx}{\sqrt{2\pi}},
\end{equation}
where $Q_2(N)$ is the second cumulant of $\log|P_N(A,\theta)|$ (so $Q_2(N)\sim \frac{1}{2}\log N + O(1)$). Notice there is a correction to~\eqref{eq:ks_clt} -- which holds for fixed $V\in\mathbb{R}$ -- for $V$ growing at the same rate as the standard deviation of $\log|P_N(A,\theta)|$.

Conjecture~\ref{conj:zeta_meso} also reveals how the subleading term varies according to the branching structure present.  At $\theta=0$, the process is log-correlated.  As the interval grows beyond $O(1)$, instead one sees the $\operatorname{IID}$ statistics emerging.  One can interpolate between the ranges on the scale $\theta\sim(\log\log T)^\gamma$.  Resolving this discontinuity as $\theta\rightarrow0$ is work of Arguin, Dubach, and Hartung.  In~\cite{argdubhar21} they study the associated question for a random model of $\zeta(1/2+it)$ over intervals of length $\log^\theta T$, where $\theta$ is either fixed or can tend to zero at controlled speed. 

\begin{theorem}[Arguin, Dubach, Hartung~\cite{argdubhar21}]\label{thm:adh}
  Define
  \begin{equation}\label{eq:gaussian_model}
    X_T(h)\coloneqq \sum_{p\leq T}\frac{\RE(G_p p^{-ih})}{\sqrt{p}},
  \end{equation}
  where $(G_p, p\text{ primes})$ are \iid\ standard complex Gaussian random variables\footnote{One can show, see~\cite{argdubhar21}, that $(X_T(h), h\in[0,1])$ is a Gaussian log-correlated process, just as when the summands instead involve $U_p$, uniform random variables on the unit circle.}.  Then, for $\gamma\in (0,1)$ and $\theta=(\log\log T)^{-\gamma}$, we have
  \begin{equation}
    \lim_{T\rightarrow\infty}\max_{h\in[-\log^\theta T,\log^\theta T]}\frac{X_T(h)-\sqrt{1+\theta}\log\log T}{\log\log T}=-\frac{1+2\gamma}{4}
  \end{equation}
  in probability.
\end{theorem}
For this model, Arguin et al.\ are also able to show a precise upper bound for the right tail in an interval of order one of the shape given by~\eqref{eq:precise_righttail}. 

This concludes the review of the recent progress towards proving both conjectures~\ref{conj:fk_rmt} and conjecture~\ref{conj:fk_nt}.  As has been demonstrated, the identification throughout of an approximate branching structure is essential.

\section{Conjecture~\ref{conj:fk_mom} and generalized moments}\label{sec:moments}

Within this section, we focus on the third conjecture of Fyodorov and Keating, conjecture~\ref{conj:fk_mom}.  Recall from section~\ref{sec:fk_conj} that the heuristic calculations described in \cite{fyokea14} supporting conjecture~\ref{conj:fk_rmt} are based on an analysis of the random variable
\begin{equation}\label{eq:unitary_partition_function}
  g_N(\beta; A)\coloneqq\frac{1}{2\pi}\int_0^{2\pi}|P_N(A,\theta)|^{2\beta}d\theta,
\end{equation}
the $2\beta$th moment of $|P_N(A,\theta)|$ with respect to the uniform measure on the unit circle $\frac{d\theta}{2\pi}$ for a fixed $A\in \U(N)$.

The \emph{moments of moments} were hence defined as the moments of $g_N$ with respect the an average over $A\in \U(N)$,
\begin{equation}\label{eq:mom_gN}
  \mom_{\U(N)}(k,\beta)\coloneqq \int_{\U(N)}g_N(\beta;A)^kdA.  
\end{equation}

Recall from section~\ref{sec:fk_conj_recap}, that one justification for conjecture~\ref{conj:fk_mom} follows from a heuristic calculation of the moments of moments when $k$ is an integer \cite{fyohiakea12, fyokea14, kea17}, in which case one has
\begin{equation}\label{basic}
  \mom_{\U(N)}(k,\beta)=\frac{1}{(2\pi)^k}\int_0^{2\pi}\dots\int_0^{2\pi}
      \int_{\U(N)}\prod_{j=1}^k|P_N(A,\theta_j)|^{2\beta}dA\; d\theta_1\cdots d\theta_k.
\end{equation}

Before we present the recent developments towards a full understanding of \eqref{eq:fk_mom}, we first investigate a related construction for branching random walks.  As demonstrated in section~\ref{sec:fk_conj}, exploiting the connection between branching processes and log-correlated processes was instrumental to the results towards the maxima conjectures~\ref{conj:fk_rmt} and~\ref{conj:fk_nt}. 

\subsection{Branching moments of moments}\label{sec:branching_mom}

Take a binary tree of depth $n$, and a choice of leaf $l$. Load to each branch in the tree an independent centred Gaussian random variable with variance $\frac{1}{2}\log2$.  We write for the branching random walk from root to $l$
\begin{equation}\label{eq:brw_setup}
  X_n(l)\coloneqq\sum_{m=1}^nY_m(l),
\end{equation}
where $Y_m(l)\sim \cN(0,\frac{1}{2}\log2)$ are independent and identically distributed. Then  $X_n(l)\sim \cN\left(0,\tfrac{n}{2}\log 2\right)$ and the distribution has no dependence on the leaf $l$. Such labels (as for $Y_m(l)$) will play an important role later, however. The last common ancestor of two leaves $l, l^\prime$ is written $\lca(l, l^\prime)$, and is the furthest node from the root having both $l$ and $l^\prime$ as descendants. We also require the level of the last common ancestor.  Hence, the \emph{last common level} $\lcl(l_1,\dots,l_k)$ is the level of $\lca(l_1,\dots,l_k)$. 

Conjecture~\ref{conj:fk_rmt} was motivated by studying the partition function~\eqref{eq:unitary_partition_function} for $\log|P_N(A,\theta)|$.  As previously shown, $X_n(l)$ is a log-correlated process. It is natural here, therefore, to investigate the associated partition function (or moment generating function) for $X_n(l)$:
\begin{align}
  \frac{1}{2^{n}}\sum_{l=1}^{2^n}e^{2\beta X_n(l)}&=\frac{1}{2^{n}}\sum_{l=1}^{2^n}e^{2\beta \sum_{m=1}^nY_m(l)}.\label{eq:final_branching_model}
\end{align}
From a statistical mechanical perspective, the parameter $2\beta$ acts as inverse temperature.  

In particular, we are interested in the moments of the partition function \eqref{eq:final_branching_model}, 
\begin{align}
  \mathbb{E}\left[\left(\frac{1}{2^n}\sum_{l=1}^{2^n}e^{2\beta X_n(l)}\right)^k\right]&=\frac{1}{2^{kn}}\sum_{l_1=1}^{2^n}\cdots\sum_{l_k=1}^{2^n}\mathbb{E}\left[e^{2\beta(X_n(l_1)+\cdots+X_n(l_k))}\right],\label{branching_model_mom}
\end{align}
where the expectation in \eqref{branching_model_mom} is with respect to the Gaussian random variables.  These are the \emph{moments of moments} for the branching random walk.  They are the analogues of \eqref{def:mom_lit_review}. Such moments of partition functions were also considered for more general branch weightings by Derrida and Spohn~\cite{derspo88}. 

In~\cite{baikea21b}, the following results are established.
\begin{theorem}[Bailey and Keating~\cite{baikea21b}]\label{thm:branching_mom}
  Take $n, k\in\mathbb{N}$ and $\beta\in\mathbb{R}$. If $\beta\neq0$ then
  \begin{equation}
    \mathbb{E}\left[\left(\frac{1}{2^n}\sum_{l=1}^{2^n}e^{2\beta X_n(l)}\right)^k\right]\sim
    \begin{dcases}
      \rho_{k,\beta}2^{k\beta^2n},&\text{if }k<1/\beta^2,\\
      \sigma_{k,\beta}n2^n,&\text{if }k=1/\beta^2,\\
      \tau_{k,\beta}2^{(k^2\beta^2-k+1)n},&\text{if }k>1/\beta^2,
    \end{dcases}
  \end{equation}
  as $n\rightarrow\infty$, for some positive constants $\rho_{k,\beta}, \sigma_{k,\beta},$ and $\tau_{k,\beta}$ depending only on $k,\beta$.  Clearly, if $\beta=0$ then the expectation evaluates to $1$. 
\end{theorem}

For small values of $k$, one can calculate exact and explicit formulae for the moments of moments; such examples can be found in the appendix of~\cite{baikea21b}.

Furthermore, for integer values of the moment parameters, the branching moments of moments are polynomials.

\begin{corollary}[Bailey and Keating~\cite{baikea21b}]\label{cor:branching_mom}
When $k,\beta\in\mathbb{N}$, \eqref{branching_model_mom} is a polynomial in $2^n$ of degree $k^2\beta^2-k+1$. 
\end{corollary}
Thus, the branching moments of moments exhibit asymptotic behaviour identical to that predicted by conjecture~\ref{conj:fk_mom} for the random matrix moments of moments, if the identification $N=2^n$ is made.  However, the leading-order coefficients in each case are (unsurprisingly) provably different.

The proof of theorem~\ref{thm:branching_mom} is based on an inductive argument. The simplest case $k=1$ follows from a moment generating function calculation,
\begin{equation}\label{eq:k=1b}
  \frac{1}{2^n}\sum_{l=1}^{2^n}\mathbb{E}\left[\prod_{j=1}^ne^{2\beta Y_j(l)}\right]=\frac{1}{2^n}\sum_{l=1}^{2^n}\prod_{j=1}^n\mathbb{E}\left[e^{2\beta Y_j(l)}\right]=2^{\beta^2n}.
\end{equation}

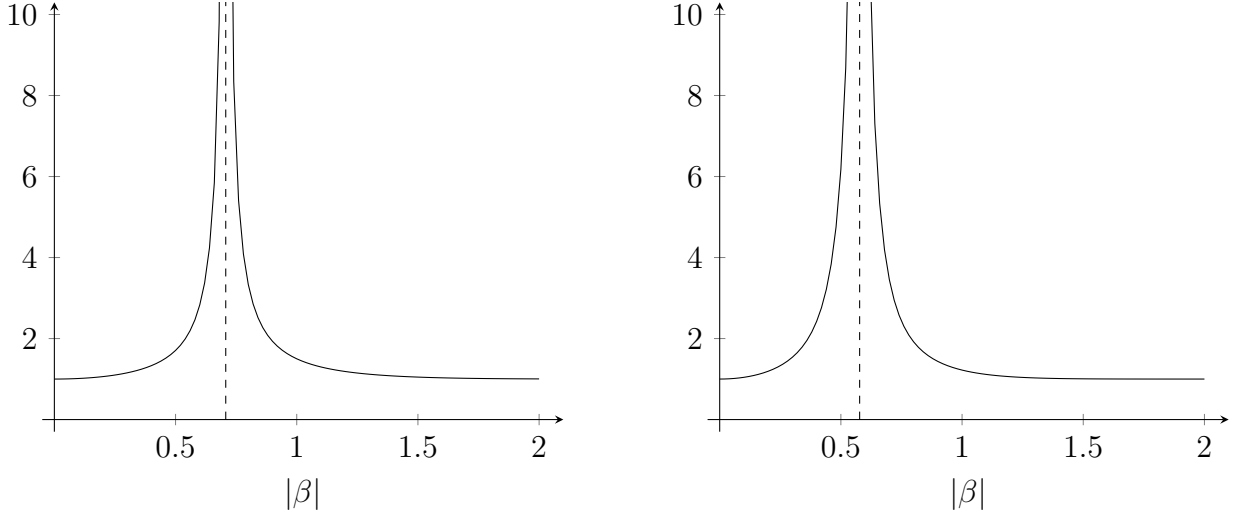
\begin{figure}
  \centering
  \begin{subfigure}[t]{.47\textwidth}
    \centering
    \begin{tikzpicture}[
      declare function={
        func(\x)= (\x < 0.7071) * (1/(2-2^(2*x*x))) + and(\x >= 0.7071, \x<= 0.7071) * (1/2) + (\x > 0.7071) * ((2^(2*x*x)-1)/(2^(2*x*x)-2));
      }
      ]
    \begin{axis}[
        axis x line=middle,
        axis y line=middle,
        ymin=-0.3,
        ymax=10.3,
        xmin=-0.05,
        xmax=2.1,
        domain=0:2,
        samples=101,
        xlabel near ticks,
        xlabel = $\abs{\beta}$
      ]
      \addplot[black] {func(x)};
      \addplot[black, dashed,domain=0:10.5] (0.7071,x);
    \end{axis}
  \end{tikzpicture}
  \caption[]{Plot of the leading coefficient of $\mom_n(2,\beta)$ as $\beta$ varies.  The dashed line is at $x=\frac{1}{\sqrt{2}}$, the transition point for $\mom_n(2,\beta)$.}\label{fig:mom(2,b)}
  \end{subfigure}
  \hfill%
  \begin{subfigure}[t]{.47\textwidth}
    \centering
    \begin{tikzpicture}[
        declare function={
          func(\x)= (\x < 0.5774) * (3*(2^(2*x*x))/((2^(6*x*x)-4)*(2^(2*x*x)-2))) + and(\x >= 0.5774, \x<= 0.5774) * (0) + (\x > 0.5774) * (1+(3*(2^(6*x*x)-2))/((2^(4*x*x)-2)*(2^(6*x*x)-4)));
        }
      ]
      \begin{axis}[
          axis x line=middle,
          axis y line=middle,
          ymin=-0.3,
          ymax=10.3,
          xmin=-0.05,
          xmax=2.1,
          domain=0:2,
          samples=101,
          xlabel near ticks,
          xlabel = $\abs{\beta}$
        ]
        \addplot[black] {func(x)};
        \addplot[black, dashed,domain=0:10.5] (0.5774,x);
      \end{axis}
    \end{tikzpicture}
    \caption[Plot of the leading coefficient of $\mom_n(3,\beta)$ as $\beta$ varies.]{Plot of the leading coefficient of $\mom_n(3,\beta)$ as $\beta$ varies.  The dashed line is at $x=\frac{1}{\sqrt{3}}$, the transition point for $\mom_n(3,\beta)$.}\label{fig:mom(3,b)}
  \end{subfigure}
  \caption[]{Figures showing the leading coefficients of $\mom_n(2,\beta)$ and $\mom_n(3,\beta)$ as $\beta$ varies.}\label{fig:branching_loc}
\end{figure}

When $k=2$, one has to handle the (typical) case of when the paths are not independent. To do so, one splits the moments of moments at the level of the last common ancestor.  Using the branching structure, one is able explicitly to determine that
\begin{align}
  \mom_n(2,\beta)
  &=2^{2\beta^2n-1}\frac{2^{(2\beta^2-1)n}-1}{2^{2\beta^2-1}-1}+2^{(4\beta^2-1)n}.\label{eq:k2_sum}
\end{align}
The statement of theorem~\ref{thm:branching_mom} for $k=2$ hence follows by analysing~\eqref{eq:k2_sum} as $\beta$ varies. Inductively, one can handle general $k\in\mathbb{N}$.  A similar argument reveals the polynomial structure of corollary~\ref{cor:branching_mom}. 
It is also interesting to examine the leading order coefficients $\rho_{k,\beta}$, $\sigma_{k,\beta}$, and $\tau_{k,\beta}$. Through the explicit computations of the branching moments of moments in~\cite{baikea21b}, in all cases computed\footnote{With the exception of the simplest case $k=1$.}, the leading order coefficient fails to be analytic in $\beta$ in the limit $n\rightarrow\infty$.  This property is shown in figure~\ref{fig:branching_loc} for $k=2,3$. 

\subsection{Unitary moments of moments}\label{sec:unitary_mom}

Common to all approaches described hereafter, one assumes that the moment parameter $k$ in \eqref{eq:mom_gN} is an integer.  Such an assumption allows one to exploit the rich structure of the resulting average over the Haar measure (after an application of Fubini),
\begin{equation}\label{eq:k-fold_ave}
  \mathbb{E}\left[\prod_{j=1}^k|P_N(A,\theta_j)|^{2\beta}\right].
\end{equation}
Here and henceforth, $\mathbb{E}[\cdot]$ represents the average over $\U(N)$.

It would be interesting to understand~\eqref{eq:mom_gN} for general $k$. Indeed, conjecture~\ref{conj:fk_rmt} implicitly assumes that $k\geq 1$ (but not that $k$ is integral).  If the analogy with number theory holds, then one should expect a correction to conjecture~\ref{conj:fk_rmt} for $k\in(0,1)$ and $\beta=1$ after the result of Harper~\cite{har19b}, cf. theorem~\ref{thm:har}.

We now present the progress towards a proof of conjecture~\ref{conj:fk_mom}.  At the time of writing, the conjecture has been established to leading order for $k\in\mathbb{N}$ and positive $\beta$. The leading order coefficient in the critical ($k\beta^2=1$) and supercritical ($k\beta^2>1$) regimes in general are unknown, but there are results for $k=2$, $\beta>0$ as well as for $k,\beta\in\mathbb{N}$.  

\subsubsection*{Results towards conjecture~\ref{conj:fk_mom}}

The case $k=1,\beta\in \mathbb{N}$ in conjecture~\ref{conj:fk_mom} follows immediately from the moment formula of Keating and Snaith~\cite{keasna00a} (cf. theorem~\ref{thm:ks},~also \cite{bakfor97}).   Specifically,
\begin{equation}\label{k=1}
  \mom_{\U(N)}(1,\beta)={\mathbb{E}}[|P_N(A,\theta)|^{2\beta}]=\prod_{0\le i,j\le \beta-1}\left(1+\frac{N}{i+j+1}\right),
\end{equation}
which additionally is clearly a polynomial in $N$ of degree $\beta^2$.

Bump and Gamburd~\cite{bumgam06} later gave an alternative proof using symmetric function theory.  In this second approach, the leading order coefficient of $\mom_{\U(N)}(1,\beta)$ was obtained by counting certain semistandard Young tableaux. These parallel stories of symmetric function theory and complex analysis continue for higher values of $k$.

A proof of conjecture~\ref{conj:fk_mom} when $k=2, \beta\in\mathbb{N}$ follows directly from formulae given by Keating et al.~\cite{krrr18}. It differs from the proof given by Claeys and Krasovsky~\cite{clakra15} (discussed below), which establishes the conjecture for $k=2$ and all $\beta$, but without identifying a polynomial structure when $\beta\in\mathbb{N}$.

For $A\in \U(N)$, the \textit{secular coefficients} of $A$, written\footnote{Clearly, $\Sc_n$ also depends on the specific matrix, $A$.} $\Sc_n(N)$, are the coefficients of its characteristic polynomial 
\begin{equation}\label{def:seccoeff}
  \det(I+xA)=\sum_{n=0}^N\Sc_n(N)x^n.
\end{equation}
The following theorem is proved Keating et al.~\cite{krrr18}.
\begin{theorem}[Keating et al.~\cite{krrr18}]\label{thm:krrr1}
  For $A\in \U(N)$, define
  \begin{equation}
    I_\eta(m;N)\coloneqq\int_{\U(N)}\Big|\sum_{\substack{j_1+\cdots+j_\eta=m\\0\leq j_1,\dots,j_\eta\leq N}}\Sc_{j_1}(N)\cdots \Sc_{j_\eta}(N)\Big|^2dA.
  \end{equation}
  If $c=m/N, c\in[0,\eta]$, then $I_\eta(m;N)$ is a polynomial in $N$ and
  \begin{equation}\label{eq:krrr1}
    I_\eta(m;N)=\alpha_\eta(c)N^{\eta^2-1}+O_\eta(N^{\eta^2-2}),
  \end{equation}
  where 
  \begin{equation}\label{krrr:leadingcoeff}
    \alpha_\eta(c)=\sum_{0\leq l<c}\binom{\eta}{l}^2(c-l)^{(\eta-l)^2+l^2-1}p_{\eta,l}(c-l),
  \end{equation}
  with $p_{\eta,l}(c-l)$ being polynomials in $(c-l)$. 
\end{theorem}

Immediately, theorem~\ref{thm:krrr1} shows that $\mom_{\U(N)}(2,\beta)$ is a polynomial in $N$ for $\beta\in\mathbb{N}$, and asymptotically is 
\begin{equation}
  \mom_{\U(N)}(2,\beta)\sim \alpha_{2\beta}(\beta)N^{4\beta^2-1}+O_{\beta}(N^{4\beta^2-2}).
\end{equation}
Theorem~\ref{thm:krrr1} was proved by two methods: symmetric function theory and complex analysis. The former determines an equivalent structure for $\alpha_\eta(c)$ to that given in \eqref{krrr:leadingcoeff} coming from a standard lattice point count, which proves that $I_\eta(m;N)$ is a polynomial in $N$ and makes it clear that $\alpha_{2\beta}(\beta)\neq 0$. By using complex analysis the result regarding the leading order in $N$ can be established and the form for $\alpha_\eta(c)$ given in \eqref{krrr:leadingcoeff} is found.

Claeys and Krasovsky instead use Toeplitz determinants to understand $\mom_{\U(N)}(2,\beta)$. The $N\times N$ Toeplitz determinant for the symbol $f$ is defined by
\begin{equation}
  D_N(f)\coloneqq \det(\hat{f}_{j-k})_{j,k=1}^N
\end{equation}
where $f$ is a real-valued, $2\pi$-periodic, integrable function with Fourier coefficients
\begin{equation}
  \hat{f}_j\coloneqq \frac{1}{2\pi}\int_0^{2\pi}f(\theta)e^{-ij\theta}d\theta.
\end{equation}
The connection between Toeplitz determinants and random matrix averages is the following Heine-Szeg\" o identity~\cite{szego39}
\begin{equation}\label{eq:heine_szego}
  D_N(f)=\frac{1}{(2\pi)^NN!}\int_0^{2\pi}\cdots \int_0^{2\pi}\prod_{j=1}^Nf(\theta_j)|\Delta(e^{i\theta_1},\dots,e^{i\theta_N})|^2d\theta_1\cdots d\theta_N.
\end{equation}
Thus, one can write~\eqref{eq:k-fold_ave} as $D_N(f)$ for the particular symbol $f(z)=\prod_{j=1}^k|z-e^{i\theta_j}|^{2\beta}$.  The symbol $f$ has $k$ Fisher-Hartwig singularities at $e^{i\theta_1},\dots,e^{i\theta_k}$.

In the simplest case, $k=1$, there is a single Fisher-Hartwig singularity.  Without loss of generality, due to the rotational invariance of the Haar measure, one writes the symbol as $f^*(z)=|z-1|^{2\beta}$ in this case. Then, one can show (see for example (1.13) in~\cite{clakra15}),
\begin{equation}\label{eq:clakra_top}
  \log D_N(f^*)=\beta^2\log N+\log\frac{\mathcal{G}^2(1+\beta)}{\mathcal{G}(1+2\beta)}+O\left(\frac{1}{N}\right),
\end{equation}
provided that $\RE(\beta)>-1/2$. One sees that this agrees with the asymptotic form for $\mom_{\U(N)}(1,\beta)$ calculated by Keating and Snaith.

For $\mom_{\U(N)}(2,\beta)$, Claeys and Krasovsky must handle two Fisher-Hartwig singularities. Their results are uniform in that they describe the transition between the two singularities being distinct and when they are permitted to merge. There is also a connection to a solution of a certain non-linear second-order ordinary differential equation, known as Painlev\'e $\operatorname{V}$.

\begin{theorem}[Claeys and Krasovsky~\cite{clakra15}]\label{thm:claeys}
  Take $\beta>-1/4$.  Let
  \begin{equation}
    f(z)=|z-e^{i\theta_1}|^{2\beta}|z-e^{-i\theta_2}|^{2\beta}.
  \end{equation}
  Then for $0<t_1<\pi$ and as $N\rightarrow\infty$,
  \begin{equation}\label{eq:claeys_krasovsky}
    \int_0^{t_1}D_N(f)dt=
    \begin{dcases}
      c_1(t_1,\beta) N^{2\beta^2}(1+o(1)) &\text{if }2\beta^2<1\\
      c_2 N\log N(1+o(1)) &\text{if }2\beta^2=1\\
      c_3(\beta) N^{4\beta^2-1}(1+o(1)) &\text{if }2\beta^2>1.
    \end{dcases}
  \end{equation}
  The constants $c_1, c_2, c_3$ are explicitly given in~\cite{clakra15}, and are additionally related to a solution to the Painlev\'e $\operatorname{V}$ differential equation.
\end{theorem}

Thus, \eqref{eq:claeys_krasovsky} agrees\footnote{Within their proof it is clear how to deal with the constraint on $t_1$, see the comments after (1.47) in~\cite{clakra15}.} with conjecture~\ref{conj:fk_mom} when $k=2$.  Theorem~\ref{thm:claeys} is proved using Riemann-Hilbert techniques.

After the publication of~\cite{baikea19}, Fahs extended the results of Claeys and Krasovsky from two merging singularities to $k\in\mathbb{N}$ merging singularities.  Fahs' result\footnote{Adapted from~\cite{fah21} for notational consistency.} captures the leading order behaviour in each regime, as well as at the transition point $k=1/\beta^2$, though is not precise enough to capture information about the leading order coefficients beyond $k=2$. 

\begin{theorem}[Fahs~\cite{fah21}]\label{thm:fahs}
  Let $k\in\mathbb{N}$ and set
  \begin{equation}
    f(z)\coloneqq\prod_{j=1}^k|z-e^{i\theta_j}|^{2\beta},
  \end{equation}
  with $\beta\geq 0$ and $0\leq\theta_1<\theta_2<\cdots<\theta_k\leq 2\pi$. Then as $N\rightarrow\infty$,
  \begin{equation}
    \log D_N(f)=k\beta^2\log N-2\beta^2\sum_{1\leq i<j\leq k}\log\left(\sin\left|\frac{\theta_i-\theta_j}{2}\right|+\frac{1}{N}\right)+O(1),
  \end{equation}
  where the error term is uniform for $0\leq \theta_1<\cdots<\theta_k\leq 2\pi$. 
\end{theorem}
This verifies conjecture~\ref{conj:fk_mom} for $k\in\mathbb{N}$ and $\beta\geq 0$. 

To prove theorem~\ref{thm:claeys} and theorem~\ref{thm:fahs}, the authors first relate the Toeplitz determinants to a system of polynomials orthogonal on the unit circle.  Then, both characterize these polynomials as a Riemann--Hilbert problem and use the associated techniques in order to determine the respective results. The topic of Riemann--Hilbert problems is outside the scope of this review, but the interested reader may seek out~\cite{bot21, deift99, deizho93} for further details, and~\cite{blebleits01} for an overview of connections to random matrix theory.

There has also been recent progress on understanding the behaviour of $\mom_{\U(N)}(k,\beta)$ at the critical point $k\beta^2=1$. In~\cite{keawon20}, Wong and Keating study so-called `critical-subcritical' moments of $\GMC$s, in dimensions $d\leq 2$.  For the random matrix models, $d=1$. `Critical-subcritical' here refers to the ordered parameter pair $(k,\beta)$.  `Critical' $k$th moments occur at $k=1/\beta^2$. The $\GMC$ measure is defined in the `subcritical' regime for $\beta^2<1$ by~\eqref{eq:GMC_again}. Wong and Keating determine the asymptotic leading order, including coefficient, for $\mom_{\U(N)}(k,1/\sqrt{k})$ provided that $k\in\mathbb{N}$.

Recall that the result of Remy shows that the $k$th moment of $\GMC$ exists in the (subcritical) regime $k<1/\beta^2$, cf.~\eqref{eq:remy}. Wong and Keating can extend to the (critical) case where $k=1/\beta^2$, i.e. beyond where $\GMC$ moments exist. By performing an argument akin to~\eqref{remy_conv_to_gmcB}--\eqref{eq:gmc_mom}, they apply their $\GMC$ result to moments of moments. 

By theorems~\ref{thm:claeys} and~\ref{thm:fahs}, the leading behaviour in $N$ of $\mom_{\U(N)}(k,1/\sqrt{k})$ is known for $k\in\mathbb{N}$, but the leading coefficient was only identified for $k=2$ (cf.~\eqref{eq:claeys_krasovsky}, also~\cite{clakra15}). Through analysing the critical moments of $\GMC$ measures as one approaches $k\beta^2=1$ from the subcritical $\GMC$ regime ($k\beta^2<1$), Wong and Keating identify the asymptotic leading order coefficient of $\mom_{\U(N)}(k,1/\sqrt{k})$. 

\begin{theorem}[Keating and Wong~\cite{keawon20}]\label{thm:keawon}
  Take $k\in\mathbb{N}$, $k\geq 2$.  Then as $N\rightarrow\infty$,
  \begin{equation}\label{eq:keawon}
    \mom_{\U(N)}\left(k,\tfrac{1}{\sqrt{k}}\right)\sim \frac{k-1}{\Gamma\left(1-\frac{1}{k}\right)^k}\left(\frac{\mathcal{G}^2\left(1+\frac{1}{\sqrt{k}}\right)}{\mathcal{G}\left(1+\frac{2}{\sqrt{k}}\right)}\right)^k N\log N. 
  \end{equation}
\end{theorem}

Although the proof of theorem~\ref{thm:keawon} requires $k$ to be integral (so that one can perform the usual switching of order of integration in~\eqref{eq:mom_gN}), one expects that~\eqref{eq:keawon} should hold for general $k>1$. Wong and Keating also give precise conjectures for the leading order coefficient at the critical point for the symplectic, orthogonal, and \CBE{} cases\footnote{See~\eqref{eq:mom_general} and~\eqref{eq:mom_cbe} for the relevant definitions.}, with appropriate parameter restrictions.

\subsubsection*{Integer moments of moments}

Prior to the extension of the result of Claeys and Krasovsky by Fahs, the present authors employed formulae developed in \cite{keasna00a, cfkrs03, bumgam06, keaodg08, krrr18} to verify conjecture~\eqref{conj:fk_mom} for integer moment parameters.  We separately discuss these results and extensions here.  As shown, the results of Claeys and Krasovsky, Fahs, and Keating and Wong~\cite{clakra15, fah21, keawon20} have now established\footnote{Up to an explicit leading order coefficient in the supercritical regime.} conjecture~\ref{conj:fk_mom} for $k\in\mathbb{N}$. The techniques of these section require a further restriction on $\beta$.  However, this results in more access to the leading order coefficient, potentially lower order terms, and allows one to appeal to new methods.  

In particular, the two themes of complex function theory and symmetry function theory play an important role.  Conjecture~\eqref{conj:fk_mom} can be reformulated in terms of particular symmetric polynomials and a lattice point count function.  This gives a polynomial bound on $\mom_{\U(N)}(k,\beta)$ at integer values of $k$, $\beta$, and $N$.  Additionally, a representation in terms of multiple contour integrals is employed; this furnishes an expression for $\mom_{\U(N)}(k,\beta)$ as an entire function of $N$ and yields the following theorem.

\begin{theorem}[Bailey and Keating~\cite{baikea19}]\label{thm:mom_unitary}
  Let $k,\beta\in\mathbb{N}$.  Then 
  \begin{equation}\label{eq:momkbeta}
    \mom_{\U(N)}(k,\beta)=\gamma_{k,\beta}N^{k^2\beta^2-k+1}+O(N^{k^2\beta^2-k}),
  \end{equation}
  where $\gamma_{k,\beta}$ can be written explicitly in the form of an integral. 
\end{theorem}

Using a combinatorial sum equivalent to the multiple contour integrals due to~\cite{cfkrs03}, one can deduce the following result.

\begin{theorem}[Bailey and Keating~\cite{baikea19}]\label{thm:unitary_polynomial}
  Let $k, \beta\in\mathbb{N}$. Then $\mom_{\U(N)}(k,\beta)$ is a polynomial in $N$.
\end{theorem}

This establishes the validity of conjecture~\ref{conj:fk_mom} for $k,\beta\in\mathbb{N}$, as well as revealing the polynomial structure of the moments.  A selection of examples of the polynomials are given in section~\ref{sec:examples}.

It had been previously established~\cite{krrr18, bumgam06} that in the cases $k=1,2$ and $\beta\in\mathbb{N}$, one can rephrase~\eqref{eq:k-fold_ave} in terms of symmetric function theory.

Bump and Gamburd rederived theorem~\ref{thm:ks} provided\footnote{Whilst their proof relies on $\beta\in\mathbb{N}$, one can analytically extend to $\RE(2\beta)>-1$.} $\beta\in\mathbb{N}$.  For ease of presentation we restate this as a separate theorem in the form in which they proved it, even though it coincides precisely with theorem~\ref{thm:ks}.

\begin{theorem}[Keating \& Snaith~\cite{keasna00a}; Bump \& Gamburd~\cite{bumgam06}]~\label{thm:bumgam}
  Let $\beta\in\mathbb{N}$.  Then
  \begin{equation}
    \mathbb{E}[|P_N(A,\theta)|^{2\beta}]=
    \prod_{j=0}^{N-1}\frac{j!(j+2\beta)!}{(j+\beta)!^2}.
  \end{equation}
\end{theorem}
In order to prove theorem~\ref{thm:bumgam}, Bump and Gamburd re-express the moments of characteristic polynomials as an average over Schur polynomials.  They derive the following proposition\footnote{The statement of proposition~\ref{prop:bumgam} is more generally stated in~\cite{bumgam06}, giving a further equality for \eqref{eq:bumgam1} as a permutation sum. The representation of the moments as a permutation sum had also already been proved by Conrey et al.~\cite{cfkrs03}, though again using a different method.}.

\begin{proposition}[Bump and Gamburd~\cite{bumgam06}]\label{prop:bumgam}
  Take $K, L, N\in\mathbb{N}$ and $\alpha_1,\dots,\alpha_{K+L}\in\mathbb{C}$, then
  \begin{equation}\label{eq:bumgam1}
    \int_{\U(N)}\prod_{l=1}^L\det(I+\alpha_l^{-1}A^*)\prod_{k=1}^K\det(I+\alpha_{L+k}A)dA=\frac{s_{\langle N^L\rangle}(\alpha_1,\dots,\alpha_{K+L})}{\prod_{l=1}^L\alpha_l^N}.
  \end{equation}
\end{proposition}

In~\eqref{eq:bumgam1}, the Schur polynomial $s_\lambda(\underline{x})$ was defined in definition~\ref{schur_function}. 

Theorem~\ref{thm:bumgam} then follows from proposition~\ref{prop:bumgam} by setting $L=K=\beta$ and $\alpha_1=\cdots=\alpha_{2\beta}=1$ (using the rotational invariance of the Haar measure). Thus
\begin{equation}
  \int_{\U(N)}|P_N(A,\theta)|^{2\beta}dA=s_{\langle N^\beta\rangle}(\overbrace{1,\dots,1}^{2\beta}).
\end{equation}
Using~\eqref{eq:schur_def}, the Schur polynomial $s_\lambda(1^n)$ is equal to the number of semistandard Young tableaux of shape $\lambda$ with entries in $\{1,\dots,n\}$. By the following lemma, the `Hook-content formula' (see~\cite{stanley99}), one has the statement of the theorem. 

\begin{lemma}~\label{lemma:ssyt}
  Take a partition $\lambda$ and $n\in\mathbb{N}$.  Then
  \begin{equation}
    |\SSYT_n(\lambda)|=\prod_{1\leq i<j\leq n}\frac{\lambda_i-\lambda_j+j-i}{j-i},
  \end{equation}
  which manifestly is a polynomial in $\lambda_i-\lambda_j$. 
\end{lemma}

The following result (proposition 2.1 in~\cite{baikea19}) is an extension of proposition~\ref{prop:bumgam} to handle~\eqref{eq:k-fold_ave}. 

\begin{proposition}[Bailey and Keating~\cite{baikea19}]\label{prop:symmetric}
  For $N, k, \beta\in\mathbb{N}$, we have
  \begin{equation}
    \mathbb{E}_{A\in \U(N)}\left[\prod_{j=1}^k|P_N(A,\theta_j)|^{2\beta}\right]=\frac{s_{\langle N^{k\beta}\rangle}(e^{-i\underline{\theta}})}{\prod_{j=1}^ke^{-iN\beta\theta_j}},
  \end{equation} 
  where $\langle \alpha^n\rangle=(\underbrace{\alpha,\dots,\alpha}_n)$ and 
  \begin{equation}\label{schurvector}
    e^{i\underline{\theta}}=(\overbrace{e^{i\theta_1},\dots,e^{i\theta_1}}^{2\beta},\overbrace{e^{i\theta_2},\dots,e^{i\theta_2}}^{2\beta},\dots,\overbrace{e^{i\theta_{k}},\dots,e^{i\theta_{k}}}^{2\beta}).
  \end{equation}
\end{proposition} 

Therefore, by proposition~\ref{prop:symmetric} and~\eqref{eq:schur_def}, one can rewrite $\mom_{\U(N)}(k,\beta)$ at $k, \beta\in\mathbb{N}$ in terms of Schur functions,
\begin{align}
  \mom_{\U(N)}(k,\beta)
  &=\frac{1}{(2\pi)^k}\int_0^{2\pi}\dots\int_0^{2\pi}\sum_{T\in SSYT_{2k\beta}(\langle N^{k\beta}\rangle)}\prod_{j=1}^k e^{-i\theta_j(\tau_j-N\beta)}d\theta_j\label{eq:integrate_schur_theta}
  \end{align}
where
\begin{equation}\label{eq:ssyt_restriction}
  \tau_j=t_{2(j-1)\beta+1}+\cdots+t_{2j\beta}\quad\text{for }j=1,\dots,k,
\end{equation}
and $t_n=t_n(T)$ is the number of entries in the tableau $T$ equal to $n$. By computing the $\theta$ integrals in \eqref{eq:integrate_schur_theta}, one finds that
\begin{equation}\label{eq:poly_bound}
  \mom_{\U(N)}(k,\beta)=\sideset{}{^*}\sum_{T\in \SSYT_{2k\beta}(\langle N^{k\beta}\rangle)}1,
\end{equation}
where the sum is now over $T\in\SSYT_{2k\beta}(\langle N^{k\beta}\rangle)$ subject to the additional restriction~\eqref{eq:ssyt_restriction}.   When specialized to the case of $k=1,\beta\in\mathbb{N}$, this approach agrees with the result of Bump and Gamburd. Combining lemma~\ref{lemma:ssyt} and \eqref{eq:poly_bound} provides a (not sharp) polynomial bound of $N^{k^2\beta^2}$ on $\mom_{\U(N)}(k,\beta)$.

To achieve the predicted asymptotic form from theorem~\ref{thm:mom_unitary}, one instead turns to another representation of the average over $\U(N)$, due to Conrey et al.\ (lemma 2.1 in~\cite{cfkrs03}).

\begin{lemma}[Conrey et al.~\cite{cfkrs03}]\label{lemma:mci}
  For $\alpha_j\in\mathbb{C}$,
  \begin{align*}
    \int_{\U(N)}&\prod_{j=m+1}^n\det(I-Ae^{\alpha_j})\prod_{j=1}^m\det(I-A^*e^{-\alpha_j})dA\\
    &=\frac{(-1)^{n(n-1)/2}}{(2\pi i)^nm!(n-m)!}\prod_{q=m+1}^{n}e^{N\alpha_q}\oint\cdots\oint\frac{e^{-N\sum_{l=m+1}^nz_l}\Delta(z_1,\dots,z_{n})^2dz_1\cdots dz_n}{\prod_{1\leq l\leq m<q\leq n}\left(1-e^{z_q-z_l}\right)\prod_{l=1}^n\prod_{q=1}^n(z_l-\alpha_q)},
  \end{align*}
  where the contours enclose the poles at $\alpha_1,\dots,\alpha_n$ and $\Delta(z_1,\dots,z_n)=\prod_{i<j}(z_j-z_i)$ is the Vandermonde determinant.
\end{lemma}
This multiple contour integral is nearly identical to the conjectural form for shifted moments of the zeta function, as discussed in section~\ref{sec:mom_other}. Choosing the $\alpha_j$ correctly and analysing the multiple contour integral for large $N$, one verifies the statement of theorem~\ref{thm:mom_unitary}. Theorem~\ref{thm:unitary_polynomial} requires yet another representation of~\eqref{eq:k-fold_ave} (again due to Conrey et al.~\cite{cfkrs03}), involving a permutation sum.

Following the publication of~\cite{baikea19}, Assiotis and Keating~\cite{asskea20} extended the symmetric function theoretic analysis using \emph{Gelfand-Tsetlin} patterns, still with the restriction that $k, \beta\in\mathbb{N}$.  They are able to recover theorem~\ref{thm:mom_unitary} using this combinatorial approach, yielding an alternative interpretation of the leading order coefficient $\gamma_{k,\beta}$.  This permits a further connection to a particular Painlev\'e differential equation.  Such an approach is well-adaptable to the case of symplectic and orthogonal moments of moments, which is discussed further in the next section.

It is also interesting to remark that Barhoumi-Andr{\'e}ani~\cite{bar20} has recently proposed a general framework rooted in symmetric function theory for understanding various averages of unitary characteristic polynomials.  Through this lens, one can rephrase and rederive many classical results in unitary random matrix theory.  This includes moments of moments, but also encompasses other interesting averages.  We refer to section 5.4 of~\cite{bar20} for a summary table of applicable problems.  

\subsection{Generalized moments and applications}\label{sec:mom_other}

One can generalize \eqref{eq:mom_gN} to the other compact matrix groups:
\begin{equation}\label{eq:mom_general}
  \mom_{G(N)}(k,\beta)\coloneqq\int_{G(N)}\left(\frac{1}{2\pi}\int_0^{2\pi}|P_{G(N)}(A,\theta)|^{2\beta}d\theta\right)^kdA, 
\end{equation}
where $G(N)\in\{\U(N), \Sp(2N), \SO(N), \operatorname{O}(N), \SU(N)\footnote{$\SU(N)$ is the set of all $N\times N$ unitary matrices with unit determinant.}\}$. In this section we focus on $G(N)=\Sp(2N)$ and $G(N)=\SO(2N)$, though the ideas presented generalize to the other compact groups. This is primarily for number theoretic reasons. Recall that matrices from $\Sp(2N)$ and $\SO(2N)$ have eigenvalues which come in complex conjugate pairs, hence there are $N$ `independent' eigenvalues.

\subsubsection*{Symplectic and Orthogonal groups}

Assiotis and Keating~\cite{asskea20} rederived the asymptotic result of~\cite{baikea19}, theorem~\ref{thm:mom_unitary}, via Gelfand-Tsetlin patterns.  These ideas are extended\footnote{Since there is a generalization of lemma~\ref{lemma:mci} to averages over $\Sp(2N)$ or $\SO(2N)$, the results described in this section could also be arrived at via the method of proof of theorem~\ref{thm:mom_unitary}.} by Assiotis et al.~\cite{assbaikea20} to compute the asymptotic behaviour of \eqref{eq:mom_general}.

\begin{theorem}[Assiotis, Bailey, Keating~\cite{assbaikea20}]\label{thm:mom_symplectic}
  Let $G(N)=\Sp(2N)$. Let $k,\beta \in \mathbb{N}$. Then, $\mom_{\Sp(2N)}\left(k,\beta\right)$ is a \emph{polynomial function} in $N$. Moreover, 
  \begin{align}\label{eq:mom_symplectic}
    \mom_{\Sp(2N)}\left(k,\beta\right)=\gamma^{(\Sp)}_{k,\beta}N^{k\beta(2k\beta+1)-k}+O\left(N^{k\beta(2k\beta+1)-k-1}\right),
  \end{align}
  where the leading order term coefficient $\gamma^{(\Sp)}_{k,\beta}$ is the volume of a convex region and is strictly positive.
\end{theorem}

\begin{theorem}[Assiotis, Bailey, Keating~\cite{assbaikea20}]\label{thm:mom_orthogonal}
  Let $G(N)=\SO(2N)$.  Let $k, \beta\in\mathbb{N}$.  Then, $\mom_{\SO(2N)}(k,\beta)$ is a \emph{polynomial function} in $N$.  Moreover, 
  \begin{align}
    \mom_{\SO(2N)}(1,1)&=2(N+1)\\
    \intertext{otherwise,}
    \mom_{\SO(2N)}(k,\beta)&=\gamma^{(\SO)}_{k,\beta}N^{k\beta(2k\beta-1)-k}+O\left(N^{k\beta(2k\beta-1)-k-1}\right),\label{eq:mom_orthogonal}
  \end{align}
  where the leading order term coefficient $\gamma^{(\SO)}_{k,\beta}$ is given as a sum of volumes of convex regions and is strictly positive. 
\end{theorem}

(For a precise description of the convex regions involved in the definition of $\gamma^{(\Sp)}_{k,\beta}$ and $\gamma^{(\SO)}_{k,\beta}$, see section 4.2 and 5.2 of~\cite{assbaikea20}.) By comparing theorems~\ref{thm:mom_symplectic} and~\ref{thm:mom_orthogonal} to theorem~\ref{thm:mom_unitary}, one sees that the asymptotic behaviour differs between all three matrix groups, though the polynomial property is retained. 

In order to describe the key aspects of the proof of theorems~\ref{thm:mom_symplectic} and~\ref{thm:mom_orthogonal}, we first introduce some new notions. The first generalizes the definition of a partition, to allow for general integer entries. 
\begin{definition}[Signature]\label{def:signature}
  A signature $\lambda$ of length $n$ is a sequence of $n$ non-increasing integers $(\lambda_1\ge \lambda_2 \ge \dots \ge \lambda_n)$. The set of all such signatures is $\textsc{S}_n$. The set of the signatures with non-negative entries\footnote{Note that this is \emph{distinct} from the definition of a partition since \emph{trailing zeros} are recorded, whereas partitions $\nu$ are identified with any other partition which has the same non-zero entries.} is $\textsc{S}_n^+$.  

  Given $\lambda=(\lambda_1,\dots,\lambda_n) \in \textsc{S}_n^+$, define $\lambda^-\coloneqq(\lambda_1,\dots,\lambda_{n-1},-\lambda_n)$. If $\lambda_1=\cdots=\lambda_n=L$ then one write $\lambda=\langle L^n\rangle$.
\end{definition}

\begin{figure}
  \centering
  \begin{subfigure}[t]{0.48\textwidth}
    \centering
    \begin{tikzpicture}[scale=0.6]
      \node at (0,0) {$\lambda_{5}$};
      \node at (2,0) {$\lambda_{4}$};
      \node at (4,0) {$\lambda_{3}$};
      \node at (6,0) {$\lambda_{2}$};
      \node at (8,0) {$\lambda_{1}$};

      \node at (-1,1) {$\nu_{6}$};
      \node at (1,1) {$\nu_{5}$};
      \node at (3,1) {$\nu_{4}$};
      \node at (5,1) {$\nu_{3}$};
      \node at (7,1) {$\nu_{2}$};
      \node at (9,1) {$\nu_{1}$};

      \draw (-0.5,0.5) node[rotate=-45] {$\leq$};
      \draw (0.5,0.5) node[rotate=45] {$\leq$};
      \draw (1.5,0.5) node[rotate=-45] {$\leq$};
      \draw (2.5,0.5) node[rotate=45] {$\leq$};
      \draw (3.5,0.5) node[rotate=-45] {$\leq$};
      \draw (4.5,0.5) node[rotate=45] {$\leq$};
      \draw (5.5,0.5) node[rotate=-45] {$\leq$};
      \draw (6.5,0.5) node[rotate=45] {$\leq$};
      \draw (7.5,0.5) node[rotate=-45] {$\leq$};
      \draw (8.5,0.5) node[rotate=45] {$\leq$};
    \end{tikzpicture}
    \caption{Interlacing with $\lambda\in\textsc{S}_5$, $\nu\in \textsc{S}_{6}$}\label{fig:interlacing1}
  \end{subfigure}\hfill
  \begin{subfigure}[t]{0.48\textwidth}
    \centering
    \begin{tikzpicture}[scale=0.6]
      \node at (2,0) {$\lambda_{4}$};
      \node at (4,0) {$\lambda_{3}$};
      \node at (6,0) {$\lambda_{2}$};
      \node at (8,0) {$\lambda_{1}$};

      \node at (3,1) {$\nu_{4}$};
      \node at (5,1) {$\nu_{3}$};
      \node at (7,1) {$\nu_{2}$};
      \node at (9,1) {$\nu_{1}$};

      \draw (2.5,0.5) node[rotate=45] {$\leq$};
      \draw (3.5,0.5) node[rotate=-45] {$\leq$};
      \draw (4.5,0.5) node[rotate=45] {$\leq$};
      \draw (5.5,0.5) node[rotate=-45] {$\leq$};
      \draw (6.5,0.5) node[rotate=45] {$\leq$};
      \draw (7.5,0.5) node[rotate=-45] {$\leq$};
      \draw (8.5,0.5) node[rotate=45] {$\leq$};
    \end{tikzpicture}
    \caption{Interlacing with $\lambda\in\textsc{S}_4$, $\nu\in \textsc{S}_{4}$}\label{fig:interlacing2}
  \end{subfigure}\hfill
  \caption[Examples of interlacing of signatures]{Examples of interlacing of signatures.  Figure~\ref{fig:interlacing1} shows signatures whose length differs by $1$, and figure~\ref{fig:interlacing2} shows two signatures of the same length. Note that in both cases, the numbering is from right to left, as to be in-keeping with future definitions. The inequalities are explicitly shown, though it is not standard to do so.}\label{fig:interlacing}
\end{figure}
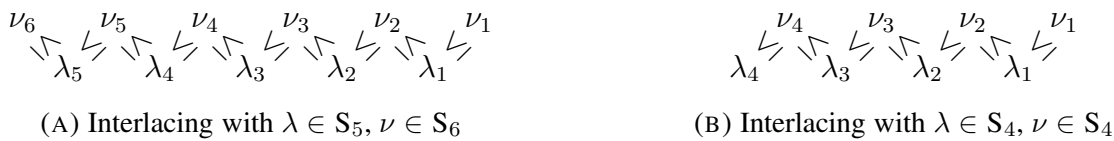

The next definition describes how two signatures interact. 
\begin{definition}[Interlacing]\label{def:interlacing}
 Signatures $\lambda \in \textsc{S}_n$ and $\nu \in \textsc{S}_{n+1}$ interlace, written $\lambda\prec \nu$, if:
  \begin{align}\label{interlacing_eqs}
    \nu_1\ge \lambda_1 \ge \nu_2 \ge \cdots \ge \nu_n \ge \lambda_n \ge \nu_{n+1}.
  \end{align}
  Similarly, $\lambda \in \textsc{S}_n$ and $\nu \in \textsc{S}_n$ interlace (retaining the notation $\lambda \prec \nu$) if:
  \begin{align}
    \nu_1\ge \lambda_1 \ge \nu_2 \ge \cdots \ge \nu_n \ge \lambda_n.
  \end{align}
  It is common to draw interlacing signatures, so to emphasize their interaction, see figure~\ref{fig:interlacing}. In this review, our convention is to draw signatures from right to left. 
\end{definition}

The next definition introduces a `full' \emph{Gelfand-Tsetlin} pattern.  These are required for unitary moments of moments, see~\cite{asskea20}.  Hereafter we will instead focus on `half' \emph{Gelfand-Tsetlin} patterns. 

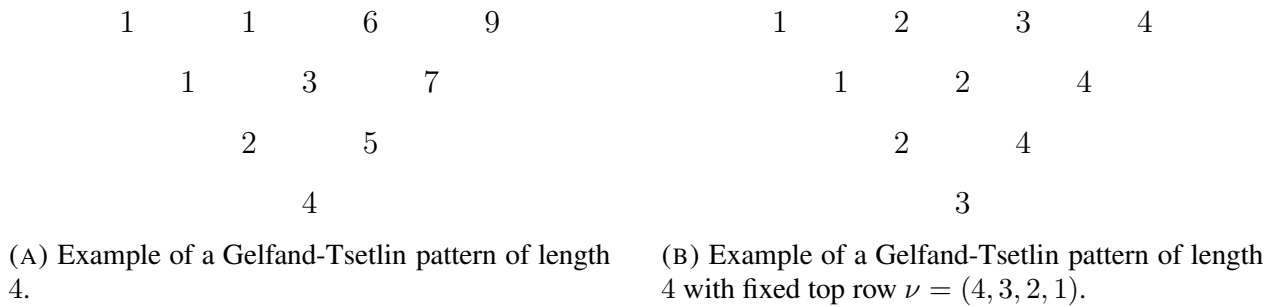
\begin{figure}
  \centering
  \begin{subfigure}[t]{0.48\textwidth}
    \centering
    \begin{tikzpicture}[scale=0.8]
      \node at (0,0) {$4$};
      \node at (-1,1) {$2$};
      \node at (1,1) {$5$};
      \node at (-2,2) {$1$};
      \node at (0,2) {$3$};
      \node at (2,2) {$7$};
      \node at (-3,3) {$1$};
      \node at (-1,3) {$1$};
      \node at (1,3) {$6$};
      \node at (3,3) {$9$};
    \end{tikzpicture}
    \caption[Example of a Gelfand-Tsetlin pattern.]{Example of a Gelfand-Tsetlin pattern of length $4$.}\label{fig:gt1}
  \end{subfigure}\hfill
  \begin{subfigure}[t]{0.48\textwidth}
    \centering
    \begin{tikzpicture}[scale=0.8]
      \node at (0,0) {$3$};
      \node at (-1,1) {$2$};
      \node at (1,1) {$4$};
      \node at (-2,2) {$1$};
      \node at (0,2) {$2$};
      \node at (2,2) {$4$};
      \node at (-3,3) {$1$};
      \node at (-1,3) {$2$};
      \node at (1,3) {$3$};
      \node at (3,3) {$4$};
    \end{tikzpicture}
    \caption[Example of a Gelfand-Tsetlin pattern with fixed top row.]{Example of a Gelfand-Tsetlin pattern of length $4$ with fixed top row $\nu=(4,3,2,1)$.}\label{fig:gt2}
  \end{subfigure}\hfill
  \caption[Examples of Gelfand-Tsetlin Patterns]{Examples of Gelfand-Tsetlin Patterns.}\label{fig:full_gt}
\end{figure}

\begin{definition}[Gelfand-Tsetlin pattern]
  A non-negative Gelfand-Tsetlin pattern of length/depth $n$ is a sequence of signatures $\left(\lambda^{(i)}\right)_{i=1}^n$ such that $\lambda^{(i)}\in \textsc{S}_i^+$ and
  \begin{equation}
    \lambda^{(1)}\prec \lambda^{(2)}\prec \cdots \prec \lambda^{(n-1)}\prec \lambda^{(n)}.
  \end{equation}
  $GT_n^+$ is the set of all such patterns. Given a signature $\nu\in S_n^+$, it is often useful to additionally consider those Gelfand-Tsetlin pattern with fixed top row $\nu$.  The set of such patterns is written $GT_n^+(\nu)$.

  It is common to draw Gelfand-Tsetlin patterns as a triangular array, essentially a generalization of interlacing diagrams, see figure~\ref{fig:full_gt}. 
\end{definition}

For a signature $\nu\in \textsc{S}_n^+$, there is a well-known bijection between semistandard Young tableaux of shape $\nu$ with entries in $\{1,\dots,n\}$ and non-negative Gelfand-Tsetlin patterns of length $n$ with fixed top row $\nu$,  see for example~\cite{asskea20,gorrah19}. It is this bijection that was exploited by Assiotis and Keating~\cite{asskea20} when rederiving the asymptotic result of theorem~\ref{thm:mom_unitary}.  When handling symplectic and orthogonal moments of moments, we instead require \emph{half} Gelfand-Tsetlin patterns. 

\begin{definition}[Half (Gelfand-Tsetlin) pattern]\label{def:halfpat}
  Let $n$ be a positive integer. A half (Gelfand-Tsetlin) pattern of length $n$ is given by a sequence of interlacing signatures $\left(\lambda^{(i)}\right)_{i=1}^{n}$ such that $\lambda^{(2i-1)},\lambda^{(2i)}\in \textsc{S}_{i}$ with the interlacing condition:
  \begin{align*}
    \lambda^{(1)}\prec \lambda^{(2)}\prec \cdots \prec\lambda^{(n-1)}\prec \lambda^{(n)}.
  \end{align*}
  The first entries on the odd rows, namely $\lambda_{i}^{(2i-1)}$, are the \emph{odd starters}.
\end{definition}

We now arrive at the definition of a symplectic (Gelfand-Tsetlin) pattern, see figure~\ref{fig:symplpattern} for an illustration. 

\begin{definition}[Symplectic patterns]\label{def:symplpat}
  Let $n$ be a positive integer. A $(2n)$--symplectic Gelfand-Tsetlin pattern $P=\left(\lambda^{(i)}\right)_{i=1}^{2n}$ is a half pattern of length $2n$ all of whose entries are non-negative integers. For fixed complex numbers $(x_1,\dots,x_n)$ we associate to the pattern $P$ a weight $w_{sp}(P)$ (dependence on $x_1,\dots,x_n$ is suppressed from the notation and will be clear from context in what follows) given by:
  \begin{align*}
    w_{sp}(P)=\prod_{i=1}^{n}x_i^{\sum_{j=1}^{i}\lambda_j^{(2i)}-2\sum_{j=1}^{i}\lambda_j^{(2i-1)}+\sum_{j=1}^{i-1}\lambda_j^{(2i-2)}},
  \end{align*}
  with $\lambda^{(0)}\equiv 0$. Given $\nu \in \textsc{S}_n^+$, $SP_{\nu}$ is the set of all $(2n)$--symplectic Gelfand-Tsetlin patterns with top row $\lambda^{(2n)}=\nu$. 
\end{definition}

\begin{figure}[!htb]
  \centering
  \begin{subfigure}[t]{.48\textwidth}
    \centering
    \begin{tikzpicture}
      \node at (0,0) {$1$};
      \node at (1,1) {$2$};
      \node at (0,2) {$1$};
      \node at (2,2) {$2$};
      \node at (1,3) {$2$};
      \node at (3,3) {$3$};
      \draw (-0.5,-0.5) -- (3,-0.5);
      \node at (1,-1) {$w_{sp}(P)=x_2$};
    \end{tikzpicture}
    \caption[A symplectic Gelfand-Tsetlin pattern of length $4$]{An example of a (4)-symplectic Gelfand-Tsetlin pattern $P$, with its corresponding weight $w_{sp}(P)$ below for some complex numbers $x_1, x_2$ as appearing in definition~\ref{def:symplpat}.}\label{fig:symplpattern}
  \end{subfigure}\hfill
  \begin{subfigure}[t]{.48\textwidth}
    \centering
    \begin{tikzpicture}
      \node at (0,0) {$-1$};
      \node at (1,1) {$1$};
      \node at (0,2) {$0$};
      \node at (2,2) {$2$};
      \node at (1,3) {$2$};
      \node at (3,3) {$2$};
      \node at (0,4) {$-2$};
      \node at (2,4) {$2$};
      \node at (4,4) {$4$};
      \draw (-0.5,-0.5) -- (3,-0.5);
      \node at (1.5,-1) {$w_{o}(P)=(x_1x_2x_3^2)^{-1}$};
    \end{tikzpicture}
    \caption[An orthogonal Gelfand-Tsetlin pattern of length $5$]{An example of a (5)-orthogonal Gelfand-Tsetlin pattern $P$, with its corresponding weight $w_{o}(P)$ below for some complex numbers $x_1, x_2, x_3$ as appearing in definition~\ref{def:orthopat}.}\label{fig:orthopattern}
  \end{subfigure}
  \caption{Figures giving examples of symplectic and orthogonal Gelfand-Tsetlin patterns.}\label{fig:symp_orth_pattern_ex}
\end{figure}

We now give the combinatorial definition of the symplectic Schur polynomial as a sum of weights over symplectic patterns. This should be seen in the context of the definition of a (unitary) Schur polynomial, see definition~\ref{schur_function}. 
\begin{definition}[Symplectic Schur polynomial]\label{CombinatorialFormulaSymplectic}
  Let $\nu \in \textsc{S}_n^+$. The symplectic Schur polynomial is associated to $\nu$ is
  \begin{align}
    sp^{(2n)}_{\nu}\left(x_1,\dots,x_n\right)=\sum_{P \in SP_{\nu}}^{}w_{sp}(P).
  \end{align}
\end{definition}

Orthogonal patterns are similarly defined, though slightly more involved than the symplectic case since some of the elements are now permitted to be negative.  We write $\sgn(x)=1$ if $x\geq 0$ and $\sgn(x)=-1$ if $x<0$. 

\begin{definition}[Orthogonal patterns]\label{def:orthopat}
  Let $n\in\mathbb{N}$. A $(2n-1)$--orthogonal Gelfand-Tsetlin pattern $P=\left(\lambda^{(i)}\right)_{i=1}^{2n-1}$ is a half pattern of length $2n-1$ all of whose entries are either all integers or all half-integers\footnote{It transpires that for the problems at hand, the entries of $(2n-1)$--orthogonal Gelfand-Tsetlin patterns are always all integers.} and which moreover satisfy:
  \begin{enumerate}[label=(\roman*)]
  \item All entries except odd starters are non-negative.
  \item The odd starters satisfy $|\lambda_i^{(2i-1)}|\le \min \{\lambda_{i-1}^{(2i-2)},\lambda_i^{(2i)}\}$ for $i=2,\dots,n-1$ and moreover \hbox{$|\lambda_1^{(1)}|\le \lambda_1^{(2)}$} and $|\lambda_n^{(2n-1)}|\le \lambda_{n-1}^{(2n-2)}$.
  \end{enumerate}
  For fixed complex numbers $(x_1,\dots,x_n)$ we associate to the pattern $P$ a weight $w_{o}(P)$ given by:
  \begin{align*}
    w_{o}(P)=\prod_{i=1}^{n}x_i^{\textnormal{sgn}(\lambda_i^{(2i-1)})\textnormal{sgn}(\lambda_{i-1}^{(2i-3)})\left[\sum_{j=1}^{i}|\lambda_j^{(2i-1)}|-2\sum_{j=1}^{i-1}|\lambda_j^{(2i-2)}|+\sum_{j=1}^{i-1}|\lambda_j^{(2i-3)}|\right]},
  \end{align*}
  with $\lambda^{(0)},\lambda^{(-1)}\equiv 0$. Given $\nu \in \textsc{S}_n$, $OP_{\nu}$ is the set of all $(2n-1)$--orthogonal Gelfand-Tsetlin patterns with top row $\lambda^{(2n-1)}=\nu$. 
\end{definition}
See figure~\ref{fig:orthopattern} for an example of an orthogonal Gelfand-Tsetlin pattern. 

As in the symplectic case, we have the following combinatorial definition of the orthogonal Schur polynomial as a sum of weights over orthogonal patterns.
\begin{definition}[Orthogonal Schur polynomial]\label{CombinatorialFormulaOrthogonal}
  Let $\nu \in \textsc{S}_n^+$. The orthogonal Schur polynomial associated to $\nu$ is
  \begin{align}
    o^{(2n)}_{\nu}\left(x_1,\dots,x_n\right)=\sum_{P \in OP_{\nu}\cup OP_{\nu^-}}^{}w_{o}(P).
  \end{align}
\end{definition}

The connection between \eqref{eq:mom_general} and symplectic (resp. orthogonal) Gelfand-Tsetlin patterns follows via the symplectic (resp. orthogonal) Schur polynomials and the representation theory of $\Sp(2N)$ (resp. $\SO(2N)$), see~\cite{bumgam06}.  

To prove the polynomial part of theorems~\ref{thm:mom_symplectic} and~\ref{thm:mom_orthogonal}, one uses another (equivalent) representation of averages of characteristic polynomials, due to Conrey et al.~\cite{cfkrs03}. This is the same technique employed in the unitary case. 

To prove the asymptotic part of theorems~\ref{thm:mom_symplectic} and~\ref{thm:mom_orthogonal}, one translates \eqref{eq:mom_general} (via a generalization of proposition~\ref{prop:symmetric}) to an average over the respective Schur polynomials. By applying the bijection to symplectic or orthogonal half Gelfand-Tsetlin patterns, one has hence determined that \eqref{eq:mom_general} is equal to a count of (symplectic or orthogonal) half Gelfand-Tsetlin patterns with constraints.  By finally translating these discrete structures to a continuous setting and applying lattice point count asymptotics, one arrives at the stated results.

Naturally, one might be interested in Fyodorov-Keating type conjectures for the symplectic and orthogonal cases.  The outline given in sections~\ref{sec:fk_progress} suggests that, to leading order at least, the maxima conjecture would still hold\footnote{This may be unsurprising since away from the symmetry point, unitary statistics dominate in both cases, cf.~\cite{keaodg08, keasna00b}.}.

Notice that in passing, theorems~\ref{thm:mom_symplectic} and~\ref{thm:mom_orthogonal} prove the asymptotic growth and polynomial structure of certain restricted combinatorial counts, as theorem~\ref{thm:mom_unitary} did in the unitary case.  For example, one could read the argument of~\cite{assbaikea20} after the statement of proposition 4.2 agnostic to the random matrix motivation. We defer to the next section a discussion on the moments of moments conjecture.

\subsubsection{Related results}

Recently, Assiotis extended the results of~\cite{assbaikea20} to handle moments of moments of characteristic polynomials of the Circular $\beta$ ensemble (\CBE{}) (cf.~\eqref{eq:cbe}) for $\beta>0$. There is a natural generalization of \eqref{eq:mom_general} to the entire \CBE,
\begin{equation}\label{eq:mom_cbe}
  \mom_{\CBE}(k,q)\coloneqq \mathbb{E}_{\CBE}\left[\left(\frac{1}{2\pi}\int_0^{2\pi}|P_N(\theta)|^{2q}d\theta\right)^k\right],
\end{equation}
where, temporarily, we have changed the notation for the moment parameters.  The expectation in \eqref{eq:mom_cbe} is with respect to the $\beta$-dependent \CBE{} measure \eqref{eq:cbe}, and $P_N$ is the characteristic polynomial for the appropriate ensemble\footnote{Specifically, $P_N(\theta)=\prod_{j=1}^N(1-e^{-i(\theta-\theta_j)})$, where $e^{i\theta_1},\dots,e^{i\theta_N}$ are distributed according to~\eqref{eq:cbe}.}. 

\begin{theorem}[Assiotis~\cite{ass20}]\label{thm:ass_cbe}
  Let $\beta>0$ and $k,q\in\mathbb{N}$.  If moreover $\beta$ is such that
  \begin{itemize}
  \item $\beta<4q^2$, if $k=2$
  \item $\beta\leq 2$, if $k\geq 3$, then
  \end{itemize}
  \begin{equation}
    \lim_{N\rightarrow\infty}\frac{1}{N^{\frac{2}{\beta}(kq)^2-k+1}}\mom_{\CBE}(k,q)=\gamma^{(\beta)}_{k,q}.
  \end{equation}
  Further, the leading order coefficient is strictly positive and finite.  As in~\cite{asskea20, assbaikea20}, it can be described as an integral of a weight over constrained, continuous lacing arrays\footnote{See~\cite{ass20} for the precise definition.}.
\end{theorem}

Clearly, when $\beta=2$ in theorem~\ref{thm:ass_cbe}, this result coincides with theorem~\ref{thm:mom_unitary} (and the description of $\gamma^{(2)}_{k,q}$ specializes to the one found by Assiotis and Keating~\cite{asskea20}, and matches -- in a non-obvious way -- the description of $\gamma_{k,\beta}$ found in~\cite{baikea19}). Theorem~\ref{thm:ass_cbe} is proved using generalizations of the ideas in~\cite{assbaikea20}; in particular a result of Matsumoto~\cite{mat07} generalizes proposition~\ref{prop:bumgam} to the \CBE{} using Jack polynomials.  It is expected that the result should hold for $\beta<2kq^2$, for (possible non-integer) $k>1$. As suggested by~\cite{keawon20} (see the discussion after theorem~\ref{thm:keawon}), at $\beta=2kq^2$ there should be a transition point and the leading order in $N$ of the moments of moments should be like $N\log N$, as it is in the unitary $\beta=2$ case.

Recall that the asymptotic statement of theorem~\ref{thm:mom_unitary} may be rederived by using asymptotics of Toeplitz determinants\footnote{As noted, these techniques additionally permit the $\beta$ parameter to be a positive real number.}, see section~\ref{sec:unitary_mom}.  Therefore, it would be interesting to extend this approach developed by Claeys and Krasovsky in~\cite{clakra15} and Fahs in~\cite{fah21} to the more general moments of moments.  For the orthogonal and symplectic cases, this would require uniform asymptotics for determinants of the form Toeplitz $\pm$ Hankel as the singularities merge.  Such an approach would almost certainly have the benefit of not requiring the integrality conditions of theorems~\ref{thm:mom_symplectic} and~\ref{thm:mom_orthogonal}. It would also be interesting to describe the full moments of moments behaviour, akin to~\eqref{eq:fk_mom}.  

Recently, Claeys et al.~\cite{cgmy20} proved such uniform asymptotics, using a Riemann-Hilbert analysis.  Their work allows for the associated Fisher-Hartwig singularities to vary with the matrix size, and extends previous work on Toeplitz $\pm$ Hankel determinants~\cite{deiitskra13, basehr02, basehr01, bairai01, bastra92}. They first write averages over the orthogonal and symplectic ensembles\footnote{Their analysis includes $\SO(2N), \SO(2N+1), \Sp(2N)$ as well as $O^{-}(2N+1)$ and $O^{-}(2N+2)$, the complementary cosets of the orthogonal group of matrices with determinant $-1$. It is useful to record that the joint probability density for $\Sp(2N)$ is the same as for $O^{-}(2N+2)$ -- see for example~\cite{meckes19}.} in terms of unitary group averages and certain orthogonal polynomials on the unit circle.

By analysing these asymptotically (provided the singularities do not approach the symmetry points faster than the scale of mean eigenvalue spacing), they derive the relevant uniform asymptotics. In the case of two merging singularities, the leading order coefficient can be determined explicitly using theorem~\ref{thm:claeys}. More generally, the asymptotics are only known up to a multiplicative constant.  To prove stronger results, one would need to prove a stronger version of theorem~\ref{thm:fahs}.  

In a related work, Forkel and Keating~\cite{forkea21} proved the analogue of \eqref{eq:char_pol_gmc} for symplectic and orthogonal characteristic polynomials in the $L^2-$range.  Using unpublished work\footnote{The result extends to the orthogonal and symplectic groups results of Hughes et al.~\cite{hugkeaoco00} showing that the real and imaginary part of the logarithm of unitary characteristic polynomials jointly converge to a pair of Gaussian fields on the unit circle, cf.~\eqref{eq:expansion_log}. See appendix A of~\cite{forkea21} for a statement of the results.} of Assiotis and Keating, they prove that
\begin{equation}
  f_{N,\lambda, \nu}(\theta)\coloneqq |P_{G(N)}(A,\theta)|^{2\lambda}e^{2i\nu\IM(\log P_{G(N)}(A,\theta))},
\end{equation}
suitably normalized, converges to a $\GMC$ measure $\mu_{\lambda,\nu}$.  Here, $G(N)\in{\operatorname{O}(N), \Sp(2N)}$.

Their parameter range corresponds to the $L^2-$phase (cf.~\eqref{eq:gmc_phase_l2}).  If the symmetry points of the circle are excluded, then they allow $\lambda>-1/4$, $\lambda^2-\nu^2<1/2$.  Using the result of Claeys et al.~\cite{cgmy20}, Forkel and Keating can extend to the entire circle at the expense of a reduced parameter set:  $\lambda^2-\nu^2<1/2$, $\lambda\geq 0$, $4\lambda^2<1$. For an application \`a la~\eqref{remy_conv_to_gmcB}, one would like $\nu=0$ and at least $\lambda\in(0,1)$.  A natural next step would be to extend to the $L^1-$phase, as Nikula et al.~\cite{niksakweb20} did in the unitary case.

Finally, we discuss recent work of Najnudel, Paquette, and Simm~\cite{najpaqsim20}. It is now well-established that characteristic polynomials can be normalized to converge to $\GMC$ measures.  Let us briefly set the notation for the \CBE{} characteristic polynomial
\begin{equation}
  P^\beta_N(z)=\prod_{j=1}^N(1-ze^{-i\theta_j}),
\end{equation}
where $e^{i\theta_1},\dots,e^{i\theta_N}$ distributed according to the \CBE{} measure~\eqref{eq:cbe}. If $\beta=2$ then $P^2_N$ coincides with $P_{U(N)}$. Then we recall that the secular coefficients $\Sc_n(N)$ (cf.~\eqref{def:seccoeff}) of $P^\beta_N$ are exactly the polynomial coefficients 
\begin{equation}
  P^\beta_N(z)=\sum_{n=0}^N \Sc_n(N) z^n.
\end{equation}
Clearly $\Sc_n(N)$ also has a $\beta$-dependence, but we suppress it in favour of a cleaner notation. In~\cite{congam06, krrr18}, unitary secular coefficients are related to various number theoretic averages.  Again for $\beta=2$, moments of secular coefficients have been related to certain combinatorial objects: namely magic squares~\cite{diagam06}. Haake et al.~\cite{hkssz96} determined that $\mathbb{E}[\Sc_n(N)]=0$ and $\mathbb{E}[|\Sc_n(N)|^2]=1$ (where the averages are taken over $U(N)$).  Diaconis and Gamburd extended the analysis to higher moments using the combinatorial approach.  Their result gives the limiting distribution for the $n$th secular coefficient as $N$ grows: it is a polynomial of degree $n$ in independent Gaussian random variables. However, such as statement cannot handle (for example) the middle secular coefficient $\Sc_{\left\lfloor\frac{N}{2}\right\rfloor}(N)$. 

In~\cite{najpaqsim20}, the authors use \emph{Holomorphic multiplicative chaos} to show that in the $\CUE$ the central secular coefficient tends to zero as $N\rightarrow\infty$.
\begin{theorem}[Najnudel, Paquette, Simm~\cite{najpaqsim20}]\label{thm:najpaqsim}
  Set $w_N=\log(1+N)^{-1/4}$.  Then
  \begin{equation}
    \left\{\frac{\Sc_{\left\lfloor\frac{N}{2}\right\rfloor}(N)}{w_N}\right\}_{N\geq 1}\quad\text{and}\quad\left\{\frac{w_N}{\Sc_{\left\lfloor\frac{N}{2}\right\rfloor}(N)}\right\}_{N\geq 1}
  \end{equation}
  are both tight families of random variables.

  If $n_N\rightarrow\infty$ such that $2n_N\leq N$ then $\Sc_{n_N}(N)\overset{d}{\rightarrow}0$ as $N\rightarrow\infty$.
\end{theorem}

The relationship of theorem~\ref{thm:najpaqsim} to this review, beyond being an important result in random matrix theory in its own right, is the introduction of holomorphic multiplicative chaos.  Define on the unit disk $G^\mathbb{C}(z)=\sum_{k=1}^\infty \frac{z^k}{\sqrt{k}}\mathcal{N}_k$, for $\mathcal{N}_k$ independent standard complex normals.  On the circle $|z|=1$, this is the usual log-correlated Gaussian field. Set $\theta\coloneqq\sqrt{2/\beta}$. Then, if $\phi$ is a trigonometric polynomial and for $z$ on the unit disk $z\mapsto\phi(z)$ is the extension to the disk, then
\begin{equation}\label{eq:hmc}
  (\operatorname{HMC}_{\theta},\phi)\coloneqq \lim_{\vareps\rightarrow 1} \frac{1}{2\pi}\int_0^{2\pi}e^{\theta G^\mathbb{C}(\vareps e^{i\gamma})}\overline{\phi(\vareps\gamma)}d\gamma
\end{equation}
is holomorphic multiplicative chaos $\operatorname{HMC}$. Such a random distribution first implicitly appeared in~\cite{chhnaj19}. If $\phi(x)=e^{inx}$ for $n=0,1,2,\dots$, then the limit \eqref{eq:hmc} is simply the $n$th coefficient in expansion of $\exp(\theta G^\mathbb{C}(z))$ at $z=0$. Set $c_n$ to be the $n$th Fourier coefficient of the $\operatorname{HMC}_{\theta}$.  Using a result of Chhaibi and Najnudel~\cite{chhnaj19}, which connects $\GMC$ measures to the \CBE{}, the authors show that $\Sc_n(N)$ almost surely converges to $c_n$, as $N\rightarrow\infty$ (for $\beta>0$). This is shown via the help of Verblunsky coefficients, cf.~\eqref{eq:orthogonal_polynomials}. Hence, one can work directly with the $\operatorname{HMC}_{\theta}$ and its `secular coefficients' $c_n$ (which usefully have an approximate Martingale structure). 

Najnudel et al.\ also prove interesting results regarding the limiting distributions of the secular coefficients in the case $\beta>4$, freezing-style properties associated with the $\operatorname{HMC}_{\theta}$, as well showing that the combinatorial magic-square results of Diaconis and Gamburd~\cite{diagam06} extend to the $\beta>0$ regime.  We direct the interested reader to their paper~\cite{najpaqsim20}.  

\subsubsection*{Modelling families of $L$-functions}
One reason for specializing to $\SO(2N)$ and $\Sp(2N)$ is that these groups, along with $\U(N)$, are conjectured to model various number theoretic families~\cite{keasna00a, keasna00b, katsar99a, katsar99b, confar00, cfkrs05}.  The canonical example, as demonstrated throughout section~\ref{sec:fk_conj}, is using unitary characteristic polynomials to model $\zeta(1/2+it)$.  Explicitly, the set
\begin{equation}
  \{\zeta(\tfrac{1}{2}+it)|t>0\}
\end{equation}
is said to display \emph{unitary} symmetries in that, for example, the moments
\begin{equation}
  \frac{1}{T}\int_T^{2T}|\zeta(\tfrac{1}{2}+it)|^{2\beta}dt
\end{equation}
are well-modelled by averages over $A\in \U(N)$ of $\det(I-Ae^{-i\theta})$.

Examples of $L$-functions with symplectic and orthogonal symmetries are as follows. Recall that a quadratic Dirichlet $L$-function (cf.~\eqref{dirichlet_l_function}) is
\begin{equation}\label{dirichlet_l_function2}
  L(s,\chi_d)\coloneqq \sum_{n=1}^\infty\frac{\chi_d(n)}{n^s}, 
\end{equation}
for some real, quadratic, Dirichlet character $\chi_d$.  For a fixed fundamental discriminant $d$, the statistics of the non-trivial zeros of $L(s,\chi_d)$ high up the critical line again display \emph{unitary} symmetries and hence form a unitary family,
\begin{equation}
  \{L(\tfrac{1}{2}+it,\chi_d)|\; d\text{ a fixed, fundamental discriminant}, t\geq 0\},
\end{equation}
see Rudnick and Sarnak~\cite{rudsar96}.  However, one can instead take the value of $L(s,\chi_d)$ at a fixed symmetry point $s=1/2$ and average over $d$.  One then should recover \emph{symplectic} symmetries (see for example~\cite{confar00, keasna00b}).  To this end, define the family
\begin{equation}\label{ex:symplectic_family}
  \{L\left(\tfrac{1}{2}, \chi_d\right) |\ d\text{ a fundamental discriminant, } \chi_d(n)=\left(\tfrac{d}{n}\right)\}.
\end{equation}
This is an example of a \emph{symplectic family}. To demonstrate this, one can turn to moments. It is conjectured that (see for example~\cite{confar00})
\begin{equation}\label{conj:dirichlet_l_function}
  \frac{1}{D^*}\sideset{}{^*}\sum_{|d|\leq D}L(\tfrac{1}{2},\chi_d)^{2\beta}\sim c_{L_D}(\beta)a_{L_D}(\beta)\left(\log D\right)^{\frac{2\beta(2\beta+1)}{2}},
\end{equation}
as $D\rightarrow\infty$. The sum is only over fundamental discriminants $d$; $D^*$ is the length of the sum; and $a_{L_D}(\beta)$ has a similar form to \eqref{zeta_arithmetic} (see Conrey et al.~\cite{cfkrs05} for example for the full formulation). The conjecture is based on work of Jutila~\cite{jut81} and Soundararajan~\cite{sou00}, and so the values of $c_{L_D}(\beta)$ are known for $\beta=1,2,3$ and conjectured for $\beta=4$.

Keating and Snaith~\cite{keasna00b} determined the asymptotic behaviour of the $2\beta$th moments of symplectic characteristic polynomials at the symmetry point,
\begin{equation}
\int_{\Sp(2N)}|P_{\Sp(2N)}(A,0)|^{2\beta}dA\sim c_{Sp}(\beta)N^{\frac{2\beta(2\beta+1)}{2}},
\end{equation}
where $c_{Sp}(\beta)$ is the leading order moment coefficient depending on $\beta$,
\begin{equation}\label{eq:sympl_moment_coeff}
  c^{Sp}(\beta)=\frac{1}{\prod_{j=1}^{2\beta} (2j-1)!!}.
\end{equation}
Unlike $\U(N)$, one no longer has rotational invariance of the measure.  The `symplectic' behaviour is captured at the symmetry points: $\pm 1$. See~\cite{keaodg08} for an asymptotic description of the moments (also for the orthogonal case) for a fixed point away from $\pm 1$, where there is a transition back to unitary statistics on the scale of mean eigenvalue spacing. 

As with the comparison between $\zeta(s)$ and unitary polynomials, one associates $N$, the matrix size, with the logarithm of the `height' of the family: $N\sim\log D$.  With this dictionary in place, the symplectic form of \eqref{conj:dirichlet_l_function} is evident and suggests that $c_{L_D}(\beta)=c_{Sp}(\beta)$.  All known values of $c_{L_D}(\beta)$ indeed satisfy such a relation. 

Finally we give an example of an orthogonal family.  In this case, one considers two categories: even and odd\footnote{The parity comes from the sign of the functional equation for the $L$-function, see for example~\cite{cfkrs05}.}.  The even families are related to the matrices from $\SO(2N)$, and the odd to those from $\SO(2N+1)$.  However, since the `odd' $L$-functions take the value $0$ at their symmetry point\footnote{This is a consequence of their functional equation, see for example~\cite{cfkrs05}.}, we only consider `even' families.   In general, the $L$-functions displaying orthogonal symmetries are  more complicated than the Dirichlet $L$-functions, see~\eqref{dirichlet_l_function}.  Arguably the simplest example is derived from $L$-functions attached to elliptic curves.  Recall these were defined by \eqref{elliptic_euler_prod}.

In fact, we are interested in certain twists by a Dirichlet character $\chi_d(n)$.  Define
\begin{equation}
  L_E(s,\chi_d)\coloneqq\sum_{n=1}^\infty \frac{a_n \chi_d(n)}{n^s}
\end{equation}
for $\RE(s)>1$ and by analytic continuation otherwise. The coefficients $a_n$ are the weightings from the $L$-function associated to the elliptic curve $E$, cf.~\eqref{elliptic_euler_prod}.  Hence, fix $E$ an elliptic curve over $\mathbb{Q}$ such that the sign of its functional equation is $+1$ (i.e. provided the $L$-function is even).  Then, one can define the family
\begin{equation}\label{ex:orthogonal_family}
\{L_E(1, \chi_d) |\ d\text{ a fundamental discriminant, } \chi_d(n)=\left(\tfrac{d}{n}\right)\}.
\end{equation}
This is an example of an \emph{orthogonal family}. The corresponding Riemann hypothesis for $L_E(s,\chi_d)$ places the critical line at $\RE(s)=1$, rather than the conventional $1/2$.  This is why the $L$-function is evaluated at $1$ in \eqref{ex:orthogonal_family}. However, this difference is merely due to the conventional normalization one chooses when defining elliptic curve $L$-functions (cf.~\eqref{elliptic_euler_prod}); it is simple to redefine $L_E$ so that its critical line is shifted to $\RE(s)=1/2$.  

It is conjectured that (see for example~\cite{confar00})
\begin{equation}\label{conj:elliptic_l_function}
  \frac{1}{D^*}\sideset{}{^*}\sum_{|d|\leq D}L_E(1,\chi_d)^{2\beta}\sim c_{L_E}(\beta)a_{L_E}(\beta)\left(\log D\right)^{\frac{2\beta(2\beta-1)}{2}},
\end{equation}
as $D\rightarrow\infty$. The sum is only over fundamental discriminants $d$; $D^*$ is the length of the sum; and $a_{L_E}(\beta)$ again has a similar form to \eqref{zeta_arithmetic} (see Conrey et al.~\cite{confarzir05}). Once again, small moments, and hence small values of $c_{L_E}$, have been computed, see~\cite{confar00}.  

Keating and Snaith showed that the asymptotic behaviour of the $2\beta$th moments of special orthogonal characteristic polynomials at the symmetry point is asymptotically
\begin{equation}
\int_{\SO(2N)}|P_{\SO(2N)}(A,0)|^{2\beta}dA\sim c_{SO}(\beta)N^{\frac{2\beta(2\beta-1)}{2}},
\end{equation}
where $c_{SO}(\beta)$ is the leading order moment coefficient depending on $\beta$,
\begin{equation}\label{eq:ortho_moment_coeff}
  c_{SO}(\beta)= \frac{2^{2\beta}}{\prod_{j=1}^{2\beta-1} (2j-1)!!}.
\end{equation}
As with the symplectic case, one no longer has rotational invariance of the orthogonal measure compared to unitary moments. 

Like in the case of Dirichlet $L$-functions, by comparing $N$ with $\log D$, the orthogonal symmetry is evident.  Additionally, this furnishes the conjecture that $c_{L_E}(\beta)$ is equal to $c_{SO}(\beta)$~\cite{keasna00b}.

Given these families, one can define the associated moments of moments.  For~\eqref{ex:symplectic_family} one has
\begin{equation}\label{mom_symplectic}
  \mom_{L_{\chi_d}}(k,\beta)\coloneqq\frac{1}{D^*}\sideset{}{^*}\sum_{|d|\leq D}\left(\int_{0}^{1}L(\tfrac{1}{2}+it,\chi_d)^{2\beta} dt\right)^k,
\end{equation}
where the sum is only over fundamental discriminants $d$, and $D^*$ denotes the number of terms in the sum. We would expect that the asymptotic behaviour of \eqref{mom_symplectic} should be given by theorem~\ref{thm:mom_symplectic}.

For~\eqref{ex:orthogonal_family}, the moments of moments are 
\begin{equation}\label{mom_orthogonal}
  \mom_{L_{E}}(k,\beta)\coloneqq\frac{1}{D^*}\sideset{}{^*}\sum_{|d|\leq D}\left(\int_{0}^{1}L_E(1+it,\chi_d)^{2\beta} dt\right)^k,
\end{equation}
where again the sum is only over fundamental discriminants $d$, $D^*$ denotes the number of terms in the sum, and $1$ is the symmetry point for $L_E$. Theorem~\ref{thm:mom_orthogonal} should determine the asymptotic behaviour of \eqref{mom_orthogonal}. 

We now turn to moments of moments of $\zeta(1/2+it)$. For $T>0$ and $\RE(\beta)>-1/2$
\begin{align}\label{def:mom_zeta}
  \mom_{\zeta_T}(k,\beta)\coloneqq \frac{1}{T}\int_0^T\left(\int_{t}^{t+1}|\zeta(\tfrac{1}{2}+ih)|^{2\beta}dh\right)^kdt.
\end{align}
Choosing to integrate over intervals of length $1$ in the integrand of \eqref{def:mom_zeta} may be easily extended to any interval of $O(1)$, with the appropriate normalization.  Harper~\cite{har19b} determined an asymptotic upper bound on \eqref{def:mom_zeta} for a particular parameter range.

\begin{theorem}[Harper~\cite{har19b}]
Let $k\in[0,1]$.  Then, uniformly for large $T$
\begin{equation}\label{eq:mom_harper}
  \frac{1}{T}\int_T^{2T}\left(\int_{t-1/2}^{t+1/2}|\zeta(\tfrac{1}{2}+ih)|^2dh\right)^kdt\ll \left(\frac{\log T}{1+(1-k)\sqrt{\log\log T}}\right)^k.
\end{equation}
\end{theorem}
Harper argues that this upper bound is likely to be best possible. Notice that for $k\in(0,1)$, the growth is slower than one would conjecture using~\eqref{eq:fk_mom}.  It would be interesting to prove if the equivalent result holds on the random matrix side. 

When $k\in\mathbb{N}$, one may rewrite \eqref{def:mom_zeta} as
\begin{equation}\label{def:mom_zeta_integer}
   \mom_{\zeta_T}(k,\beta)=\frac{1}{T}\int_{0}^{1}\cdots\int_{0}^{1}\int_0^T\prod_{j=1}^k|\zeta(\tfrac{1}{2}+i(t+h_j))|^{2\beta}dh_1\cdots dh_kdt.
\end{equation}
Conrey et al.~\cite{cfkrs05} provide the following conjectural form\footnote{The conjecture in~\cite{cfkrs05} is more general in that it permits more general families of $L$-functions, non-uniform shifts, as well as integrating against different weight functions.}.
\begin{conj}[Conrey et al.~\cite{cfkrs05}]\label{conj:cfkrs}
  Take $k, \beta\in\mathbb{N}$, and $\underline{h}=(h_1,\dots,h_k)$ with $h_j\in\mathbb{R}$.  Then
  \begin{equation}\label{eq:cfkrs}
    \frac{1}{T}\int_{0}^T\prod_{j=1}^k|\zeta(\tfrac{1}{2}+i(t+h_j))|^{2\beta}dt=\frac{1}{T}\int_0^TP_{k,\beta}(\log\tfrac{t}{2\pi},\underline{h})dt+O(T^{-\delta}),
  \end{equation}
  for some $\delta>0$, where 
  \begin{equation}\label{def:p}
    P_{k,\beta}(x;\underline{h})\coloneqq\frac{(-1)^{k\beta}}{(k\beta)!^2}\frac{1}{(2\pi i)^{2k\beta}}\oint\cdots\oint\frac{G(z_1,\dots,z_{2k\beta})\Delta(z_1,\dots,z_{2k\beta})^2}{e^{\frac{x}{2}\sum_{j=1}^{k\beta}z_{k\beta+j}-z_j}\prod_{j=1}^{2k\beta}\prod_{l=1}^{k}(z_j-ih_l)^{2\beta}}dz_1\cdots dz_{2k\beta}
  \end{equation}
  in which $\Delta(x_1,\dots,x_n)=\prod_{i<j}(x_j-x_i)$ is the Vandermonde determinant, and the contours are small circles surrounding the poles.  Additionally,
  \begin{equation}\label{def:g_func}
    G(z_1,\dots,z_{2k\beta})\coloneqq A_{k\beta}(z_1,\dots,z_{2k\beta})\prod_{1\leq i\leq k\beta < j\leq 2k\beta}\zeta(1+z_i-z_j)
  \end{equation}
  which includes the Euler product
  \begin{equation}\label{def_A}
    A_{k\beta}(\underline{z})\coloneqq\prod_p \prod_{1\leq l\leq k\beta < m\leq 2k\beta} (1-p^{z_m-z_l-1})\int_0^1\prod_{j=1}^{k\beta}\left(1-\frac{e^{2\pi i\theta}}{p^{\frac{1}{2}+z_j}}\right)^{-1}\left(1-\frac{e^{-2\pi i\theta}}{p^{\frac{1}{2}-z_{k\beta+j}}}\right)^{-1}d\theta. 
  \end{equation}
\end{conj}

In light of this, define 
\begin{equation}\label{mom_p_def}
  \mom_{P_{k,\beta}}(T)\coloneqq  \frac{1}{T}\int_{0}^{1}\cdots\int_{0}^{1}\int_0^TP_{k,\beta}(\log\tfrac{t}{2\pi},\underline{h})dtdh_1\cdots dh_k.
\end{equation}
Under conjecture~\ref{conj:cfkrs}, $\mom_{P_{k,\beta}}(T)$ should approximate $\mom_{\zeta_T}(k,\beta)$ up to a power saving in $T$. One may note that the structure of \eqref{def:p} is very similar to the multiple contour integral appearing in lemma~\ref{lemma:mci}, used to prove theorem~\ref{thm:mom_unitary}.  The main result of~\cite{baikea21a} is the following. 
\begin{theorem}[Bailey and Keating~\cite{baikea21a}]\label{thm:mom_zeta}
  For $k,\beta\in\mathbb{N}$, as $T\rightarrow\infty$
  \begin{equation}
    \mom_{P_{k,\beta}}(T)=\alpha_{k,\beta}\gamma_{k,\beta}\left(\log\tfrac{T}{2\pi}\right)^{k^2\beta^2-k+1}(1+O\left(\log^{-1}T\right)),
  \end{equation}
  where the term $\alpha_{k,\beta}=A_{k\beta}(0,\dots,0)$ contains the arithmetic information and $A_{k\beta}$ is as defined in \eqref{def_A}, and $\gamma_{k,\beta}$ denotes the remaining, non-arithmetic, contribution (for further details see~\cite{baikea21a}).
\end{theorem}

The proof of theorem~\ref{thm:mom_zeta} follows by shifting the contours of \eqref{def:p} and then applying complex analytic techniques to determine the correct asymptotic behaviour in $\log T$. It is also evident from the proof that the result will still hold (albeit with a simple modification to the leading order coefficient) if the ranges of the inner integrals of \eqref{mom_p_def} are over any $O(1)$ interval. Additionally, since conjecture~\ref{conj:cfkrs} generalizes to other $L$-functions, the proof techniques of~\cite{baikea21b} would additionally apply more generally. 

Finally, the theory of moments of moments is not confined to just studying the classical compact matrix groups.  There are natural extensions to other matrix ensembles and here we mention some related results.

For Hermitian ensembles, Jonnadula, Keating, and Mezzadri~\cite{jonkeamez21a, jonkeamez21b} proved a version of proposition~\ref{prop:bumgam} using multivariate orthogonal polynomials. They subsequently calculated the asymptotics of the moments of the associated characteristic polynomials, and , in the $\GUE$ case, identified a subtle dependence on the parity of the matrix size. Bourgade, Mody, and Pain~\cite{boumodpai21} have recently established a local law and central limit theorem for general $\beta$-ensembles using loop-equations.  These are akin to the Gaussian $\beta$-ensembles, but with a generic potential. There a log-correlated field also naturally arises.  Similarly, if one considers Wigner matrices\footnote{(Real) Wigner matrices generalize the $\GOE$; they are $N\times N$ random Hermitian matrices such that the entries above the diagonal are independent centred random variables some fixed common variance.}, then in~\cite{boumod19} Bourgade and Mody show that in the bulk the characteristic polynomial\footnote{This result holds for both the real and imaginary part.} also has logarithmic correlations.  Originally this result was conditional on the equivalent result holding for the $\GOE$ (resp. $\GUE$ for complex Wigner), but following~\cite{boumodpai21} now holds unconditionally.

Number theoretically, one is also not restricted to studying averages of $L$-functions as evidenced by the recent work of de la Bret\`eche and Fiorilli~\cite{brefio20} on moments of moments of primes in arithmetic progressions. 

\subsection{Explicit examples}\label{sec:examples}

In~\cite{baikea19}, explicit examples of the polynomials $\mom_{\U(N)}(k,\beta)$ were computed for small, integer values of $k$ and $\beta$.  There are numerous methods available for computing moments of moments, as outlined previously in this section.  Specifically, one can compute the first couple of moments by hand using either the combinatorial Schur function approach\footnote{Equivalently, the Gelfand-Tsetlin pattern approach of~\cite{asskea20, assbaikea20}.} or via the multiple contour integral \eqref{lemma:mci}.  However, as soon as $k, \beta$ increase (and in particular, the $k$ parameter) this technique becomes infeasible.

Further progress is possible using the ratios theorem of Conrey, Farmer and Zirnbauer~\cite{confarzir05, confarzir08} with the aid of computer algebra software. Keating and Scott~\cite{keasco15} produced initial code based on the ratios conjecture which produced the first four novel\footnote{Recall that the Keating-Snaith result, theorem~\ref{thm:ks}, gives a full description of $\mom_{\U(N)}(1,\beta)$.} unitary moments of moments.  However, this technique also proves to be inefficient, especially as $k$ increases.

The method used to compute the polynomials presented here was instead the Toeplitz determinant representation\footnote{We thank Chris Hughes for first highlighting this approach to us.}. By~\eqref{eq:heine_szego}, one can write~\eqref{eq:k-fold_ave} as a Toeplitz determinant $D_N(f)$ for a particular symbol $f$, depending on $\theta_1,\dots,\theta_k$. After computing the circle integrals of $D_N(f)$, we recover the moments of moments polynomials.  A selection are given here, for a more complete list see~\cite{baikea19}. 
\begin{align*}
  \mom_{\U(N)}(1,1)&=N+1\\
  \mom_{\U(N)}(2,1)&=\frac{1}{6}(N+3)(N+2)(N+1)\\
  \mom_{\U(N)}(1,2)&=\frac{1}{12}(N+1)(N+2)^2(N+3)\\
  \mom_{\U(N)}(2,2)&=\frac{1}{163459296000}(N+7)(N+6)(N+5)(N+4)(N+3)(N+2)(N+1)\\
  &\quad\times(298N^8+9536N^7+134071N^6+1081640N^5+5494237 N^4+18102224N^3\\
  &\qquad+38466354N^2+50225040N+32432400).
\end{align*}

Similar techniques can be adapted to calculate moments of moments for the symplectic and special orthogonal groups (see \eqref{eq:mom_symplectic} and \eqref{eq:mom_orthogonal}), and such computations can be found in~\cite{assbaikea20}.

\section{Acknowledgements}\label{sec:acknowledgements}

ECB is grateful to the Heilbronn Institute for Mathematical Research for support.  JPK is pleased to acknowledge support from ERC Advanced Grant 740900 (LogCorRM). We would like to thank the anonymous reviewers for their helpful suggestions and questions. 


\end{document}